\documentclass[twocolumn,superscriptaddress,showpacs,nofootinbib]{revtex4}

\usepackage{bm,color}

\usepackage{graphicx}
\usepackage{amsmath}
\usepackage{amssymb}
\usepackage{amsfonts}
\usepackage{bm}
\usepackage{color}

\def\be{\begin{equation}}
\def\ee{\end{equation}}
\def\bea{\begin{eqnarray}}
\def\eea{\end{eqnarray}}

\begin{document}

\title{K-essence Lagrangians of polytropic and
logotropic unified dark matter \\
and dark energy models}

\author{Pierre-Henri Chavanis}
\email{chavanis@irsamc.ups-tlse.fr}
\affiliation{Laboratoire de Physique
Th\'eorique, Universit\'e de Toulouse,
CNRS, UPS, France}

\begin{abstract}

We determine the k-essence Lagrangian of a relativistic barotropic fluid. The
equation of state of the fluid can be specified in different manners depending
on whether the pressure is expressed in terms of the energy density (model I),
the rest-mass density (model II), or the pseudo
rest-mass density for a complex scalar field in the Thomas-Fermi approximation
(model
III).
In the nonrelativistic limit, these three formulations coincide. In the
relativistic regime, they lead to different models that we study exhaustively.
We provide
general results valid for an arbitrary equation of state and show how the
different models are connected to each other. For illustration, we specifically
consider polytropic and logotropic dark fluids that  have been proposed as
unified
dark matter and dark energy models.  We recover the
Born-Infeld action of the
Chaplygin gas in models I and III and obtain the explicit expression of the
reduced action of the logotropic dark fluid in models II and III. We also derive
the two-fluid representation of the Chaplygin and logotropic models. Our general
formalism can be applied to many other situations such as
Bose-Einstein condensates with a $|\varphi|^4$ (or more general)
self-interaction, dark matter superfluids, and mixed models.

\end{abstract}

\pacs{95.30.Sf, 95.35.+d, 95.36.+x, 98.62.Gq,
98.80.-k}


\maketitle


\section{Introduction}
\label{sec_intro}

Baryonic matter constitutes only $5\%$ of the content of the universe today. The rest
of the universe is made of approximately $25\%$ dark matter (DM) and  $70\%$
dark energy
(DE) \cite{planck2014,planck2016}.
DM can explain the flat rotation curves of the
spiral galaxies. It is also necessary to form the large-scale structures of the
universe. DE does not cluster but is responsible for the
late time acceleration of the universe revealed by the observations of type Ia
supernovae, the cosmic microwave background (CMB) anisotropies, and galaxy
clustering. Although there have been many theoretical attempts to explain DM and
DE, we still do not have a robust model for these dark components that can pass all the
theoretical and observational tests.

The most natural and simplest model
is the $\Lambda$CDM model which treats DM as a nonrelativistic cold pressureless
gas and DE as a cosmological constant $\Lambda$ possibly representing vacuum energy \cite{carroll,ss}. The effect of the cosmological constant is equivalent to that of a
fluid with a constant energy density $\epsilon_\Lambda=\Lambda c^2/8\pi G$ and a
negative pressure $P_\Lambda=-\epsilon_\Lambda$. Therefore, the $\Lambda$CDM
model 
is a two-fluid
model comprising  DM with an equation of state $P_{\rm dm}=0$ and DE with an equation of
state $P_{\rm de}=-\epsilon_{\rm de}$. When combined with the energy conservation equation, the equation of state $P_{\rm dm}=0$ implies that the DM density
decreases with the scale factor as $\epsilon_{\rm dm} \propto a^{-3}$ and the
equation of state
$P_{\rm de}=-\epsilon_{\rm de}$ implies that the DE density is constant:
$\epsilon_{\rm de}=\epsilon_\Lambda$.
Therefore, the total energy density of the universe (DM $+$ DE) evolves as
\begin{equation}
\epsilon=\frac{\epsilon_{\rm dm,0}}{a^3}+\epsilon_\Lambda,
\label{intro1}
\end{equation}
where $\epsilon_{\rm dm,0}$ is the present density of DM.\footnote{For
simplicity of presentation we ignore the contribution of baryonic matter in the
present discussion. We 
also assume that the universe is spatially flat in agreement with the inflation
paradigm \cite{guthinflation} and the measurements of the CMB
anisotropies \cite{planck2014,planck2016}.} DM dominates at early times when
the density is high and DE dominates at late times when the density is low. The
scale factor increases algebraically as $a\propto t^{2/3}$ during the DM era
(Einstein-de Sitter regime) and exponentially as $a\propto
e^{\sqrt{\Lambda/3}t}$ during the DE era (de Sitter regime). At the present
epoch, both components are important in the energy budget of the universe.

Although the $\Lambda$CDM model is perfectly consistent with current
cosmological  observations, it faces two main problems. The first
problem is to explain the tiny value of the cosmological constant 
$\Lambda=1.00\times 10^{-35}\, {\rm s}^{-2}$.
Indeed, if DE can be attributed to vacuum fluctuations, quantum field theory
predicts that $\Lambda$ should correspond to the Planck scale which lies $123$
orders of
magnitude above the observed value.  This is called the cosmological constant
problem \cite{weinbergcosmo,paddycosmo}. The second problem is to explain why DM and DE are of similar
magnitudes today although they scale differently with the universe's
expansion. This is the cosmic coincidence problem \cite{stein1,zws},
frequently triggering anthropic explanations. The CDM model also faces important
problems at the scale of DM halos such as the core-cusp problem
\cite{moore}, the missing satellite problem
\cite{satellites1,satellites2,satellites3}, and the ``too big to fail'' problem
\cite{boylan}. This leads to the so-called small-scale crisis of CDM 
\cite{crisis}.

For these reasons, other types of matter with negative pressure  that
can behave like a cosmological constant at late time have
been considered as candidates of DE: fluids of topological defects (domain
walls,
cosmic strings) \cite{vs,kt,fm,fg,pen}, $x$-fluids with a linear equation of
state \cite{tw,sw,csn}, quintessence -- an evolving
self-interacting scalar field (SF) minimally coupled to gravity --
\cite{cds} (see earlier
works in \cite{pr,ratra,wett1,wett2,fhsw,cdf,fj,clw,fj2}),\footnote{SFs
have been
used in a variety of inflationary models  \cite{linde}  to describe the transition from the exponential
(de Sitter) expansion of the early universe to a decelerated expansion. It was
therefore natural to try to understand the present acceleration of the
universe, which has an exponential behaviour too, in terms of SFs
\cite{star,srss}. However, one has
to deal now with the opposite situation, i.e., describing the transition from a
decelerated expansion to an exponential (de Sitter) expansion.}
k-essence fields -- SFs with  a noncanonical kinetic term \cite{chiba,ams,ams2} 
that 
were initially introduced to describe inflation (k-inflation) \cite{adm,gm} --
and
even phantom or ghost fields \cite{caldwell,ckw} which predict
that the energy density of
the universe may ultimately increase with time.
Quintessence can
be viewed as a dynamical vacuum energy following the old idea that the
cosmological term could evolve \cite{bronstein,berto1,berto2,taha}.  However,
these models still face the cosmic
coincidence problem\footnote{A class of k-essence models  \cite{ams,ams2} has
been claimed to solve the coincidence problem by linking the onset of DE
domination to the epoch of DM domination.} because they treat DM and
DE as distinct entities. Accounting for similar magnitude of DM and DE today
requires very particular (fine-tuned) initial conditions. For some kind of
potential terms, which have their justification in supergravity
\cite{braxmartin}, this problem can be solved by the so-called tracking solution
\cite{zws,swz}. The
self-interacting SF evolves in such a way
that it approaches a cosmological constant behaviour exactly today \cite{braxmartin}.
However, this is  achieved  at the expense of
fine-tuning the potential parameters. This unsatisfactory state of affairs motivated a search for
further alternatives.

In the standard $\Lambda$CDM model and in quintessence CDM models, DM and DE 
are two distinct entities introduced to explain the clustering of matter and the
cosmic acceleration, respectively. However, DM and DE could be two different
manifestations of a common structure, a dark fluid. In this respect,
Kamenshchik {\it et al.}
\cite{kmp}  have
 proposed a simple unification of DM and DE
in the form of a perfect fluid with an exotic equation of state known 
as the Chaplygin gas, for which
\begin{equation}
P=-\frac{A}{\epsilon/c^2},
\label{intro2}
\end{equation}
where  $A$ is a positive constant. This gas exhibits a
negative pressure, as required to explain the acceleration of
the universe today, but the squared speed of sound is positive ($c_s^2=P'(\epsilon)c^2=Ac^4/\epsilon^2>0$). This is a very important property because many fluids with negative pressure obeying a barotropic equation of state suffer from
instabilities at small scales due to an imaginary speed of sound
\cite{fm,fg}.\footnote{To explain
the accelerated expansion taking place today, the universe must be dominated by a fluid of negative
pressure violating the strong energy condition, i.e.,
$P<-\epsilon/3$. For a fluid
with a linear equation of state $P=\omega\epsilon$, where $\omega$ is a
constant, we need $\omega<-1/3$ to have an acceleration. But, in that case,
$c_s^2=\omega c^2<0$, yielding instabilities at small scales \cite{fm}. This is
the usual problem for a fluid description of domain walls
($\omega=-2\epsilon/3$) and cosmic strings ($\omega=-\epsilon/3$). Quintessence
models with a
standard kinetic term do not have this problem because the speed of sound is
equal 
to the speed of light \cite{grishchuk}. K-essence models \cite{chiba,ams,ams2}
with a nonstandard kinetic term
are different in this respect \cite{gm}, but they still have a positive squared
speed of sound  (note that $c_s$ can exceed the speed of light).} This is not
the case for the Chaplygin gas.

Integrating the energy
conservation equation with the
Chaplygin equation of state (\ref{intro2}) leads to
\begin{equation}
\epsilon/c^2= \sqrt{\frac{A}{c^2} + \frac{B}{a^{6}} },
\label{intro3}
\end{equation}
where $B$ is an integration constant. Therefore, the Chaplygin gas smoothly
interpolates between pressureless DM ($P\simeq 0$, $\epsilon \sim a^{-3}$,
$c_s\simeq 0$) at high redshift and a cosmological constant ($P=-\epsilon$,
$\epsilon \rightarrow \sqrt{Ac^2}$, $c_s\rightarrow c$) as $a$ tends 
to infinity. There is also an intermediate phase which can be described by a
cosmological constant mixed with a stiff matter fluid ($P\sim \epsilon$,
$\epsilon\sim a^{-6}$, $c_s\sim c$) \cite{kmp}. In the Chaplygin gas model, DM
and DE
are different manifestations of a single underlying substance (dark fluid) that
is called  ``quartessence'' \cite{makler}.  These models where the fluid
behaves as DM at early times and as DE at late times are also called unified
models for DM and DE (UDM) 
\cite{makler}. This
dual
behavior avoids fine-tuning problems since  the Chaplygin gas model can be
interpreted as an
entangled mixture of DM and DE. In this cosmological context, Kamenshchik {\it
et al.} \cite{kmp}  
introduced a real SF representation of the Chaplygin gas and
determined its potential $V(\varphi)$ explicitly.

The Chaplygin gas model has an interesting history that we briefly sketch below.
Chaplygin \cite{chaplygin} introduced his equation of state $P=-A/\rho$ in 1904 as a convenient
soluble model to
study the lifting force on a plane wing in aerodynamics.  The same model was
rediscovered later by Tsien \cite{tsien} and von Karman \cite{karman}. It was
also realized that certain deformable solids can be
described by the Chaplygin equation of state \cite{stan}. The
integrability of the corresponding Euler equations resides in the fact that
they have a large symmetry group (see \cite{bj,jp,jackiw,hh} for a modern
description). Indeed, the Chaplygin gas model possesses further
space-time symmetries beyond those
of the Galileo group \cite{bj}. In addition, the
Chaplygin gas is the
only fluid which admits a supersymmetric
generalization \cite{hoppesuper,bsuper,jpsuper,bjsuper,hassaine}. The Chaplygin
equation of state involves a
negative pressure  which is required
to account for the accelerated expansion of the universe.\footnote{Negative pressures
arise in different domains of physics such as exchange forces in atoms, stripe
states in the quantum Hall effect, Bose-Einstein condensates with an attractive
self-interaction etc.} It is possible to develop a Lagrangian description of the
nonrelativistic Chaplygin gas \cite{bj,jp,jackiw,jevicki,ogawa,bh} leading to an
action of the form 
\begin{equation}
\label{chintro}
{\cal L}_{\rm Chap}=-(2A)^{1/2}\sqrt{
\dot\theta+\frac{1}{2}(\nabla
\theta)^2},
\end{equation}
where $\theta$ is the potential of the flow.
The relativistic generalization of the Chaplygin gas model leads to a
Born-Infeld-type \cite{bi}  theory for a real SF
\cite{jp,jackiw,ogawa,hoppe,btv}. The Born-Infeld action
\begin{equation}
\label{biintro}
{\cal L}_{\rm BI}=-(Ac^2)^{1/2}\sqrt{1-\frac{1}{c^2}\partial_{\mu}\theta
\partial^{\mu}\theta}
\end{equation}
possesses
additional symmetries beyond the Lorentz and Poincar\'e invariance and
has an interesting connection with string/M theory \cite{polchinski}.
The Chaplygin gas model can
be motivated by a brane-world interpretation (see \cite{rubakov} for a 
review on brane world models).
Indeed, the ``hidden'' symmetries and the associated transformation laws
for the
Chaplygin and Born-Infeld models may be given a coherent setting \cite{jackiw}
by considering
the Nambu-Goto action \cite{ng} for a $d$-brane in $(d+1)$ spatial dimensions
moving in a
$(d+1,1)$-dimensional spacetime.
The Galileo-invariant (nonrelativistic) Chaplygin gas action (\ref{chintro}) is
obtained in the
light-cone
parametrization and the Poincar\'e-invariant (relativistic) Born-Infeld action
(\ref{biintro}) is obtained in the Cartesian
parametrization \cite{jp,jackiw}.\footnote{A $d$-brane is a $d$-dimensional
extended object. For example, a
($d=2$)-brane is a membrane. $d$-branes arise in string theory for the following reason.  Just as the
action of a relativistic point particle is proportional to the world line it
follows, the action of a relativistic string is proportional to the area of the
sheet that it traces by traveling through spacetime. The close
connection between a relativistic  membrane [($d=2$)-brane]  in three spatial dimensions and
planar fluid mechanics  was known to
J. Goldstone (unpublished) and developed by Hoppe and  Bordemann \cite{bh,hoppe}
(the Chaplygin equation of state $P=-A/\rho$ appears explicitly in \cite{bh} and
the Born-Infeld Lagrangian associated with the action of a membrane appears
explicitly in \cite{hoppe}). These results were generalized to arbitrary
$d$-branes by Jackiw and Polychronakos
\cite{jp}. The same Lagrangian appears
as the leading term in Sundrum's \cite{sundrum} effective
field theory approach to large extra dimensions. The Born-Infeld Lagrangian can
be viewed as a k-essence Lagrangian involving a nonstandard kinetic term. The
Chaplygin equation of state is obtained from the
stress-energy tensor $T_{\mu\nu}$ derived from this Lagrangian. Therefore, the
Chaplygin gas
is the hydrodynamical description of a SF with the Born-Infeld Lagrangian. A
more general k-essence model is the string theory inspired tachyon Lagrangian
with a potential $V(\theta)$
\cite{garousi,sen1,sen2,sen3,gibbons,frolov,felder,paddytachyon,pc,
bagla,gkmp,feinstein}. 
It can be shown
that every tachyon condensate model can be
interpreted as a $3+1$ brane moving in a $4+1$ bulk \cite{btvB1,btvB2}. The
Born-Infeld Lagrangian is recovered when
$V(\theta)$ is replaced by a constant \cite{frolov}.
} A fluid with a Chaplygin equation of state is also necessary to stabilize the
branes \cite{rs} in black hole bulks \cite{bhtz,hp}. This is how Kamenshchik {\it
et al.} \cite{kmpBH} came across this fluid and had the idea to consider its
cosmological implications \cite{kmp}. Bilic {\it et al.} \cite{btv} generalized
the Chaplygin gas model in the inhomogeneous case and showed that the real SF
that occurs in the Born-Infeld action  can be
interpreted as the phase of a complex self-interacting SF described by the
Klein-Gordon (KG) equation. This SF may be given a hydrodynamic representation
in terms of an irrotational barotropic flow with the Chaplygin equation of
state. This explains the connection of the Born-Infeld action with fluid
mechanics in the Thomas-Fermi (TF) approximation.  Bilic {\it et al.} \cite{btv}
determined the
potential $V(|\varphi|^2)$ of the complex SF associated with the Chaplygin gas.
This potential is different from the potential $V(\varphi$) of the  real SF
introduced by Kamenshchik {\it et al} \cite{kmp} which is valid for an
homogeneous SF in a cosmological context.

A generalized Chaplygin gas
model (GCG)  has been introduced. It has an equation of
state\footnote{This generalization was mentioned by Kamenshchik {\it et al.}
\cite{kmp}, Bilic {\it et al.} \cite{btv} and Gorini {\it et al.}
\cite{gkmsf}, and was specifically worked
out by Bento {\it et al.}
\cite{bentoGCG}. Equation (\ref{intro5}) can be viewed as a polytropic equation
of
state $P=K(\epsilon/c^2)^{\gamma}$ with a polytropic index $\gamma=-\alpha$ and
a polytropic constant $K=-A$. A further generalization of the GCG model has been
proposed. It has an equation of state
\begin{equation}
P = \omega\epsilon - \frac{A}{(\epsilon/c^2)^{\alpha}},
\label{intro4}
\end{equation}
where $\omega$ is a constant. This is called the Modified Chaplygin Gas (MCG) model \cite{benaoum}. It can be viewed as the sum of a linear
equation of state and a polytropic equation of state. This generalized
polytropic equation of state has been studied in detail in a cosmological
framework in Refs. \cite{cosmopoly1,cosmopoly2,cosmopoly3}. 
The potential $V(\varphi)$ of its real SF representation generalizing the result
of Kamenshchik {\it et al.}
\cite{kmp} has been determined explicitly in these papers.}
\begin{equation}
P = - \frac{A}{(\epsilon/c^2)^{\alpha}}
\label{intro5}
\end{equation}
with $A>0$ and a generic $\alpha$ constant in
the range $0 \le \alpha \le 1$ in order to ensure the condition of stability
$c_s^2\ge 0$ and the 
condition of causality $c_s\le c$ (the quantity $c_s^2/c^2$ goes from $0$ to
$\alpha$ when $a$ goes from $0$ to $+\infty$). Combined with
the energy
conservation equation, we obtain
\begin{equation}
\epsilon/c^2=\left\lbrack
\frac{A}{c^2}+\frac{B}{a^{3(\alpha+1)}}\right\rbrack^{\frac{1}{\alpha+1}}.
\label{intro6}
\end{equation}
This model interpolates between
a universe dominated by dust and de Sitter eras via an intermediate phase described
by a linear equation of state $P\sim \alpha \epsilon$  \cite{kmp,bentoGCG}. The
original Chaplygin gas 
model is recovered for $\alpha=1$.  Bento {\it et al.} \cite{bentoGCG} argued
that the GCG model corresponds to a generalized Nambu-Goto action which can be
interpreted as a
perturbed $d$-brane in a $(d+1, 1)$ spacetime. Bilic {\it et
al.} \cite{bilicNL} mentioned that the generalized Nambu-Goto action  lacks any
geometrical interpretation, but that the generalized Chaplygin equation of state can
be obtained from a moving brane in
Schwarzschild-anti-de-Sitter bulk \cite{neves}.

For $\alpha=0$, the generalized Chaplygin equation of state (\ref{intro5})
reduces 
to a constant negative pressure
\begin{equation}
P=-A.
\label{intro6cst}
\end{equation}
In that case, the  speed of sound
vanishes  identically ($c_s^2=0$). It can
be shown \cite{avelinoZ,sandvik} that this model is equivalent to the $\Lambda$CDM model not only to 0th
order in perturbation theory (background) but to all orders, even in the
nonlinear clustering regime (contrary to the initial claim made in
Ref. \cite{fabrisZ}). Therefore,
the $\Lambda$CDM model can either be considered as a two-fluid model involving
a DM fluid with $P_{\rm
dm}=0$ and a DE fluid with $P_{\rm de}=-\epsilon_{\rm de}$, or as a single dark
fluid with a constant negative pressure $P=-\epsilon_\Lambda$ \cite{cosmopoly2}.
In this sense, it
may be regarded as the simplest UDM model  one can possibly conceive in which DM
and DE appear as different manifestations of a single dark fluid. As a result,
the GCG model includes the original Chaplygin gas model
($\alpha=1$) and the  $\Lambda$CDM model ($\alpha=0$) as particular cases.

The GCG model has been successfully confronted with various
phenomenological tests such as high precision Cosmic Microwave Background
Radiation data \cite{bbs1,bbs2,bbs3,cf,afbc}, type Ia supernova (SNIa) data
\cite{fgsI,daj,makler,ajd,gkmsf,cfgs}, age estimates of
high-$z$ objects \cite{ajd} and
gravitational lensing \cite{sbert}. Although the GCG model is
consistent with
observations related to the background cosmology (the Hubble law is almost
insensitive to $\alpha$) \cite{fgsI,makler,cfgs}, Sandvik {\it et al.} 
\cite{sandvik}  showed that it produces unphysical oscillations or even an
exponential blow-up which are not seen in the observed
matter power spectrum calculated at the present time.  This is caused by
the behaviour of the sound speed through the GCG fluid. At early times, the
GCG behaves as DM and  its sound speed vanishes. In that case, the GCG
clusters like pressureless dust. At late times, when the GCG behaves as DE, its
sound speed becomes relatively large yielding
unphysical features in the matter power spectrum. To avoid such unphysical
features, the value of $\alpha$ must be extremely close to zero
($|\alpha|<10^{-5}$), so that the GCG model becomes indistinguishable from the 
$\Lambda$CDM model. Similar conclusions were reached by Bean and Dor\'e
\cite{bd}, Carturan and Finelli \cite{cf} and Amendola {\it et al.} \cite{afbc}
who
studied  the effect of the GCG on density perturbations and on cosmic
microwave background (CMB) anisotropies and found that the GCG
strongly increases the amount of integrated Sachs-Wolfe effect. Therefore, CMB data are more selective than SN Ia data to constrain $\alpha$. The  GCG is essentially ruled out except for a tiny region
of parameter space very close to the $\Lambda$CDM limit. This conclusion is not restricted to the GCG model but is actually valid for all UDM models.\footnote{If a solution to these problems cannot be provided, this would appear as an
evidence for an independent origin of DM and DE (i.e., they are two distinct
substances) \cite{sandvik}.} Some solutions to this problem have been
suggested (see a short 
review in Sec. XVI of \cite{graal}) but there is no definite consensus at the
present time.

These results show that a viable UDM model should be as close as possible to the $\Lambda$CDM
model, but sufficiently different from it in order  to solve its problems. This
is the basic idea that led us to introduce the 
logotropic model in Ref. \cite{epjp} (see also \cite{lettre,jcap,pdu,logosf}).
The logotropic dark fluid has an equation of state
\begin{equation}
P=A\ln\left (\frac{\rho_m}{\rho_*}\right ),
\label{intro7}
\end{equation}
where $\rho_m$ is the rest-mass density. This equation of state can be obtained
from the polytropic (GCG) equation of state  $P=K\rho_m^{\gamma}$ by considering
the
limit $\gamma\rightarrow 0$ and $K\rightarrow \infty$ with $A=K\gamma$ constant
(see Sec. 3 of \cite{epjp} and Appendix A of \cite{graal}). 
This yields
\begin{equation}
P = Ke^{\gamma\ln\rho_m}\simeq K(1+\gamma\ln\rho_m+...)\simeq K+A\ln\rho,
\label{eq104}
\end{equation}
which is equivalent to Eq. (\ref{intro7}) up to a constant term.\footnote{This
procedure is not well-defined mathematically because it yields an infinite
additional constant $K\rightarrow +\infty$. This constant disappears if we take
the gradient of 
the pressure as in \cite{epjp}. However, in general, an infinite constant term
remains. Therefore, the above procedure simply suggests a connection between the
polytropic and logotropic equations of state, but this connection is rather
subtle.} Since the $\Lambda$CDM
model (viewed as a UDM model) is equivalent to a polytropic gas of index
$\gamma=0$ (constant pressure), one can say that the logotropic model which has
$\gamma\rightarrow 0$  is the simplest extension of the $\Lambda$CDM model.  It
is argued in Ref. \cite{epjp} that $\rho_*=\rho_P={c^5}/{(\hbar G^2)}=5.16\times
10^{99}\, {\rm g/m^3}$ is the Planck density.  It is also argued that 
$A/c^2=B \rho_\Lambda=2.10\times 10^{-26}\,
{\rm g\, m^{-3}}$ (where $B=3.53\times 10^{-3}$ and  $\rho_\Lambda=5.96\times
10^{-24} \, {\rm g\, m^{-3}}$)  is a fundamental constant of physics that
supersedes the cosmological constant $\Lambda$. The logotropic model is able to
account for the
transition between a DM era and a DE era and is indistinguishable from the
$\Lambda$CDM model, for what concerns the evolution of the cosmological
background, up to $25$ billion years in the future  when it becomes phantom
\cite{epjp,lettre,jcap,pdu,logosf}. 
Very interestingly, the logotropic model implies that DM halos should have a
constant surface density  and it predicts its universal value $\Sigma_0^{\rm
th}=0.01955  c\sqrt{\Lambda}/G=133\, M_{\odot}/{\rm pc}^2$
\cite{epjp,lettre,jcap,pdu,logosf} without adjustable parameter. This
theoretical value is in good agreement with the value $\Sigma_0^{\rm
obs}=141_{-52}^{+83}\, M_{\odot}/{\rm pc}^2$ obtained from the observations
\cite{donato}. The logotropic model also predicts the value of the constant
$\Omega_{\rm dm,0}$ that is usually interpreted as the present proportion of DM 
in the Universe: it gives $\Omega_{\rm dm,0}^{\rm th}\simeq {1}/({1+e})=0.269$
\cite{pdu,logosf} in good agreement
with the measured value $\Omega_{\rm dm,0}^{\rm
obs}=0.260$.\footnote{There is no DM and no DE in the
logotropic model, just a single dark fluid. In that case, $\Omega_{\rm dm,0}$
represents the constant that appears in the asymptotic expression 
$\epsilon/\epsilon_0\sim \Omega_{\rm dm,0}/a^3$ of the energy density versus
scale
factor relation for $a\ll 1$.}
Unfortunately, the logotropic  model suffers from the same
problems as the GCG model regarding the presence of unphysical oscillations  in
the matter power spectrum \cite{fa,logosf}. It is not clear how these problems
can be circumvented (see the discussion in \cite{logosf}). Anyway, the
logotropic dark fluid (LDF) remains an interesting UDM model, especially because
of its connection with the polytropic (GCG) model.

The aim of the present paper is to develop the Lagrangian formulation of the
polytropic (GCG) and logotropic models. We point out that the equation of state
can be specified in different manners, yielding three sorts of models. In model
I, the pressure is a function $P(\epsilon)$ of the energy density; in model II,
the pressure is a function $P(\rho_m)$ of the rest-mass density; in model III,
the pressure is a function $P(\rho)$ of the pseudo rest-mass density associated
with a complex SF (in the sense given below). In the nonrelativistic regime,
these three formulations coincide. However, in the relativistic regime, 
they lead to different models. In this paper, we describe these models in detail
and show their interrelations. For example, given $P(\epsilon)$, we show how one
can obtain $P(\rho_m)$ and $P(\rho)$, and reciprocally. We also explain how one
can obtain the k-essence Lagrangian (action) for each model. We first provide
general results that can
be applied to an arbitrary barotropic equation of state. Then, for illustration,
we obtain explicit analytical results for a polytropic and a logotropic equation
of state. We recover the Born-Infeld action of the Chaplygin gas and determine
the expression of the action of the GCG of  type I, II and III. We also
explicitly obtain the logotropic action in models II and III. We show that it
can be recovered from the polytropic (GCG) action in the limit
$\gamma\rightarrow 0$ and $K\rightarrow \infty$ with $A=K\gamma$ constant.

The paper is organized as follows. In Sec. \ref{sec_nra}, we consider a
nonrelativistic complex self-interacting SF which may represent the wavefunction
of a Bose-Einstein condensate (BEC) described by the Gross-Pitaevskii (GP)
equation. We determine its Madelung hydrodynamic representation and show that it
is equivalent to an irrotational quantum fluid with a quantum potential and a
barotropic equation of state $P(\rho)$ determined by the self-interaction
potential. In the TF limit, it reduces to an irrotational classical barotropic
fluid. We determine its Lagrangian and ``reduced'' Lagrangian for an arbitrary
equation of state. The reduced Lagrangian has the form of a k-essence
Lagrangian ${\cal L}(x)$ with $x=\dot\theta+(1/2)(\nabla\theta)^2$, where
$\theta$ is the 
potential of the velocity field. In Sec. \ref{sec_ra}, we consider a
relativistic complex
self-interacting SF  described by the KG equation which may represent the
wavefunction of a relativistic BEC. We determine its de Broglie
hydrodynamic representation and show that it is equivalent to an irrotational 
quantum fluid with a covariant quantum potential and a barotropic equation of
state $P(\rho)$ determined by the self-interaction potential. In the TF limit,
it reduces to an irrotational classical barotropic fluid. We determine its
Lagrangian and reduced Lagrangian for an arbitrary equation of state. Its
reduced Lagrangian has the form of a  $k$-essence Lagrangian ${\cal
L}(X)$ with $X=(1/2)\partial_\mu\theta\partial^\mu\theta$, where the real SF
$\theta$ is
played by the phase of the complex SF. In Sec. \ref{sec_ges}, we introduce three
types of equations of state (models I, II and III) and explain their physical
meanings. We provide general equations allowing us to connect one model to the
other and to determine the reduced Lagrangian ${\cal L}(X)$ for an arbitrary
equation of state. In Secs. \ref{sec_poly} and \ref{sec_logo}, we illustrate our
general results by applying them to a polytropic (GCG) equation of state and a
logotropic equation of state. We recover the Born-Infeld action of the Chaplygin
gas and determine the reduced action of the GCG and of the logotropic gas. In
Appendix \ref{sec_gicg}, we establish general identities valid for a
nonrelativistic cold gas. In Appendix \ref{sec_lrsf}, we consider a general
k-essence Lagrangian and specifically discuss the case of a canonical
and tachyonic SF. In Appendix
\ref{sec_mtu}, we define the equation of state of model I and detail how one
can
obtain the potential $V(\varphi)$ of a homogeneous real SF in an expanding
universe. In  Appendix \ref{sec_mtd}, we define the equation of state of model
II and
detail how one can obtain the corresponding internal energy. In Appendix
\ref{sec_mtt}, we define the equation of state of model III (see also Sec.
\ref{sec_ra}) and detail how the basic equations of the problem can be
simplified in the case of a homogeneous SF 
in a cosmological context. In Appendix \ref{sec_sum}, we
discuss the analogies and the differences between
the internal energy and the potential of a complex SF in the TF limit.
In Appendix \ref{sec_stu}, we list some studies devoted to polytropic and
logotropic equations of state of type I, II and III. In Appendix
\ref{sec_stress}, we detail the Lagrangian structure and the conservation laws
of a nonrelativistic and relativistic SF. Applications and
generalizations of the results of this paper will be presented
in future works \cite{prep}.

\section{Nonrelativistic theory}
\label{sec_nra}

\subsection{The Gross-Pitaevskii equation}
\label{sec_ggpp}

We consider a complex SF $\psi({\bf r},t)$ whose evolution is governed by the GP equation
\begin{eqnarray}
\label{mfgp9}
i\hbar \frac{\partial\psi}{\partial t}=-\frac{\hbar^2}{2m}\Delta\psi
+m h(|\psi|^2)\psi.
\end{eqnarray}
This equation describes, for example, the wave function of a BEC at
$T=0$
\cite{prd1}. For the sake of
generality, we have introduced an arbitrary
nonlinearity determined by the effective potential $h(|\psi|^2)$ instead of the
quadratic potential $h(|\psi|^2)=g|\psi|^2=(4\pi a_s\hbar^2/m^3)
|\psi|^2$, where $a_s$ is the s-scattering length of the bosons,  arising from
pair contact
interactions in the usual GP equation \cite{gross1,gross2,gross3,pitaevskii2}
(see, e.g., the discussion in Ref. \cite{prd1}). In
this manner, we can describe a larger class of systems.\footnote{We could also
consider the case of self-gravitating BECs. In that case, one has to introduce
a mean field gravitational potential $\Phi({\bf r},t)$ in the GP equation which
is produced by the particles themselves through a Poisson equation (see Ref.
\cite{prd1} for more details).} The GP equation (\ref{mfgp9}) can also be
derived from the Klein-Gordon (KG) equation
\begin{eqnarray}
\label{kg}
\square\varphi+\frac{m^2c^2}{\hbar^2}\varphi+2\frac{dV}{d|\varphi|^2}\varphi=0,
\end{eqnarray}
which governs the evolution of a complex SF  $\varphi({\bf r},t)$ with a
self-interaction potential $V(|\varphi|^2)$. The GP equation (\ref{mfgp9}) is
obtained from the KG equation (\ref{kg})  in the nonrelativistic limit
$c\rightarrow +\infty$ after making the Klein transformation
\begin{eqnarray}
\label{klein}
\varphi({\bf r},t)=\frac{\hbar}{m}e^{-imc^2t/\hbar}\psi({\bf r},t).
\end{eqnarray}
In that case, the
effective potential $h(|\psi|^2)$ that appears in the GP
equation is related to the self-interaction potential $V(|\varphi|^2)$ present
in the
KG equation by  (see Refs. \cite{playa,chavmatos} and Appendix C of Ref.
\cite{epjp})
\begin{eqnarray}
\label{kggp}
h(|\psi|^2)=\frac{dV}{d|\psi|^2}\quad {\rm with}\quad 
|\psi|^2=\frac{m^2}{\hbar^2}|\varphi|^2.
\end{eqnarray}
As a result, the GP equation (\ref{mfgp9}) can be rewritten as
\begin{eqnarray}
\label{mfgp9new}
i\hbar \frac{\partial\psi}{\partial t}=-\frac{\hbar^2}{2m}\Delta\psi
+m \frac{dV}{d|\psi|^2}\psi.
\end{eqnarray}

{\it Remark:} The GP equation (\ref{mfgp9new}) can also be
derived from the KG equation governing the evolution of a real SF but, in that
case, the potential $V(|\psi|^2)$ that appears in the GP equation does not
coincide with the potential $V(\varphi)$ present in the KG equation.
Indeed, $V(|\psi|^2)$ is an effective potential   
obtained after averaging $V(\varphi)$ over the fast
oscillations of the SF (see Secs. II and III of \cite{phi6} and Appendix A of
\cite{tunnel} for details).

\subsection{The Madelung transformation}
\label{sec_mad}

We can write the GP equation (\ref{mfgp9}) under the form of hydrodynamic equations by using the Madelung \cite{madelung} transformation. To that purpose, we write the
wave function as
\begin{equation}
\label{mad1}
\psi({\bf r},t)=\sqrt{{\rho({\bf r},t)}} e^{iS({\bf r},t)/\hbar},
\end{equation}
where  $\rho({\bf r},t)$ is the density and $S({\bf r},t)$ is the action. They are given by
\begin{equation}
\label{mad2}
\rho=|\psi|^2\qquad {\rm and}\qquad S=-i\frac{\hbar}{2}\ln\left
(\frac{\psi}{\psi^*}\right ).
\end{equation}
Following Madelung, we introduce the
velocity field
\begin{equation}
\label{mad5}
{\bf u}=\frac{\nabla S}{m}.
\end{equation}
Since the velocity derives from a potential, the flow is irrotational: $\nabla\times {\bf
u}={\bf 0}$. Substituting Eq. (\ref{mad1}) into Eq. (\ref{mfgp9}) and separating
the real and the imaginary parts, we obtain
\begin{equation}
\label{mad6}
\frac{\partial\rho}{\partial t}+\nabla\cdot (\rho {\bf u})=0,
\end{equation}
\begin{equation}
\label{mad7}
\frac{\partial S}{\partial t}+\frac{1}{2m}(\nabla S)^2+m h(\rho)+Q=0,
\end{equation}
where
\begin{equation}
\label{mad8}
Q=-\frac{\hbar^2}{2m}\frac{\Delta
\sqrt{\rho}}{\sqrt{\rho}}=-\frac{\hbar^2}{4m}\left\lbrack
\frac{\Delta\rho}{\rho}-\frac{1}{2}\frac{(\nabla\rho)^2}{\rho^2}\right\rbrack
\end{equation}
is the quantum potential. It takes into account the Heisenberg uncertainty
principle. Eq. (\ref{mad6}) is similar to the equation of continuity in
hydrodynamics. It accounts for the local conservation of mass $M=\int \rho\,
d{\bf r}$. Eq. (\ref{mad7}) has a form similar to the classical
Hamilton-Jacobi equation with an additional quantum potential. It  can also be interpreted as a
 quantum Bernoulli equation for
a potential flow. Taking
the gradient of Eq. (\ref{mad7}), and using the well-known  identity of
vector analysis $({\bf u}\cdot \nabla){\bf u}=\nabla ({{\bf u}^2}/{2})-{\bf
u}\times (\nabla\times {\bf u})$ which reduces to $({\bf u}\cdot \nabla){\bf
u}=\nabla ({{\bf u}^2}/{2})$ for an irrotational flow, we obtain
\begin{equation}
\label{mad9}
\frac{\partial {\bf u}}{\partial t}+({\bf u}\cdot \nabla){\bf u}=-\nabla
h-\frac{1}{m}\nabla Q.
\end{equation}
Since  $h=h(\rho)$ we can introduce a function $P=P(\rho)$ satisfying $\nabla h=(1/\rho)\nabla P$. Eq. (\ref{mad9}) can then be rewritten as 
\begin{equation}
\label{mad10}
\frac{\partial {\bf u}}{\partial t}+({\bf u}\cdot \nabla){\bf
u}=-\frac{1}{\rho}\nabla P-\frac{1}{m}\nabla
Q.
\end{equation}
This equation is similar to the Euler equation with a pressure 
force  $-(1/\rho)\nabla P$ and a quantum force $-\frac{1}{m}\nabla Q$. Since
$P({\bf r},t)=P\lbrack \rho({\bf r},t)\rbrack$ is a function of the density, the
flow is barotropic. The equation of state $P(\rho)$ is determined by the
potential $h(\rho)$ through the relation
\begin{equation}
\label{mad11}
h'(\rho)=\frac{P'(\rho)}{\rho}.
\end{equation}
Equation
(\ref{mad11}) can be integrated into
\begin{equation}
\label{mad11b}
P(\rho)=\rho h(\rho)-V(\rho)=\rho
V'(\rho)-V(\rho)=\rho^2\left\lbrack
\frac{V(\rho)}{\rho}\right\rbrack',
\end{equation}
where $V$ is a primitive of $h$. This notation is consistent with
Eq. (\ref{kggp}) which can be rewritten as
\begin{eqnarray}
\label{kggpv}
h(\rho)=V'(\rho),
\end{eqnarray}
where $V(\rho)$ is the potential in the KG equation (\ref{kg}) or in the GP
equation (\ref{mfgp9new}). Eq. (\ref{mad11b}) determines the pressure $P(\rho)$
as a function of the potential $V(\rho)$. Inversely, the potential is determined
as a function of the pressure by\footnote{We note that the potential is defined
from the pressure up to a term of the form $A\rho$, where $A$ is a constant. If
we add a term $A\rho$ in the potential $V$, we do not change the pressure. On
the other hand, if we add a constant term $C$ in the potential $V$, this adds a
term $-C$ in the pressure. However, for nonrelativistic systems, this
constant term has no observable effect  since only the gradient of the pressure
matters.}
\begin{eqnarray}
V(\rho)=\rho\int\frac{P(\rho)}{\rho^2}\, d\rho.
\label{etoile}
\end{eqnarray}
The speed of sound is $c_s^2=P'(\rho)=\rho
V''(\rho)$. The GP equation (\ref{mfgp9}) is equivalent to
the hydrodynamic equations (\ref{mad6}), (\ref{mad7}) and (\ref{mad10}). We
shall call them the
quantum Euler equations. Since there is no viscosity, they describe a
superfluid. In the TF approximation
$\hbar\rightarrow 0$, they reduce to the classical Euler equations.

{\it Remark:} We show in Appendix \ref{sec_gicg} that the effective potential
$h$ appearing in the GP equation can be
interpreted, in the hydrodynamic equations, as an enthalpy (or as a chemical
potential by unit of mass $h=\mu/m$) and that its primitive
$V(\rho)$, equal to the potential in the KG equation, can be interpreted as 
the internal energy density $u$. Thus, we have
\begin{eqnarray}
u(\rho)=V(\rho),\qquad h(\rho)=\frac{P(\rho)+V(\rho)}{\rho}.
\label{addq}
\end{eqnarray}
On the other hand, if we define the energy by 
\begin{equation}
E({\bf r},t)=-\frac{\partial S}{\partial t},
\end{equation}
the Hamilton-Jacobi equation (\ref{mad7}) can be
rewritten as
\begin{equation}
E({\bf r},t)=\frac{1}{2}m{\bf u}^2+m h(\rho)+Q.
\end{equation}

\subsection{Lagrangian of a quantum barotropic gas}
\label{sec_lag}

The action of the complex SF associated with the GP equation (\ref{mfgp9new}) is given by
\begin{equation}
S=\int L\, dt
\label{lh1a}
\end{equation}
with the Lagrangian 
\begin{equation}
L=\int \biggl\lbrace\frac{i\hbar}{2m} \left (\psi^*\frac{\partial\psi}{\partial
t}-\psi\frac{\partial\psi^*}{\partial t}\right
)-\frac{\hbar^2}{2m^2}|\nabla\psi|^2
-V(|\psi|^2)\biggr\rbrace\, d{\bf
r}.
\label{lh1}
\end{equation}
We can view the Lagrangian (\ref{lh1}) as a functional of $\psi$, $\dot\psi$
and $\nabla\psi$. The least action principle
$\delta S=0$, which is
equivalent to the Euler-Lagrange equation
\begin{eqnarray}
\label{lh2}
\frac{\partial}{\partial t}\left (\frac{\delta
L}{\delta\dot\psi}\right)+\nabla\cdot \left (\frac{\delta
L}{\delta\nabla\psi}\right)-\frac{\delta L}{\delta\psi}=0,
\end{eqnarray}
returns the GP equation (\ref{mfgp9new}).  The Hamiltonian (energy) is
obtained from the
transformation
\begin{eqnarray}
H=\int \frac{i\hbar}{2m} \left (\psi^*\frac{\partial\psi}{\partial
t}-\psi\frac{\partial\psi^*}{\partial t}\right )\, d{\bf r}-L
\label{lh3}
\end{eqnarray}
yielding
\begin{eqnarray}
H=\frac{\hbar^2}{2m^2}\int|\nabla\psi|^2\, d{\bf
r}+\int V(|\psi|^2)\, d{\bf r}.
\label{lh4}
\end{eqnarray}
The first term is the kinetic energy
$\Theta=-\frac{\hbar^2}{2m^2}\int\psi^*\Delta\psi\, d{\bf r}$ and
the second term is the self-interaction energy $U$. Since the Lagrangian does
not explicitly depend on time, the Hamiltonian (energy) is
conserved.\footnote{Similarly, the invariance of the
Lagrangian with respect to spatial translation implies the conservation of
linear momentum. In relativity theory, these apparently separate conservation
laws are aspects of a single conservation law, that of the energy-momentum
tensor (see Sec. \ref{sec_ra}).}
The GP equation (\ref{mfgp9new}) can be written as
\begin{eqnarray}
\label{lh4g}
i\hbar\frac{\partial \psi}{\partial t}=m\frac{\delta H}{\delta \psi^*},\qquad
i\hbar\frac{\partial \psi^*}{\partial t}=-m\frac{\delta H}{\delta \psi},
\end{eqnarray}
which can be interpreted as Hamilton equations (see Appendix \ref{sec_cpsi}).

Using the Madelung transformation, we can rewrite the
Lagrangian in terms of hydrodynamic variables. According to
Eqs. (\ref{mad1}) and (\ref{mad2}) we have
\begin{eqnarray}
\label{lh6}
\frac{\partial S}{\partial t}=-i\frac{\hbar}{2}\frac{1}{|\psi|^2}\left
(\psi^*\frac{\partial\psi}{\partial t}-\psi\frac{\partial\psi^*}{\partial
t}\right )
\end{eqnarray}
and
\begin{eqnarray}
\label{lh7}
|\nabla\psi|^2=\frac{1}{4\rho}(\nabla\rho)^2+\frac{\rho}{\hbar^2}
(\nabla S)^2.
\end{eqnarray}
Substituting these identities into Eq. (\ref{lh1}), we get
\begin{equation}
\label{lh8}
L=-\int \biggl \lbrace \frac{\rho}{m}\frac{\partial S}{\partial
t}+\frac{\rho}{2m^2}(\nabla S)^2
+\frac{\hbar^2}{8m^2}\frac{(\nabla\rho)^2}{\rho}+V(\rho)\biggr\rbrace\, d{\bf
r}.
\end{equation}
We can view the Lagrangian (\ref{lh8}) as a functional of $S$, $\dot S$,
$\nabla S$, $\rho$, $\dot\rho$, and $\nabla\rho$. The Euler-Lagrange equation
for the
action
\begin{eqnarray}
\label{lh9}
\frac{\partial}{\partial t}\left (\frac{\delta L}{\delta\dot
S}\right)+\nabla\cdot \left (\frac{\delta L}{\delta\nabla S}\right)-\frac{\delta
L}{\delta S}=0
\end{eqnarray}
returns the equation of continuity (\ref{mad6}). The Euler-Lagrange equation for
the
density
\begin{eqnarray}
\label{lh10}
\frac{\partial}{\partial t}\left (\frac{\delta L}{\delta\dot
\rho}\right)+\nabla\cdot \left (\frac{\delta L}{\delta\nabla
\rho}\right)-\frac{\delta L}{\delta \rho}=0
\end{eqnarray}
returns the quantum Hamilton-Jacobi (or Bernoulli) equation (\ref{mad7})
leading to the quantum Euler equation (\ref{mad10}). The Hamiltonian (energy) is obtained
from
the transformation
\begin{eqnarray}
\label{lh11}
H=-\int \frac{\rho}{m}\frac{\partial S}{\partial t}\, d{\bf r}-L
\end{eqnarray}
yielding
\begin{eqnarray}
\label{lh12}
H=\int  \frac{1}{2}\rho {\bf
u}^2\,
d{\bf r}+\int \frac{\hbar^2}{8m^2}\frac{(\nabla\rho)^2}{\rho}\,
d{\bf r}
+\int V(\rho)\,
d{\bf r}.
\end{eqnarray}
This expression is equivalent to Eq. (\ref{lh4}) as can be seen by a direct
calculation 
using the Madelung transformation [see Eq. (\ref{lh7})]. The first term is the
classical kinetic energy $\Theta_c$, the second term
is the quantum kinetic energy $\Theta_Q$ (we have
$\Theta=\Theta_c+\Theta_Q$), and
the third term is the self-interaction energy $U$. The quantum kinetic energy 
can also be written as $\Theta_Q=\int\rho\frac{Q}{m}\, d{\bf
r}$ \cite{chavtotal}. The continuity equation (\ref{mad6}) and the
Hamilton-Jacobi (or Bernoulli)
equation (\ref{mad7}) can be written as
\begin{eqnarray}
\label{lh12g}
\frac{\partial \rho}{\partial t}=m\frac{\delta H}{\delta S},\qquad
\frac{\partial S}{\partial t}=-m\frac{\delta H}{\delta \rho},
\end{eqnarray}
which can be interpreted as Hamilton equations (see Appendix \ref{sec_chy}).

If we substitute  the quantum Hamilton-Jacobi (or Bernoulli) equation
(\ref{mad7}) into the Lagrangian (\ref{lh8}) and
use Eqs. (\ref{mad8}) and
(\ref{mad11b}) we find that 
\begin{eqnarray}
\label{lh12b}
L=\int P\, d{\bf r}.
\end{eqnarray}
This shows that the Lagrangian density is equal to the pressure: ${\cal L}=P$.
Actually, the Lagrangian density is equal to  ${\cal
L}=P(\rho)-\frac{\hbar^2}{4m^2}\Delta\rho$. There is an additional term which
disappears by integration. The same result is obtained by substituting the GP
equation (\ref{mfgp9new}) into the Lagrangian (\ref{lh1}) and using Eq.
(\ref{mad11b}).

\subsection{Lagrangian of a classical barotropic gas}
\label{sec_lcb}

Introducing the notation $\theta=S/m$, so that
$\psi=\sqrt{\rho}e^{im\theta/\hbar}$,\footnote{The variable $\theta$ is related
to the phase (angle) $\Theta=S/\hbar$ of the SF by $\theta=\hbar\Theta/m$.} and
taking the limit $\hbar\rightarrow 0$ in Eq. (\ref{lh8}), we
obtain the classical Lagrangian\footnote{In our framework, the limit 
$\hbar\rightarrow 0$ corresponds to the TF approximation where the quantum
potential can be
neglected.}
\begin{equation}
\label{lcb2}
L=-\int \left \lbrack \rho\dot\theta+\frac{1}{2}\rho(\nabla\theta)^2
+V(\rho)\right\rbrack\, d{\bf
r}.
\end{equation}
We can view the Lagrangian (\ref{lcb2}) as a functional of $\theta$, $\dot \theta$,
$\nabla \theta$, $\rho$, $\dot\rho$, and $\nabla\rho$. The Euler-Lagrange
equation for the
action  leads to the equation of continuity
\begin{equation}
\label{lcb5}
\frac{\partial\rho}{\partial t}+\nabla\cdot (\rho {\bf u})=0,
\end{equation}
with the velocity field
\begin{equation}
\label{lcb1}
{\bf u}=\nabla \theta.
\end{equation}
The Euler-Lagrange equation
for the
density leads to the Bernoulli (or Hamilton-Jacobi) equation
\begin{equation}
\label{lcb6}
\dot\theta+\frac{1}{2}(\nabla \theta)^2+V'(\rho)=0.
\end{equation}
Taking the gradient of Eq. (\ref{lcb6}) and using Eq. (\ref{mad11b}), we obtain the Euler equation
\begin{equation}
\label{lcb8}
\frac{\partial {\bf u}}{\partial t}+({\bf u}\cdot \nabla){\bf
u}=-\frac{1}{\rho}\nabla P.
\end{equation}

The Hamiltonian (energy) is obtained
from
the transformation
\begin{eqnarray}
\label{lcb3}
H=-\int \rho\dot\theta\, d{\bf r}-L
\end{eqnarray}
yielding
\begin{eqnarray}
\label{lcb4}
H=\int  \frac{1}{2}\rho {\bf
u}^2\,
d{\bf r}+\int V(\rho)\,
d{\bf r}.
\end{eqnarray}
The first term is the classical kinetic energy $\Theta_c$ and the second term 
is the self-interaction energy $U$.

{\it Remark:} In our presentation, we started from a quantum fluid (or from the
hydrodynamic representation of the GP equation) and finally considered the
classical limit $\hbar\rightarrow 0$. Alternatively, we can obtain the equations
of this section directly  from the classical Euler equations by assuming that
the fluid is barotropic (so that $P=P(\rho)$) and that the flow is irrotational
(so that the velocity derives from a potential: ${\bf u}=\nabla \theta$)
\cite{jackiw}. The Lagrangians (\ref{lh8}) and (\ref{lcb2}) were first obtained
by Eckart \cite{eckart}  for a classical fluid and 
from the hydrodynamic representation of the Schr\"odinger
equation (see also \cite{schakel,bj}).  The Lagrangian (\ref{lcb2}) with the
potential $V=A/(2\rho)$
corresponding to
the Chaplygin equation of state $P=-A/\rho$ (see below) also appeared in the
theory of
membranes
($d=2$) \cite{bh,bj}. It was later generalized to a $d$-brane moving in
a $(d+1,1)$ space-time \cite{jp,jackiw}.

\subsection{Reduced Lagrangian ${\cal L}(x)$}
\label{sec_rln}

We introduce the Lagrangian density
\begin{equation}
\label{lcb2h}
{\cal L}=- \left \lbrack \rho\dot\theta+\frac{1}{2}\rho(\nabla\theta)^2
+V(\rho)\right\rbrack,
\end{equation}
so that $L=\int {\cal L}\, d{\bf r}$ and $S=\int {\cal L}\, d{\bf r}dt$. Using
the Bernoulli equation (\ref{lcb6}) and the identity (\ref{mad11b}), we can
eliminate $\theta$ from the Lagrangian and obtain
\begin{equation}
\label{lcb2hb}
{\cal L}=\rho V'(\rho)-V(\rho)=P(\rho).
\end{equation}
Therefore, the Lagrangian density is equal to the pressure:
\begin{equation}
\label{leqp}
{\cal L}=P.
\end{equation}
We now eliminate $\rho$ from the Lagrangian. Introducing the notation
\begin{equation}
\label{gc3}
x=\dot\theta+\frac{1}{2}(\nabla
\theta)^2,
\end{equation}
the Bernoulli equation (\ref{lcb6}) can be written as
\begin{equation}
\label{gc4z}
x=-V'(\rho).
\end{equation}
Assuming $V''>0$, this equation can be reversed to give
\begin{equation}
\label{gc4}
\rho=F(x)
\end{equation}
with $F(x)=(V')^{-1}(-x)$.\footnote{The condition $V''>0$ is necessary for
local stablity since $c_s^2=P'(\rho)=\rho V''(\rho)$. When the system is
subjected to a potential
$\Phi$, the function $F$ determines the relation  between the density and
the potential at equilibrium 
(see Sec. 3.4 of \cite{chavtotal}).} As a result, the equation of continuity
(\ref{lcb5}) can be written in terms of $\theta$ alone as
\begin{equation}
\label{gc13}
\frac{\partial}{\partial t}\left
\lbrace F\left\lbrack \dot\theta+\frac{1}{2}(\nabla
\theta)^2\right\rbrack \right\rbrace+\nabla\cdot \left
\lbrace   F\left\lbrack \dot\theta+\frac{1}{2}(\nabla
\theta)^2\right\rbrack \nabla\theta\right \rbrace=0.
\end{equation}
On the other hand, according to  Eqs. (\ref{mad11b}) and (\ref{gc4}), we have
$P=P(x)$.
Therefore, recalling Eq. (\ref{leqp}), we get
\begin{equation}
\label{lcb11}
{\cal L}=P(x).
\end{equation}
In this manner, we have eliminated the density $\rho$ from the Lagrangian
(\ref{lcb2h}) and we have obtained a reduced Lagrangian of the form ${\cal
L}(x)$ that depends only on $x$. This kind of Lagrangian, called k-essence
Lagrangian, is discussed in
Appendix \ref{sec_lrsf}. We show below that
\begin{equation}
\label{gc10b}
\rho=F(x)=-{\cal L}'(x)=-P'(x).
\end{equation}
Using Eq. (\ref{gc10b}), we can
write Eq. (\ref{lcb5}) in terms of ${\cal L}'(x)$ like in
Eq. (\ref{no7}) of Appendix \ref{sec_lrsf}.

{\it Proof of Eq. (\ref{gc10b}):} From Eqs. (\ref{mad11b})  and (\ref{gc4z}) we
have
\begin{equation}
\label{gc7}
P'(\rho)=\rho V''(\rho)
\end{equation}
and
\begin{equation}
\label{gc8}
\frac{dx}{d\rho}=-V''(\rho).
\end{equation}
Starting from Eq. (\ref{lcb11}) and using Eqs. (\ref{gc7}) and  (\ref{gc8}) we
obtain
\begin{equation}
\label{gc12a}
{\cal L}'(x)=P'(x)=P'(\rho)\frac{d\rho}{dx}=\rho
V''(\rho)\frac{d\rho}{dx}=-\rho\frac{dx}{d\rho}\frac{d\rho}{dx}=-\rho,
\end{equation}
which establishes Eq. (\ref{gc10b}).

The preceding results are general.
In the
following sections, we consider
particular equations of state.

\subsection{Polytropic gas}
\label{sec_lac}

We first consider the polytropic equation of state \cite{chandrabook}
\begin{eqnarray}
P=K\rho^{\gamma}.
\label{lac2}
\end{eqnarray}
It can be obtained from the potential \cite{chavtotal}
\begin{eqnarray}
V(\rho)=\frac{K}{\gamma-1}\rho^{\gamma}.
\label{lac3}
\end{eqnarray}
As discussed in \cite{chavtotal} this potential is similar to the Tsallis free
energy density $-Ks_\gamma$, where the polytropic constant $K$ plays the role of
a generalized temperature and $s_\gamma=-\frac{1}{\gamma-1}(\rho^{\gamma}-\rho)$
is the Tsallis entropy density. The Lagrangian and the Hamiltonian are given by
Eqs. (\ref{lcb2}) and (\ref{lcb4}) with Eq. (\ref{lac3}). The Bernoulli equation
(\ref{lcb6}) takes the form
\begin{equation}
\label{lac5}
\dot\theta+\frac{1}{2}(\nabla
\theta)^2+\frac{K\gamma}{\gamma-1}\rho^{\gamma-1}=0.
\end{equation}
From this equation,  we obtain
\begin{equation}
\label{lac6}
\rho=\left\lbrack -\frac{\gamma-1}{K\gamma}\left (\dot\theta+\frac{1}{2}(\nabla
\theta)^2\right )\right\rbrack^{\frac{1}{\gamma-1}},
\end{equation}
which is similar to the Tsallis distribution.
The reduced Lagrangian ${\cal L}(x)$ corresponding to the polytropic gas is
\begin{equation}
\label{lac7}
{\cal L}=P=K\left\lbrack -\frac{\gamma-1}{K\gamma}\left
(\dot\theta+\frac{1}{2}(\nabla
\theta)^2\right )\right\rbrack^{\frac{\gamma}{\gamma-1}}.
\end{equation}
The equation of motion is
\begin{equation}
\label{lcb5w}
\frac{\partial}{\partial t}\left ( \left |\dot\theta+\frac{1}{2}(\nabla
\theta)^2\right |^{\frac{1}{\gamma-1}}\right )+\nabla\cdot \left (\left |\dot\theta+\frac{1}{2}(\nabla
\theta)^2\right |^{\frac{1}{\gamma-1}}\nabla\theta \right ) =0.
\end{equation}
We note that the polytropic constant $K$ does not appear in this equation.

\subsection{Isothermal gas}
\label{sec_isog}

The case $\gamma=1$, corresponding to the isothermal equation of state
\cite{chandrabook}
\begin{eqnarray}
P=K\rho,
\label{isog1}
\end{eqnarray}
must be treated specifically (here $K$ plays the role of the temperature $k_B
T/m$).  It can be 
obtained from the potential \cite{chavtotal}
\begin{eqnarray}
V(\rho)=K\rho\left\lbrack \ln(\rho/\rho_*)-1\right\rbrack.
\label{isog2}
\end{eqnarray}
As discussed in \cite{chavtotal} this potential is similar to the Boltzmann free
energy density $-Ks_B$, where $K$ plays the role of the temperature and
$s_B=-\rho[\ln(\rho/\rho_*)-1]$ is the Boltzmann entropy density. 
The Lagrangian and the Hamiltonian are given by Eqs. (\ref{lcb2}) and (\ref{lcb4}) with Eq. (\ref{isog2}). The Bernoulli equation (\ref{lcb6}) takes the form
\begin{equation}
\label{isog3}
\dot\theta+\frac{1}{2}(\nabla
\theta)^2+K\ln(\rho/\rho_*)=0.
\end{equation}
From this equation,  we obtain 
\begin{equation}
\label{isog4}
\rho=\rho_* e^{-\frac{1}{K}\left\lbrack
\dot\theta+\frac{1}{2}(\nabla\theta)^2\right\rbrack},
\end{equation}
which is similar to the Boltzmann distribution.
The reduced Lagrangian ${\cal L}(x)$ corresponding to the isothermal gas is
\begin{equation}
\label{isog5}
{\cal L}=P=K\rho_*e^{-\frac{1}{K}\left\lbrack \dot\theta+\frac{1}{2}(\nabla\theta)^2\right\rbrack}.
\end{equation}
The equation of motion is
\begin{equation}
\label{isog6}
\frac{\partial}{\partial t}\left (e^{-\frac{1}{K}\left\lbrack \dot\theta+\frac{1}{2}(\nabla\theta)^2\right\rbrack} \right )+\nabla\cdot \left (e^{-\frac{1}{K}\left\lbrack \dot\theta+\frac{1}{2}(\nabla\theta)^2\right\rbrack}\nabla\theta  \right ) =0.
\end{equation}
We note that the constant $K$ (temperature) cannot be eliminated from this equation contrary to the polytropic case.

\subsection{Chaplygin gas}
\label{sec_cl}

The Chaplygin equation of state writes \cite{chaplygin}
\begin{eqnarray}
P=\frac{K}{\rho}.
\label{cl1}
\end{eqnarray}
The ordinary Chaplygin gas corresponds to $K<0$. The case $K>0$ is called the
anti-Chaplygin gas. Eq. (\ref{cl1}) is a particular  polytropic equation of
state (\ref{lac2}) corresponding to $\gamma=-1$.\footnote{For that reason, the
polytropic equation of state (\ref{lac2}) is also called the GCG
\cite{bentoGCG}.} It 
can be obtained from the potential
\begin{eqnarray}
V(\rho)=-\frac{K}{2\rho}.
\label{cl2}
\end{eqnarray}
The Lagrangian and the Hamiltonian are given by Eqs. (\ref{lcb2}) and
(\ref{lcb4}) with Eq. (\ref{cl2}). The Bernoulli equation (\ref{lcb6}) takes the
form
\begin{equation}
\label{cl4}
\dot\theta+\frac{1}{2}(\nabla \theta)^2+\frac{K}{2\rho^2}=0,
\end{equation}
yielding 
\begin{equation}
\label{cl5}
\rho=\sqrt{\frac{-K}{2\left\lbrack \dot\theta+\frac{1}{2}(\nabla
\theta)^2\right\rbrack}}.
\end{equation}
The reduced Lagrangian ${\cal L}(x)$ corresponding to the Chaplygin gas is
\begin{equation}
\label{cl6}
{\cal L}=P=K\sqrt{\frac{2}{-K}\left\lbrack \dot\theta+\frac{1}{2}(\nabla
\theta)^2\right\rbrack}.
\end{equation}
The equation of motion is
\begin{eqnarray}
\label{cl8}
\frac{\partial}{\partial t}\left
\lbrack\frac{1}{\sqrt{|\dot\theta+\frac{1}{2}(\nabla
\theta)^2 |}}\right \rbrack+\nabla\cdot \left
\lbrack\frac{\nabla\theta}{\sqrt{|\dot\theta+\frac{1}{2}(\nabla
\theta)^2|}}\right \rbrack=0.
\end{eqnarray}
The Chaplygin constant $K$ does not appear in this equation. If we consider 
time-independent solutions, this equation reduces to
\begin{eqnarray}
\label{cl8b}
\nabla\cdot \left (\frac{\nabla\theta}{\sqrt{(\nabla
\theta)^2}}\right )=0.
\end{eqnarray}
The same equation is obtained by taking the massless limit (recalling that
$\theta=S/m$).  In the 
theory of $d$-branes, this equation means that the surface
$\theta(x_1,x_2,...,x_d)={\rm const}$ has zero extrinsic mean curvature
\cite{ogawa}. This solution exists only when $K<0$.

{\it Remark:} The Lagrangian (\ref{cl6})  was obtained by \cite{jp,jackiw} in
two 
different manners: (i) starting from the Lagrangian (\ref{lcb2})
with Eq. (\ref{cl2}) and using the Bernoulli equation (\ref{cl4}) to eliminate
$\rho$ as we have done here; (ii) for a $d$-brane moving in a $(d+1,1)$
space-time. In that second case, it can be obtained from the Nambu-Goto action
in
the light-cone
parametrization. This explains the connection between $d$-branes and the
hydrodynamics 
of the Chaplygin gas.

\subsection{Standard BEC}
\label{sec_bec}

The potential of a standard BEC 
described by the ordinary GP equation is
\begin{eqnarray}
V(|\psi|^2)=\frac{2\pi a_s\hbar^2}{m^3}|\psi|^4,
\label{lac2b}
\end{eqnarray}
where  $a_s$ is the scattering length
of the bosons (the interaction is repulsive when $a_s>0$ and attractive when
$a_s<0$). This quartic potential accounts for two-body interactions in a weakly
interacting microscopic theory of the superfluid. The
corresponding equation of state is
\begin{eqnarray}
P=\frac{2\pi a_s\hbar^2}{m^3}\rho^{2}.
\label{lac2c}
\end{eqnarray}
This is the equation of state of a polytrope of index
$\gamma=2$ and polytropic constant $K={2\pi a_s\hbar^2}/{m^3}$.
The Lagrangian and the Hamiltonian are given by Eqs. (\ref{lcb2}) and
(\ref{lcb4}) with Eq. (\ref{lac2b}). The Bernoulli equation (\ref{lcb6}) takes
the
form
\begin{equation}
\label{cl4bec}
\dot\theta+\frac{1}{2}(\nabla \theta)^2+2K\rho=0,
\end{equation}
yielding 
\begin{equation}
\label{cl5bec}
\rho=-\frac{1}{2K}\left\lbrack \dot\theta+\frac{1}{2}(\nabla
\theta)^2\right\rbrack.
\end{equation}
The reduced Lagrangian ${\cal L}(x)$ corresponding to the standard BEC is
\begin{equation}
\label{cl6bec}
{\cal L}=P=\frac{1}{4K}\left\lbrack
\dot\theta+\frac{1}{2}(\nabla
\theta)^2\right\rbrack^2.
\end{equation}
The equation of motion is
\begin{eqnarray}
\label{cl8bec}
\frac{\partial}{\partial t}\left
\lbrack\dot\theta+\frac{1}{2}(\nabla
\theta)^2\right \rbrack+\nabla\cdot \left
(\left\lbrack\dot\theta+\frac{1}{2}(\nabla
\theta)^2\right \rbrack\nabla\theta\right)=0.
\end{eqnarray}
The BEC constant $K$ does not appear in this equation. 

\subsection{DM superfluid}
\label{sec_sf}

The potential of a superfluid (BEC) with a sextic
self-interaction is
\begin{eqnarray}
V(|\psi|^2)=\frac{1}{2}K|\psi|^{6}.
\label{slac2b}
\end{eqnarray}
This potential accounts for three-body interactions in a weakly
interacting microscopic theory of the superfluid \cite{phi6}. The potential
(\ref{slac2b}) may also describe a more exotic DM superfluid \cite{ferreira}. In
that case, it
has a completely different interpretation. The corresponding
equation of state is
\begin{eqnarray}
P=K\rho^{3}.
\label{slac2c}
\end{eqnarray}
This is the equation of state of a polytrope of index
$\gamma=3$.
The Lagrangian and the Hamiltonian are given by Eqs. (\ref{lcb2}) and
(\ref{lcb4}) with Eq. (\ref{slac2b}). The Bernoulli equation (\ref{lcb6}) takes
the
form
\begin{equation}
\label{scl4bec}
\dot\theta+\frac{1}{2}(\nabla \theta)^2+\frac{3K}{2}\rho^2=0,
\end{equation}
yielding 
\begin{equation}
\label{scl5bec}
\rho=\sqrt{-\frac{2}{3K}\left\lbrack \dot\theta+\frac{1}{2}(\nabla
\theta)^2\right\rbrack}.
\end{equation}
The reduced Lagrangian ${\cal L}(x)$ corresponding to the superfluid is
\begin{equation}
\label{scl6bec}
{\cal L}=P=K\left\lbrace -\frac{2}{3K}\left\lbrack
\dot\theta+\frac{1}{2}(\nabla
\theta)^2\right\rbrack\right\rbrace^{3/2}.
\end{equation}
The equation of motion is
\begin{equation}
\label{scl8bec}
\frac{\partial}{\partial t}\left
\lbrack \sqrt{|\dot\theta+\frac{1}{2}(\nabla
\theta)^2|}\right \rbrack+\nabla\cdot \left
(\sqrt{|\dot\theta+\frac{1}{2}(\nabla
\theta)^2|}\nabla\theta\right)=0.
\end{equation}
The superfluid constant $K$ does not appear in this equation. If we consider 
time-independent solutions, this equation reduces to
\begin{eqnarray}
\label{scl8b}
\nabla\cdot \left (|\nabla \theta|\nabla\theta\right)=0.
\end{eqnarray}
This solution exists only when $K<0$. Interestingly, there is
a connection between a superfluid described by Eq. (\ref{scl8b})
and the modified Newtonian dynamics (MOND) theory (see, e.g., \cite{ferreira,bm}
for more
details).

\subsection{Logotropic gas}
\label{sec_lal}

Finally, we consider the logotropic equation of state \cite{epjp}
\begin{eqnarray}
P=A\ln\left (\frac{\rho}{\rho_*}\right ),
\label{lal1}
\end{eqnarray}
which can be obtained from the potential \cite{chavtotal} 
\begin{eqnarray}
V(\rho)=-A\ln\left (\frac{\rho}{\rho_*}\right )-A.
\label{lal2}
\end{eqnarray}
The Lagrangian and the Hamiltonian are given by Eqs. (\ref{lcb2}) and (\ref{lcb4}) with Eq. (\ref{lal2}). The Bernoulli equation (\ref{lcb6}) takes the form
\begin{equation}
\label{lal4}
\dot\theta+\frac{1}{2}(\nabla \theta)^2-\frac{A}{\rho}=0,
\end{equation}
from which we get
\begin{equation}
\label{lal5}
\rho=\frac{A}{\dot\theta+\frac{1}{2}(\nabla \theta)^2}.
\end{equation}
The reduced Lagrangian ${\cal L}(x)$ corresponding to the logotropic gas is 
\begin{equation}
\label{lal6}
{\cal L}=P=-A \ln\left\lbrack \frac{\rho_*}{A}\left
(\dot\theta+\frac{1}{2}(\nabla
\theta)^2\right )\right\rbrack.
\end{equation}
The equation of motion is
\begin{equation}
\label{lcb5wb}
\frac{\partial}{\partial t}\left ( \frac{1}{|\dot\theta+\frac{1}{2}(\nabla \theta)^2|}\right )+\nabla\cdot \left ( \frac{\nabla\theta}{|\dot\theta+\frac{1}{2}(\nabla \theta)^2|}\right ) =0.
\end{equation}
The logotropic constant $A$ does not appear in this
equation. 

{\it Remark:} We can recover these results from the polytropic equation of state
of Sec. \ref{sec_lac} by considering the limit $\gamma\rightarrow 0$,
$K\rightarrow \infty$ with $A=K\gamma$ constant \cite{logo,epjp}.  Starting from
Eq. (\ref{lac7}),
we get
\begin{eqnarray}
\label{lal7}
{\cal L}=K\left\lbrack -\frac{\gamma-1}{K\gamma}x\right\rbrack^{\frac{\gamma}{\gamma-1}}
\simeq Ke^{-\gamma\ln\left (\frac{x}{K\gamma}\right )}\nonumber\\
\simeq K\left\lbrack 1-\gamma \ln\left (\frac{x}{K\gamma}\right )+...\right\rbrack\simeq K-A\ln\left (\frac{x}{A}\right ),
\end{eqnarray}
which is equivalent to Eq. (\ref{lal6}) up to a constant term (see footnote 12).

\subsection{Summary}
\label{sec_s}

For a polytropic equation of state $P=K\rho^{\gamma}$ with $\gamma\neq 1$, the
reduced Lagrangian is
\begin{equation}
\label{s1}
{\cal L}(x)=K\left ( -\frac{\gamma-1}{K\gamma}x\right )
^{\frac{\gamma}{\gamma-1}}.
\end{equation}
It is a pure power-law ${\cal L}\propto x^{\frac{\gamma}{\gamma-1}}$. In
particular, for the Chaplygin gas ($\gamma=-1$), for the standard BEC
($\gamma=2$) and for the DM
superfluid ($\gamma=3$) we
have ${\cal L}\propto x^{1/2}$, ${\cal L}\propto x^2$ and ${\cal L}\propto
x^{3/2}$ respectively. For the unitary Fermi gas ($\gamma=5/3$) we
have ${\cal L}\propto x^{5/2}$. For an
isothermal equation of state $P=K\rho$, the
reduced Lagrangian is
\begin{equation}
\label{e2}
{\cal L}(x)=K\rho_*e^{-x/K}.
\end{equation}
For a logotropic equation of state $P=A\ln(\rho/\rho_*)$, the
reduced Lagrangian is
\begin{equation}
\label{e3}
{\cal L}(x)=-A \ln\left ( \frac{\rho_*}{A}x\right ).
\end{equation}

\section{Relativistic theory}
\label{sec_ra}

\subsection{Klein-Gordon equation}
\label{sec_csf}

We consider a relativistic  complex SF $\varphi(x^\mu)=\varphi(x,y,z,t)$ which
is a continuous
function of space and time. It can represent the wavefunction of a
relativistic BEC. The action of the SF can be written as
\begin{equation}
S=\int \mathcal{L} \sqrt{-g}\, d^4x,
\label{csf1}
\end{equation}
where $\mathcal{L}=\mathcal{L}(\varphi,
\varphi^*,\partial_\mu\varphi,\partial_\mu\varphi^*)$
is the Lagrangian density and $g={\rm det}(g_{\mu\nu})$ is the determinant of
the metric tensor.  We consider a
canonical Lagrangian density of the form
\begin{eqnarray}
{\cal L}=\frac{1}{2}g^{\mu\nu}\partial_{\mu}\varphi^*\partial_{\nu}
\varphi-V_{\rm tot}(|\varphi|^2),
\label{csf2}
\end{eqnarray}
where the first term is the kinetic energy and the second term is minus the
potential energy. The potential energy can be decomposed into a rest-mass energy
term and a self-interaction energy term:
\begin{equation}
\label{csf3}
V_{\rm
tot}(|\varphi|^2)=\frac{1}{2}\frac{m^2c^2}{\hbar^2}|\varphi|^2+V(|\varphi|^2).
\end{equation}
The least action principle $\delta S=0$ with respect to variations
$\delta\varphi$
(or $\delta\varphi^*$),
which is equivalent to the
Euler-Lagrange equation
\begin{eqnarray}
\label{lh2rel}
D_{\mu}\left\lbrack \frac{\partial {\cal
L}}{\partial(\partial_\mu\varphi)^*}\right\rbrack-\frac{\partial {\cal
L}}{\partial\varphi^*}=0,
\end{eqnarray}
yields the KG equation
\begin{equation}
\label{csf4}
\square\varphi+2\frac{dV_{\rm tot}}{d|\varphi|^2}\varphi=0,
\end{equation}
where
$\square=D_{\mu}\partial^{\mu}=\frac{1}{\sqrt{-g}}\partial_{\mu}(\sqrt{-g}\, g^{
\mu\nu} \partial_{\nu})$ is the d'Alembertian
operator in a curved spacetime. It can
be written
explicitly as
\begin{eqnarray}
\label{ak26}
\square\varphi=D_\mu\partial^{\mu}\varphi&=&g^{\mu\nu}D_\mu\partial_{\nu}
\varphi=g^{\mu\nu}(\partial_\mu\partial_\nu\varphi-\Gamma_{\mu\nu}^{\sigma}
\partial_{\sigma}\varphi)\nonumber\\
&=&\frac{1}{\sqrt{-g}}\partial_\mu(\sqrt{-g}\partial^{\mu}\varphi).
\end{eqnarray}
For
a free massless SF ($V_{\rm tot}=0$), the KG equation reduces to
$\square\varphi=0$. 

The energy-momentum (stress) tensor is given by
\begin{eqnarray}
\label{em1}
T_{\mu\nu}&=&\frac{2}{\sqrt{-g}}\frac{\delta S}{\delta g^{\mu\nu}}=\frac{2}{\sqrt{-g}}\frac{\partial (\sqrt{-g}{\cal L})}{\partial g^{\mu\nu}}\nonumber\\
&=&2\frac{\partial {\cal L}}{\partial g^{\mu\nu}}-g_{\mu\nu}{\cal L}.
\end{eqnarray}
For a complex SF, we have 
\begin{eqnarray}
\label{em1bb}
T_{\mu}^{\nu}=\frac{\partial {\cal L}}{\partial
(\partial_\nu\varphi)}\partial_\mu\varphi+\frac{\partial {\cal L}}{\partial
(\partial_\nu\varphi^*)}\partial_\mu\varphi^*-g_{\mu}^{\nu}{\cal L}.
\end{eqnarray}
For the Lagrangian (\ref{csf2}), we obtain
\begin{eqnarray}
\label{em2}
T_{\mu\nu}&=&\frac{1}{2}(\partial_{\mu}\varphi^*\partial_{\nu}\varphi+\partial_{\nu}\varphi^*\partial_{\mu}\varphi)
\nonumber\\
&-&g_{\mu\nu}\left\lbrack \frac{1}{2}g^{\rho\sigma}\partial_{\rho}\varphi^*\partial_{\sigma}
\varphi-V_{\rm tot}(|\varphi|^2)\right\rbrack.
\end{eqnarray}
The conservation of
the energy-momentum tensor,
which results from the invariance of the Lagrangian density under continuous
translations in space and time (Noether theorem \cite{noether}), writes
\begin{eqnarray}
\label{lh2r}
D_{\nu}T^{\mu\nu}=0.
\end{eqnarray}
The energy-momentum four vector is $P^{\mu}=\int T^{\mu 0}\sqrt{-g}\, d^3x$. 
Its time component $P^0$ is the energy. Each
component of $P^\mu$ is conserved
in time, i.e., it is a constant of motion. Indeed, we have
\begin{eqnarray}
\label{em1b}
{\dot P}^{\mu}=\frac{d}{dt}\int T^{\mu 0}\sqrt{-g}\, d^3x=c\int \partial_0
(T^{\mu 0}\sqrt{-g})\, d^3x\nonumber\\
=-c\int
\partial_i (T^{\mu i}\sqrt{-g})\, d^3x=0,
\end{eqnarray}
where we have used Eq. (\ref{lh2r}) with
$D_\mu V^\mu=\frac{1}{\sqrt{-g}}\partial_\mu(\sqrt{-g}V^{\mu})$ to get the third
equality.

The current of charge of a complex SF is given by
\begin{eqnarray}
\label{j1}
J^{\mu}=\frac{m}{i\hbar}\left \lbrack\varphi\frac{\partial {\cal L}}{\partial
(\partial_\mu\varphi)}-\varphi^* \frac{\partial {\cal L}}{\partial
(\partial_\mu\varphi^*)}\right\rbrack.
\end{eqnarray}
For the Lagrangian (\ref{csf2}), we obtain
\begin{eqnarray}
\label{charge1}
J_{\mu}=-\frac{m}{2i\hbar}
(\varphi^*\partial_\mu\varphi-\varphi\partial_\mu\varphi^*).
\end{eqnarray}
Using the KG equation (\ref{csf4}), one can show
that 
\begin{eqnarray}
\label{charge2}
D_{\mu}J^{\mu}=0.
\end{eqnarray}
This equation expresses the local conservation of the charge. The total charge
of the SF is
\begin{eqnarray}
\label{charge3}
Q=\frac{e}{mc}\int J^0\sqrt{-g}\, d^3x,
\end{eqnarray}
and we easily find from Eq. (\ref{charge2}) that $\dot Q=0$. 
The charge $Q$ is proportional to the number $N$
of bosons 
provided that antibosons are counted 
negatively \cite{landaulifshitz}.  Therefore, Eq. (\ref{charge2}) also
expresses the local conservation of the boson number ($Q=Ne$). This conservation
law
results via the Noether theorem from the global $U(1)$ symmetry of the
Lagrangian, i.e., from the invariance of the Lagrangian density
under a global phase transformation $\varphi\rightarrow \varphi e^{-i\theta}$
(rotation) of the
complex SF. Note that $J_{\mu}$
vanishes for a real SF so the charge and the particle number are not
conserved in that case.

The Einstein-Hilbert action of general relativity is
\begin{equation}
S_g=\frac{c^4}{16\pi G}\int R \sqrt{-g}\, d^4x,
\label{ak20}
\end{equation}
where $R$ the Ricci scalar curvature. The least action principle $\delta S_g=0$
with respect to variations $\delta g^{\mu\nu}$ yields the Einstein field
equations
\begin{equation}
G_{\mu\nu}\equiv R_{\mu\nu}-\frac{1}{2}g_{\mu\nu}R=\frac{8\pi G}{c^4}T_{\mu\nu}.
\label{ak21}
\end{equation}
The contracted Bianchi identity $D_{\nu}G^{\mu\nu}=0$ implies
the conservation of the energy momentum tensor ($D_{\nu}T^{\mu\nu}=0$).

\subsection{The de Broglie transformation}
\label{sec_db}

We can write the KG equation (\ref{csf4}) under the form of 
hydrodynamic equations by using the de Broglie
\cite{broglie1927a,broglie1927b,broglie1927c}  transformation. To that
purpose, we write the SF as
\begin{equation}
\varphi=\frac{\hbar}{m}\sqrt{\rho}e^{i S_{\rm tot}/\hbar},
\label{db1}
\end{equation}
where $\rho$ is the pseudo rest-mass density\footnote{We stress that $\rho$ is {\it not} the rest-mass density $\rho_m=nm$ (see below). It is only in the nonrelativistic regime $c\rightarrow +\infty$ that $\rho$ coincides with the rest-mass density $\rho_m$.} and $S_{\rm tot}$ is the action. They are given by
\begin{eqnarray}
\rho=\frac{m^2}{\hbar^2}|\varphi|^2\quad {\rm and}\quad S_{\rm tot}=\frac{\hbar}{2i}\ln \left (\frac{\varphi}{\varphi^*}\right ).
\label{db2}
\end{eqnarray}
For convenience, we define $\theta=S_{\rm tot}/m$ (see footnote 14) so that
Eq.
(\ref{db1}) can be rewritten as
\begin{equation}
\varphi=\frac{\hbar}{m}\sqrt{\rho}e^{i m \theta/\hbar}.
\label{db4}
\end{equation}
Substituting this expression into the Lagrangian density (\ref{csf2}), we obtain
\begin{eqnarray}
{\cal
L}=\frac{1}{2}g^{\mu\nu}\rho\partial_{\mu}\theta\partial_{\nu}
\theta+\frac{\hbar^2}{8m^2\rho}g^{\mu\nu}\partial_{\mu}\rho\partial_{\nu}
\rho-V_{\rm tot}(\rho)
\label{db5}
\end{eqnarray}
with
\begin{equation}
\label{db10}
V_{\rm
tot}(\rho)=\frac{1}{2}\rho c^2+V(\rho).
\end{equation}
The Euler-Lagrange equations for $\theta$ and $\rho$, expressing the
least action principle, are 
\begin{eqnarray}
\label{db6}
D_{\mu}\left \lbrack \frac{\partial
{\cal L}}{\partial(\partial_{\mu}\theta)}\right\rbrack-\frac{\partial
{\cal L}}{\partial\theta}=0,
\end{eqnarray}
\begin{eqnarray}
\label{db7}
D_{\mu}\left \lbrack \frac{\partial
{\cal L}}{\partial(\partial_{\mu}\rho)}\right\rbrack-\frac{\partial
{\cal L}}{\partial\rho}=0.
\end{eqnarray}
They yield
\begin{eqnarray}
D_{\mu}\left ( \rho \partial^{\mu}\theta\right )=0,
\label{db8}
\end{eqnarray}
\begin{eqnarray}
\frac{1}{2}\partial_{\mu}\theta\partial^{\mu}\theta-\frac{\hbar^2}{2m^2}\frac{
\square\sqrt { \rho } } { \sqrt { \rho } } -  V'_{\rm
tot}(\rho)=0.
\label{db9}
\end{eqnarray}
The
same equations are obtained by substituting the de Broglie transformation from
Eq.
(\ref{db4}) into the
KG equation
(\ref{csf4}),
and by separating the real and the imaginary parts. Equation (\ref{db8}) can be
interpreted as a continuity equation and Eq.
(\ref{db9}) can be interpreted as a  quantum relativistic Hamilton-Jacobi (or Bernoulli)
equation with a relativistic covariant quantum potential
\begin{eqnarray}
Q=\frac{\hbar^2}{2m}\frac{\square\sqrt{\rho}}{\sqrt{\rho}}.
\label{db11}
\end{eqnarray}
Introducing the pseudo quadrivelocity\footnote{The pseudo quadrivelocity $v_\mu$
does not 
satisfy $v_\mu v^\mu=c^2$ so it is not guaranteed to be always timelike.
Nevertheless, $v_\mu$ can be introduced as a convenient notation.}
\begin{eqnarray}
v_\mu=-\frac{\partial_{\mu}S_{\rm tot}}{m}=-\partial_\mu\theta,
\label{db12}
\end{eqnarray}
we can rewrite Eqs. (\ref{db8}) and  (\ref{db9}) as
\begin{eqnarray}
D_{\mu}\left ( \rho v^{\mu}\right )=0,
\label{db13}
\end{eqnarray}
\begin{eqnarray}
\frac{1}{2}m v_{\mu}v^{\mu}-Q-m V'_{\rm
tot}(\rho)=0.
\label{db14}
\end{eqnarray}
Taking the gradient of the quantum Hamilton-Jacobi equation (\ref{db14}) we
obtain \cite{chavmatos} 
\begin{eqnarray}
\frac{dv_\nu}{dt}\equiv v^{\mu}D_\mu v_\nu=\frac{1}{m}\partial_\nu Q+\partial_\nu V'(\rho),
\label{db15}
\end{eqnarray}
which can be interpreted as a relativistic quantum Euler equation (with the
limitations of footnote 19). The first term on the right hand side can be
interpreted as a quantum force and the second term as a pressure force
$(1/\rho)\partial_\nu P$ given by
$(1/\rho)P'(\rho)=h'(\rho)=V''(\rho)$, where $h$ is the pseudo enthalpy. We note
that 
the pressure is determined by Eqs. (\ref{mad11})-(\ref{etoile}) as in the
nonrelativistic case.

The energy-momentum tensor is given by Eq.
(\ref{em1}) or, in
the hydrodynamic representation, by 
\begin{eqnarray}
\label{tmunuh}
T_{\mu}^{\nu}=\frac{\partial {\cal L}}{\partial
(\partial_\nu\theta)}\partial_\mu\theta+\frac{\partial {\cal L}}{\partial
(\partial_\nu\rho)}\partial_\mu\rho-g_{\mu}^{\nu}{\cal L}.
\end{eqnarray}
For the Lagrangian (\ref{db5}) we obtain
\begin{eqnarray}
T_{\mu\nu}=\rho\partial_{\mu}\theta\partial_{\nu}\theta+\frac{\hbar^2}{4m^2\rho}
\partial_{\mu}\rho\partial_{\nu}\rho-g_{\mu\nu}{\cal L}.
\label{em3gen}
\end{eqnarray}

The current of charge of a complex SF is given by
\begin{eqnarray}
J^{\mu}=-\frac{\partial {\cal L}}{\partial(\partial_{\mu}\theta)}
\end{eqnarray}
For the Lagrangian (\ref{csf2}), we obtain
\begin{eqnarray}
J_{\mu}=-\frac{\rho}{m}\partial_{\mu}S_{\rm tot}=-\rho\partial_{\mu}\theta=\rho
v_{\mu}.
\label{charge4}
\end{eqnarray}
This result can also be obtained from  Eq. (\ref{charge1}) by using Eq.
(\ref{db1}) coming from the de
Broglie transformation. We then see that the continuity equation
(\ref{db8}) or
(\ref{db13}) is
equivalent to Eq. (\ref{charge2}). It
expresses the
conservation of the charge $Q$ of the SF (or the conservation of the boson
number
$N$)
\begin{eqnarray}
Q=Ne=-\frac{e}{mc}\int \rho\partial^0\theta \sqrt{-g}  \, d^3x.
\label{charge5}
\end{eqnarray}
Assuming $\partial_\mu\theta
\partial^\mu\theta>0$, we can introduce the fluid
quadrivelocity\footnote{It 
differs from the pseudo quadrivelocity $v_{\mu}$ introduced in Eq.
(\ref{db12}).}
\begin{eqnarray}
u_{\mu}=-\frac{\partial_{\mu}\theta}{\sqrt{\partial_\mu\theta
\partial^\mu\theta} }c,
\label{rtf5bw}
\end{eqnarray}
which satisfies  the  identity
\begin{eqnarray}
u_{\mu}u^{\mu}=c^2.
\label{rtf5cw}
\end{eqnarray}
Using Eqs. (\ref{charge4}) and (\ref{rtf5bw}), we can write the current as
\begin{eqnarray}
J_{\mu}=\frac{\rho}{c}\sqrt{\partial_\mu\theta
\partial^\mu\theta}\, u_{\mu}
\label{charge8w}
\end{eqnarray}
and the continuity equation  as
\begin{eqnarray}
D_{\mu}\left \lbrack \rho \sqrt{\partial_\mu\theta
\partial^\mu\theta} \, u^{\mu}\right \rbrack=0.
\label{charge9}
\end{eqnarray}
The rest-mass density $\rho_m=nm$ (which is
proportional to the charge density $\rho_e$) is defined by
\begin{eqnarray}
J_{\mu}=\rho_m  u_{\mu}.
\label{charge6w}
\end{eqnarray}
Using Eq. (\ref{rtf5cw}), we see that $\rho_m c^2=u_\mu J^\mu$.
The continuity equation
(\ref{charge2}) can be written as
\begin{eqnarray}
D_{\mu}(\rho_m u^{\mu})=0.
\label{charge7}
\end{eqnarray}
Comparing Eq. (\ref{charge8w}) with Eq. (\ref{charge6w}), we
find that the
rest-mass density $\rho_m=nm$ of the SF is given
by
\begin{eqnarray}
\rho_m=\frac{\rho}{c}  \sqrt{\partial_\mu\theta
\partial^\mu\theta}.
\label{charge10a}
\end{eqnarray}
Using the Bernoulli equation (\ref{db9}), we get
\begin{eqnarray}
\rho_m=\frac{\rho}{c}
\sqrt{\frac{\hbar^2}{m^2}\frac{\square\sqrt{\rho}}{\sqrt{\rho}}+2V_{\rm
tot}'(\rho)}.
\label{charge10b}
\end{eqnarray}

{\it Remark:} More generally, we can define the
quadrivelocity by 
\begin{eqnarray}
u^{\mu}=\frac{J^{\mu}}{\sqrt{J_{\mu}J^{\mu}}}c,
\end{eqnarray}
which satisfies the identity (\ref{rtf5cw}). Using Eq.
(\ref{charge6w}) we find that the rest-mass (or charge) density is given by
\begin{eqnarray}
\rho_m=\frac{1}{c}\sqrt{J_{\mu}J^{\mu}}.
\end{eqnarray}
We note that $J^0$
is not equal to the rest-mass density in general ($\rho_m\neq
J^0/c$) except if the SF is static in which case $u^\mu=c\, \delta^{\mu}_0$
and $J^0=\rho_m c$.

\subsection{TF approximation}
\label{sec_rtf}

In the classical or TF limit ($\hbar\rightarrow 0$), the Lagrangian from Eq.
(\ref{db5}) reduces to 
\begin{eqnarray}
{\cal
L}=\frac{1}{2}g^{\mu\nu}\rho\partial_{\mu}\theta\partial_{\nu}
\theta-V_{\rm tot}(\rho).
\label{rtf1}
\end{eqnarray}
The Euler-Lagrange equations (\ref{db6}) and (\ref{db7})
yield the equations of motion
\begin{eqnarray}
D_{\mu}\left ( \rho \partial^{\mu}\theta\right )=0,
\label{rtf4}
\end{eqnarray}
\begin{eqnarray}
\frac{1}{2}\partial_{\mu}\theta\partial^{\mu}\theta-V'_{\rm
tot}(\rho)=0.
\label{rtf5}
\end{eqnarray}
The same equations are obtained by making the TF approximation in Eq. (\ref{db9}), i.e., by neglecting the quantum potential. Equation (\ref{rtf4}) can be interpreted as a continuity equation and Eq.
(\ref{rtf5}) can be interpreted as a classical relativistic Hamilton-Jacobi (or Bernoulli)
equation. In order to determine the rest mass density, we can
proceed  as before. Assuming
$V_{\rm tot}'>0$, and using Eq. (\ref{rtf5}), we introduce the fluid
quadrivelocity
\begin{eqnarray}
u_{\mu}=-\frac{\partial_{\mu}\theta}{\sqrt{2V_{\rm
tot}'(\rho)}}c,
\label{rtf5b}
\end{eqnarray}
which satisfies the identity (\ref{rtf5cw}). Using Eqs. (\ref{charge4}) and
(\ref{rtf5b}), we can write the
current as
\begin{eqnarray}
J_{\mu}=\frac{\rho}{c}\sqrt{2V'_{\rm tot}(\rho)}\, u_{\mu}
\label{charge8}
\end{eqnarray}
and the continuity equation (\ref{rtf4}) as
\begin{eqnarray}
D_{\mu}\left \lbrack \rho \sqrt{2V_{\rm tot}'(\rho)}u^{\mu}\right \rbrack=0.
\label{charge9b}
\end{eqnarray}
Comparing Eq. (\ref{charge8}) with Eq. (\ref{charge6w}), we find that the
rest-mass density $\rho_m=nm$  is given, in the TF approximation, 
by
\begin{eqnarray}
\rho_m=\frac{\rho}{c} \sqrt{2V_{\rm tot}'(\rho)}.
\label{charge10c}
\end{eqnarray}
In general, $\rho_m\neq \rho$ except when $V$ is
constant, corresponding to the $\Lambda$CDM model (see below), and in the
nonrelativistic limit $c\rightarrow +\infty$.

The energy-momentum tensor is given by Eq. (\ref{em1}) or, in
the hydrodynamic representation, by Eq. (\ref{tmunuh}). For the Lagrangian
(\ref{rtf1}) we obtain
\begin{eqnarray}
T_{\mu\nu}=\rho\partial_{\mu}\theta\partial_{\nu}
\theta-g_{\mu\nu}{\cal L}
\label{em3}
\end{eqnarray}
or, using Eq. (\ref{rtf5b}),
\begin{eqnarray}
T_{\mu\nu}=2\rho V'_{\rm tot}(\rho) \frac{u_{\mu}u_{\nu}}{c^2}-g_{\mu\nu}{\cal
L}.
\label{em3b}
\end{eqnarray}
The   energy-momentum tensor (\ref{em3b}) can be written under the perfect fluid
form\footnote{We note that 
\begin{eqnarray}
\epsilon=\frac{u^{\mu}u^{\nu}}{c^2}T_{\mu\nu}.
\end{eqnarray}
}
\begin{eqnarray}
T_{\mu\nu}=(\epsilon+P)\frac{u_{\mu}u_{\nu}}{c^2}-P g_{\mu\nu},
\label{em4}
\end{eqnarray}
where $\epsilon$ is the energy density and $P$ is the pressure, provided that we
make the identifications
\begin{eqnarray}
P={\cal L},\qquad \epsilon+P=2\rho V'_{\rm tot}(\rho).
\label{em5}
\end{eqnarray}
Therefore, the Lagrangian plays the role of the pressure of the fluid. 
Combining  Eq. (\ref{rtf1}) with the Bernoulli equation (\ref{rtf5}), we get
\begin{eqnarray}
{\cal L}=\rho V'_{\rm tot}(\rho)-V_{\rm tot}(\rho).
\label{em6}
\end{eqnarray}
Therefore, according to Eqs. (\ref{em5}) and (\ref{em6}), the energy density and
the pressure derived from the Lagrangian (\ref{rtf1}) 
are given by 
\begin{eqnarray}
\epsilon=\rho V'_{\rm tot}(\rho)+V_{\rm
tot}(\rho)=\rho c^2+\rho V'(\rho)+V(\rho),
\label{rtf6}
\end{eqnarray}
\begin{eqnarray}
P=\rho V'_{\rm tot}(\rho)-V_{\rm
tot}(\rho)=\rho V'(\rho)-V(\rho),
\label{rtf7}
\end{eqnarray}
where we have used Eq. (\ref{db10}) to get the second equalities. Eliminating
$\rho$ between these equations, we obtain the equation of state $P(\epsilon)$.
Equation (\ref{rtf7}) for the pressure is exactly the same as
Eq. (\ref{mad11b}) obtained  in the nonrelativistic limit. Therefore, knowing
$P(\rho)$, we can obtain the SF potential $V(\rho)$ by the formula [see Eq.
(\ref{etoile})] \footnote{We note that the potential is defined from the
pressure up to a term of the form $A\rho$, where $A$ is a constant. If we add a
term $A\rho$ in the potential $V$, we do not change the pressure $P$ but we
introduce a term $2A\rho$ in the energy density. On the other hand, if we add a
constant term $C$ in the potential $V$ (cosmological constant), this adds a term
$-C$ in the pressure and a term $+C$ in the energy density. }
\begin{eqnarray}
V(\rho)=\rho\int\frac{P(\rho)}{\rho^2}\, d\rho.
\label{etoilebis}
\end{eqnarray}
The squared speed of sound is
\begin{eqnarray}
c_s^2=P'(\epsilon)c^2=\frac{\rho V''(\rho)c^2}{c^2+\rho V''(\rho)+2V'(\rho)}.
\label{rtf7x}
\end{eqnarray}

{\it Remark:}  In \cite{abrilas} we have considered a spatially homogeneous
complex SF in an expanding universe described by the Klein-Gordon-Friedmann
(KGF) equations. In the fast oscillation regime
$\omega\gg H$, where $\omega$ is the pulsation of the SF and $H$ the Hubble
constant,  we can average the KG equation over the oscillations of the SF (see
Appendix A of \cite{abrilas} and references therein) and obtain a virial
relation
leading to Eqs. (\ref{rtf6}) and (\ref{rtf7}). These equations can also be
obtained 
by transforming the KG equation into hydrodynamic equations, taking the limit
$\hbar\rightarrow 0$, and using the Bernoulli equation (see Sec. II of
\cite{abrilas}).\footnote{We cannot directly take the limit $\hbar\rightarrow 0$
in
the KG equation. This is why we have to average over the oscillations.
Alternatively, we can directly take the limit $\hbar\rightarrow 0$  in the
hydrodynamic equations associated with the KG equation. This is equivalent to
the WKB method.}  This is similar to the derivation given here. However, the
present derivation is more general since it applies to a possibly inhomogeneous
SF \cite{btv}. An interest of the results of \cite{abrilas} is to show
that
the
fast oscillation approximation in cosmology is equivalent to the TF
approximation.

\subsection{Reduced Lagrangian ${\cal L}(X)$}
\label{sec_k}

In the previous section, we have used the Bernoulli equation (\ref{rtf5}) to 
eliminate $\theta$ from the Lagrangian, leading to Eq. (\ref{em6}). Here, we
eliminate $\rho$ from the Lagrangian. Introducing the notation
\begin{eqnarray}
X=\frac{1}{2}\partial_{\mu}\theta\partial^{\mu}\theta,
\label{k1}
\end{eqnarray}
the  Bernoulli equation (\ref{rtf5}) can be written as
\begin{eqnarray}
X=V'_{\rm tot}(\rho).
\label{k2}
\end{eqnarray}
Assuming $V''_{\rm tot}>0$, this equation can be reversed to give
\begin{eqnarray}
\rho=G(X)
\label{k2inv}
\end{eqnarray}
with $G(X)=(V'_{\rm tot})^{-1}(X)$. As a result, the equation of continuity
(\ref{rtf4})
can be written 
as 
\begin{eqnarray}
D_\mu\lbrack G(X)\partial^\mu\theta\rbrack=0.
\label{k2invb}
\end{eqnarray}
According to Eqs. (\ref{rtf6}), (\ref{rtf7}) and (\ref{k2inv}), we have
$\epsilon=\epsilon(X)$ and $P=P(X)$. Therefore,
\begin{eqnarray}
{\cal
L}=P(X).
\label{k2b}
\end{eqnarray}
In this manner, we have eliminated the
pseudo rest-mass density $\rho$ from the Lagrangian (\ref{rtf1}) and we have 
obtained a reduced  Lagrangian of the form ${\cal L}(X)$  that depends only
on $X$. This kind of Lagrangian, called
k-essence Lagrangian, is discussed in Appendix \ref{sec_lrsf}. We show below
that
\begin{eqnarray}
\rho=G(X)=P'(X)={\cal L}'(X).
\label{k8b}
\end{eqnarray}
Using Eq. (\ref{k8b}), we can rewrite Eq. (\ref{k2invb}) in terms of ${\cal
L}'(X)$ as in Eq. (\ref{ak8}). We also show below that
\begin{eqnarray}
\epsilon=2XP'(X)-P.
\label{k3}
\end{eqnarray}
If we know $\epsilon=\epsilon(P)$ we can solve this differential equation to
obtain $P(X)$, hence ${\cal L}(X)$. 

{\it Proof of Eq. (\ref{k3}):} According to Eqs. (\ref{rtf7}) and (\ref{k2}),
we have
\begin{eqnarray}
P'(\rho)=\rho V''_{\rm tot}(\rho)
\label{k5}
\end{eqnarray}
and
\begin{eqnarray}
\frac{dX}{d\rho}=V''_{\rm tot}(\rho).
\label{k7}
\end{eqnarray} 
Starting from Eq. (\ref{k2b}) and using Eqs. (\ref{k5}) and (\ref{k7}) we obtain
\begin{eqnarray}
{\cal L}'(X)=P'(X)=P'(\rho)\frac{d\rho}{d X}=\rho V''_{\rm
tot}(\rho)\frac{d\rho}{d X}\nonumber\\
=\rho
\frac{dX}{d\rho}\frac{d\rho}{dX}=\rho,
\label{k8}
\end{eqnarray}
which establishes Eq. (\ref{k8b}). On the other hand, according to Eqs.
(\ref{rtf6}) and (\ref{rtf7}), we have
\begin{eqnarray}
\epsilon+P=2\rho V'_{\rm tot}(\rho).
\label{k9}
\end{eqnarray}
Using Eqs. (\ref{k2}) and (\ref{k8}), we obtain
\begin{eqnarray}
\epsilon+P=2\rho X=2X P'(X),
\label{k10}
\end{eqnarray}
which establishes Eq. (\ref{k3}).

{\it Remark:} We can obtain the preceding results in a more direct and more
general manner from a $k$-essence Lagrangian ${\cal L}(X)$ by using the results
of Appendix \ref{sec_lrsf}. The present calculations show 
how a $k$-essential Lagrangian arises from the canonical Lagrangian of a
complex SF $\varphi$ in the TF limit. In that case, the real SF $\theta$
represents
the phase
of the complex SF $\varphi$.

\subsection{Nonrelativistic limit}
\label{sec_nr}

To obtain the nonrelativistic limit 
of the foregoing equations, we first have to make the Klein transformation
(\ref{klein}) then take the limit $c\rightarrow +\infty$. In this manner,
the KG equation (\ref{csf4}) reduces to the GP equation (\ref{mfgp9}) and the
relativistic
hydrodynamic equations (\ref{db13})-(\ref{db15}) reduce to the nonrelativistic
equations (\ref{mad6})-(\ref{mad10}). These
transformations are discussed in detail in \cite{abrilph,playa,chavmatos} for
self-gravitating BECs. Here, we consider the nongravitational case and
we focus on the
nonrelativistic limit of the Lagrangien ${\cal L}(X)$ from Sec.
\ref{sec_k} leading to
the Lagrangian  ${\cal L}(x)$ from Sec. \ref{sec_rln}.

Since ${\cal L}=P$ in the two cases, we just have to find 
the relation between $X$ and $x$ when $c\rightarrow +\infty$. Making the Klein
transformation\footnote{Substituting Eqs. (\ref{mad1}) and (\ref{db1}) into Eq.
(\ref{klein}), we obtain $S_{\rm tot}=S-mc^2 t$ which is equivalent to Eq.
(\ref{nr1}).}
\begin{equation}
\label{nr1}
\theta=\theta_{\rm NR}-c^2 t
\end{equation}
in Eq. (\ref{k1}), we obtain 
\begin{eqnarray}
\label{nr2}
X&=&\frac{1}{2}\partial_\mu\theta \partial^\mu\theta\nonumber\\
&\simeq &\frac{1}{2c^2}\left (\frac{\partial\theta}{\partial t}\right
)^2-\frac{1}{2}(\nabla\theta)^2\nonumber\\
&\simeq& \frac{1}{2c^2}\left (\frac{\partial\theta_{\rm NR}}{\partial t}\right
)^2-\frac{\partial\theta_{\rm NR}}{\partial t}+\frac{c^2}{2}
-\frac{1}{2}(\nabla\theta_{\rm NR})^2.\nonumber\\
\end{eqnarray}
Taking the limit $c\rightarrow +\infty$, we find that
\begin{eqnarray}
\label{nr3}
X\sim\frac{c^2}{2}.
\end{eqnarray}
The nonrelativistic limit is then given by
\begin{eqnarray}
\label{nr3b}
\frac{c^2}{2}-X\rightarrow \dot\theta_{\rm NR}+\frac{1}{2}(\nabla\theta_{\rm NR})^2.
\end{eqnarray}
Therefore, when $c\rightarrow +\infty$, we can write
\begin{eqnarray}
\label{nr5}
X\simeq \frac{c^2}{2}-x,
\end{eqnarray}
where $x$ is defined by Eq. (\ref{gc3}).

Using Eq. (\ref{nr5}) we can easily check that the equations of
Sec. \ref{sec_k} return the equations of Sec. \ref{sec_rln} in the
nonrelativistic limit. For example, using Eq. (\ref{db10}), the relation
$X=V'_{\rm tot}(\rho)$ reduces to $x=-V'(\rho)$. On the other hand,  using
$\epsilon\sim \rho c^2$,  Eq. (\ref{k3}) reduces to
\begin{eqnarray}
\label{gc2x1}
\rho \sim  P'(x)\frac{dx}{dX}\sim - P'(x)\sim  - {\cal L}'(x),
\end{eqnarray}
which, together with Eq. (\ref{k8b}), returns Eq. (\ref{gc10b}).

\subsection{Enthalpy}
\label{sec_ent}

Using  Eqs. (\ref{mtd4a}), (\ref{rtf6}) and (\ref{rtf7}) we find that the
enthalpy is given by
\begin{equation}
h=2\frac{\rho}{\rho_m} V'_{\rm tot}(\rho).
\label{en1}
\end{equation}
Using Eq. (\ref{charge10c}), we obtain
\begin{equation}
h=\sqrt{2V'_{\rm tot}(\rho)}\, c.
\label{en2}
\end{equation}
According to Eq. (\ref{rtf5}), the enthalpy can be written as
\begin{equation}
h=c\sqrt{\partial_\mu\theta\partial^\mu\theta}=c\sqrt{2X}.
\label{en3}
\end{equation}
Substituting Eq. (\ref{db10}) into Eq. (\ref{en2}), subtracting
$c^2$, and taking the nonrelativistic limit $c\rightarrow +\infty$, we recover
Eq. (\ref{kggpv}).

\section{General equation of state}
\label{sec_ges}

In this section, we provide general results valid for an arbitrary equation of
state. We consider three different manners to specify the equation of state
depending on whether the pressure $P$ is expressed as a function of (i) the
energy density $\epsilon$; (ii) the rest-mass density $\rho_m$; (iii) the
pseudo rest-mass 
density $\rho$. In each case, we determine the
pressure $P$, the energy density $\epsilon$, the rest-mass density $\rho_m$, the
internal energy $u$, the pseudo rest-mass density $\rho$, the SF potential
$V_{\rm tot}(\rho)$ and the k-essence Lagrangian ${\cal L}(X)$.

\subsection{Equation of state of type I}
\label{sec_gesu}

We first consider an equation of state of type I (see Appendix \ref{sec_mtu})
where the pressure is given as a function of the 
energy density: $P=P(\epsilon)$.

\subsubsection{Determination of $\rho_m$, $P(\rho_m)$ and $u(\rho_m)$}

Using the results of Appendix \ref{sec_mtd}, we can obtain the rest-mass density $\rho_m=n m$ and the internal energy $u$ as follows. According to Eq. (\ref{mtd4}), we have
\begin{equation}
\label{gesu1}
\ln\rho_m=\int \frac{d\epsilon}{P(\epsilon)+\epsilon},
\end{equation}
which determines $\rho_m(\epsilon)$. Eliminating $\epsilon$ between $P(\epsilon)$ and $\rho_m(\epsilon)$ we obtain $P(\rho_m)$. On the other hand, according to Eq. (\ref{mtd2}), we have
\begin{equation}
\label{gesu2}
u=\epsilon-\rho_m(\epsilon) c^2.
\end{equation}
Eliminating $\epsilon$ between Eqs. (\ref{gesu1}) and (\ref{gesu2}), we obtain $u(\rho_m)$. We can also obtain $u(\rho_m)$ from $P(\rho_m)$, or the converse, by using Eqs. (\ref{mtd6}) and (\ref{mtd7}).

\subsubsection{Determination
of $\rho$, $P(\rho)$  and $V_{\rm tot}(\rho)$}

Using the results of Sec. \ref{sec_rtf}, we can obtain the pseudo rest-mass density $\rho$ and the SF potential $V_{\rm tot}$ as follows. According to  Eqs. (\ref{rtf6}) and (\ref{rtf7}), we have
\begin{eqnarray}
\epsilon-P=2V_{\rm
tot}(\rho),
\label{gesu3}
\end{eqnarray}
\begin{eqnarray}
\epsilon+P=2\rho V'_{\rm tot}(\rho).
\label{gesu4}
\end{eqnarray}
Differentiating Eq. (\ref{gesu3}) and using Eq. (\ref{gesu4}), we get
\begin{eqnarray}
d(\epsilon-P)=2V'_{\rm
tot}(\rho)d\rho=\frac{\epsilon+P}{\rho}d\rho.
\label{gesu5}
\end{eqnarray}
This yields
\begin{eqnarray}
\ln\rho=\int \frac{1-P'(\epsilon)}{\epsilon+P(\epsilon)}\,d\epsilon,
\label{gesu6}
\end{eqnarray}
which determines $\rho(\epsilon)$. Eliminating $\epsilon$ between $P(\epsilon)$ and $\rho(\epsilon)$ we obtain $P(\rho)$. On the other hand, according to Eq. (\ref{gesu3}), we have
\begin{eqnarray}
V_{\rm tot}=\frac{1}{2}[\epsilon-P(\epsilon)].
\label{gesu7}
\end{eqnarray}
Eliminating $\epsilon$ between Eqs. (\ref{gesu6}) and (\ref{gesu7}), we obtain $V_{\rm tot}(\rho)$.
We can also obtain $V(\rho)$ from $P(\rho)$, or the converse, by using
Eqs. (\ref{rtf7}) and (\ref{etoilebis}).

{\it Remark:} If the relation $\epsilon(P)$ is more explicit than $P(\epsilon)$, we can use
\begin{eqnarray}
\ln\rho=\int \frac{\epsilon'(P)-1}{\epsilon(P)+P}\,dP
\label{gesu8}
\end{eqnarray}
and
\begin{eqnarray}
V_{\rm tot}=\frac{1}{2}[\epsilon(P)-P],
\label{gesu9}
\end{eqnarray}
instead of Eqs. (\ref{gesu6}) and (\ref{gesu7}). The first equation gives $\rho(P)$.
Eliminating $P$ between Eqs. (\ref{gesu8}) and (\ref{gesu9}), we obtain $V_{\rm tot}(\rho)$.

\subsubsection{Lagrangian ${\cal L}(X)$}

If we know $\epsilon=\epsilon(P)$ then, according to Eq. (\ref{k3}), we have
\begin{equation}
\label{gesu10}
\ln X=2\int \frac{dP}{\epsilon(P)+P},
\end{equation}
which determines $X(P)$. If this function can be inverted we get $P(X)$ hence  ${\cal L}(X)$. If we know $P=P(\epsilon)$, we can rewrite Eq. (\ref{gesu10}) as
\begin{equation}
\label{gesu11}
\ln X=2\int \frac{P'(\epsilon)}{\epsilon+P(\epsilon)}\, d\epsilon,
\end{equation}
which determines $X(\epsilon)$.  If this function can be inverted we get
$\epsilon(X)$, 
then  $P(X)=P[\epsilon(X)]$, hence ${\cal L}(X)$.

\subsection{Equation of state of type II}
\label{sec_gesd}

We now consider an equation of state of type II (see Appendix \ref{sec_mtd})
where the pressure is given as a function of the rest-mass density:
$P=P(\rho_m)$. We can then determine the internal energy $u(\rho_m)$ from Eq.
(\ref{mtd6}). Inversely, we can specify the internal energy $u(\rho_m)$ as a
function of the rest-mass density and obtain the equation of 
state $P(\rho_m)$ from Eq. (\ref{mtd7}).

\subsubsection{Determination of $\epsilon$ and $P(\epsilon)$}

According to Eqs. (\ref{mtd2}) and (\ref{mtd7}), the energy density and the pressure are given by
\begin{eqnarray}
\label{gesd1}
\epsilon=\rho_m c^2+u(\rho_m),
\end{eqnarray}
\begin{eqnarray}
\label{gesd2}
P=\rho_m u'(\rho_m)-u(\rho_m).
\end{eqnarray}
Eliminating $\rho_m$ between Eqs. (\ref{gesd1}) and (\ref{gesd2}), we obtain 
$P(\epsilon)$.

\subsubsection{Determination of $\rho$, $P(\rho)$ and $V_{\rm tot}(\rho)$}

According to Eq.  (\ref{gesu6}), we have
\begin{equation}
\label{gesd3}
\ln\rho=\int \frac{\epsilon'(\rho_m)-P'(\rho_m)}{\epsilon(\rho_m)+P(\rho_m)}\,
d\rho_m.
\end{equation}
Then, using Eqs. (\ref{gesd1}) and (\ref{gesd2}), we obtain
\begin{equation}
\label{gesd4}
\ln\rho=\int \frac{c^2+u'(\rho_m)-\rho_m u''(\rho_m)}{\rho_m c^2+\rho_m
u'(\rho_m)}\, d\rho_m,
\end{equation}
which determines $\rho(\rho_m)$. This equation can be
integrated into
\begin{equation}
\label{gesd4b}
\rho=\frac{\rho_m}{1+\frac{1}{c^2}u'(\rho_m)},
\end{equation}
where the constant of integration has been determined in order to obtain
$\rho=\rho_m$ 
in the nonrelativistic limit. Identifying  $\epsilon+P=2\rho V'_{\rm tot}(\rho)$
from Eqs. (\ref{rtf6}) and (\ref{rtf7}) with $\epsilon+P=\rho_m(c^2+u'(\rho_m))$
from Eqs. (\ref{gesd1}) and (\ref{gesd2}) we see that Eq. (\ref{gesd4b}) is
equivalent to Eq. (\ref{charge10c}). Eliminating $\rho_m$ between $P(\rho_m)$
and
$\rho(\rho_m)$, we obtain $P(\rho)$. On the other hand, according to Eqs.
(\ref{gesu7}), (\ref{gesd1}) and (\ref{gesd2}), we get
\begin{equation}
\label{gesd5}
V_{\rm tot}=\frac{1}{2}\left\lbrack \rho_m c^2+
2u(\rho_m)-\rho_m u'(\rho_m)\right\rbrack.
\end{equation}
Eliminating $\rho_m$ between Eqs. (\ref{gesd4}) and (\ref{gesd5}) we obtain 
$V_{\rm tot}(\rho)$.  
We can also obtain $V(\rho)$ from $P(\rho)$, or the converse, by using Eqs.
(\ref{rtf7}) and (\ref{etoilebis}).

\subsubsection{Lagrangian ${\cal L}(X)$}

According to Eq. (\ref{gesu10}), we have
\begin{equation}
\label{gesd6}
\ln X=2\int \frac{P'(\rho_m)}{\epsilon(\rho_m)+P(\rho_m)}\, d\rho_m.
\end{equation}
Using Eqs. (\ref{gesd1}) and (\ref{gesd2}), we obtain
\begin{equation}
\label{gesd7}
\ln X=2\int \frac{u''(\rho_m)}{c^2+u'(\rho_m)}\, d\rho_m,
\end{equation}
which can be integrated into 
\begin{equation}
\label{gesd8}
X=\frac{1}{2c^2}\left \lbrack c^2+u'(\rho_m)\right \rbrack^2.
\end{equation}
We have determined the constant of integration so that, in the
nonrelativistic limit, $X\sim c^2/2$ (see Sec. \ref{sec_nr}). From Eq. (\ref{gesd8}) we obtain
$X(\rho_m)$. If this function can be inverted we get $\rho_m(X)$, then
$P(X)=P[\rho_m(X)]$, hence ${\cal L}(X)$.

\subsection{Equation of state of type III}
\label{sec_gest}

Finally, we consider an equation of state of type III (see Sec. \ref{sec_ra}
and Appendix \ref{sec_mtt})
where the pressure  is given as a function of the pseudo rest-mass density:
$P=P(\rho)$. We can then determine the SF potential $V(\rho)$ from Eq.
(\ref{etoilebis}).\footnote{We note that the expression of $V(\rho)$ for an
equation of state $P(\rho)$ of type III coincides with the  expression of
$u(\rho_m)$ for an
equation of state $P(\rho_m)$ of type II of the same functional form provided
that we make the replacements $u\rightarrow V$ and
$\rho_m\rightarrow \rho$ (see Appendix \ref{sec_sum}).} Inversely, we can
specify the SF
potential $V(\rho)$ and 
obtain the equation of state $P(\rho)$ from 
Eq. (\ref{rtf7}).

\subsubsection{Determination of $\epsilon$ and $P(\epsilon)$}

According to Eqs. (\ref{rtf6}) and (\ref{rtf7}), the energy density and the pressure are given by
\begin{eqnarray}
\label{gest1}
\epsilon=\rho V'_{\rm tot}(\rho)+ V_{\rm tot}(\rho),
\end{eqnarray}
\begin{eqnarray}
\label{gest2}
P=\rho V'_{\rm tot}(\rho)-V_{\rm tot}(\rho).
\end{eqnarray}
Eliminating $\rho$ between Eqs. (\ref{gest1}) and (\ref{gest2}), we obtain 
$P(\epsilon)$.

\subsubsection{Determination of $\rho_m$, $P(\rho_m)$ and  $u(\rho_m)$}

According to Eq. (\ref{mtd4}), we have
\begin{equation}
\label{gest3}
\ln\rho_m=\int \frac{\epsilon'(\rho)}{\epsilon+P(\rho)}\, d\rho.
\end{equation}
Using Eqs. (\ref{gest1}) and (\ref{gest2}), we obtain
\begin{equation}
\label{gest4}
\ln\rho_m=\int \frac{\rho V_{\rm tot}''(\rho)+2V_{\rm
tot}'(\rho)}{2\rho V'_{\rm tot}(\rho)}\, d\rho,
\end{equation}
which determines $\rho_m(\rho)$. This equation can be
integrated into
\begin{equation}
\label{gest4a}
\rho_m=\frac{\rho}{c}\sqrt{2V'_{\rm tot}(\rho)},
\end{equation}
where the constant of integration has been determined in 
order to obtain $\rho_m=\rho$ in the nonrelativistic limit. This relation is
equivalent to Eq. (\ref{charge10c}). Eliminating $\rho$
between $P(\rho)$ and $\rho_m(\rho)$ we obtain $P(\rho_m)$. On the other hand,
according to Eq. (\ref{mtd2}), we have
\begin{equation}
\label{gest5}
u=\epsilon-\rho_m c^2.
\end{equation}
Using Eqs. (\ref{gest1}) and (\ref{gest4a}), we obtain
\begin{eqnarray}
\label{gest6}
u&=&\rho V_{\rm tot}'(\rho)+V_{\rm
tot}(\rho)-\rho_m(\rho)c^2\nonumber\\
&=&\rho V_{\rm tot}'(\rho)+V_{\rm
tot}(\rho)-\rho c\sqrt{2V'_{\rm tot}(\rho)}.
\end{eqnarray}
Eliminating $\rho$ between Eqs. (\ref{gest4}) and (\ref{gest6}) we obtain 
$u(\rho_m)$. We can 
also obtain $u(\rho_m)$ from $P(\rho_m)$, or the converse, by using Eqs.
(\ref{mtd6}) and (\ref{mtd7}).

\subsubsection{Lagrangian ${\cal L}(X)$}

According to Eq. (\ref{k2}) we have
\begin{equation}
\label{gest7}
X=V'_{\rm tot}(\rho),
\end{equation}
which determines $X(\rho)$. If this function can be inverted we get $\rho(X)$, 
then $P(X)=P[\rho(X)]$, hence ${\cal L}(X)$.

\section{Polytropes}
\label{sec_poly}

In this section, we apply the general results of Sec. \ref{sec_ges} to the case of a polytropic equation of state.

\subsection{Polytropic equation of state of type I}
\label{sec_pu}

The polytropic equation of state of type I writes \cite{tooper1}
\begin{eqnarray}
P=K\left (\frac{\epsilon}{c^2}\right )^{\gamma},
\label{pu1}
\end{eqnarray}
where $K$ is the polytropic constant and $\gamma=1+1/n$ is the polytropic index.
This is the equation of state of the GCG \cite{bentoGCG}. In the
nonrelativistic regime, 
using $\epsilon\sim\rho c^2$, we recover Eq. (\ref{lac2}).

(i) For $\gamma=-1$, we obtain
\begin{eqnarray}
P=\frac{Kc^2}{\epsilon}.
\label{cgpu1}
\end{eqnarray}
This is the equation of state of the Chaplygin ($K<0$) or 
anti-Chaplygin ($K>0$) gas \cite{kmp,btv,gkmp,cosmopoly2}.

(ii) For $\gamma=2$, we obtain
\begin{eqnarray}
P=K\left (\frac{\epsilon}{c^2}\right )^{2}.
\label{becpu1}
\end{eqnarray}
This is the equation of state of the standard BEC with repulsive ($K>0$) or
attractive
($K<0$) self-interaction.\footnote{The
true equation of state of 
a relativistic BEC is given by Eq. (\ref{pt1}) or (\ref{becpt3b}) 
corresponding to a polytrope of type III with index
$\gamma=2$ \cite{colpi,shapiro,abrilas,partially,chavharko,csf}. However,
in order
to
have a unified terminology throughout the paper, we shall always associate the
polytropic index $\gamma=2$ to a BEC even if  this association  is not quite
correct for models of type I and II in the relativistic regime (see
Appendix \ref{sec_stu}). As explained in
Appendix \ref{sec_sum}, all the models coincide in the nonrelativistic limit.}
In that case, $K={2\pi a_s\hbar^2}/{m^3}$ (see Sec. \ref{sec_bec}).

(iii) For $\gamma=0$, we obtain
\begin{eqnarray}
P=K.
\label{lcdmpu1}
\end{eqnarray}
This is the equation of state of the $\Lambda$CDM ($K<0$) or  anti-$\Lambda$CDM
($K>0$) model \cite{sandvik,avelinoZ,gkmp,cosmopoly2}. In that case
$K=-\rho_\Lambda c^2$, where $\rho_\Lambda=\Lambda/(8\pi G)$ is the 
cosmological density.

(iv) For $\gamma=3$, we obtain
\begin{eqnarray}
P=K\left (\frac{\epsilon}{c^2}\right )^{3}.
\label{becpu1sup}
\end{eqnarray}
This is the equation of state of a superfluid with repulsive
($K>0$) or attractive
($K<0$) self-interaction (see Sec. \ref{sec_sf}).

(v) The case $\gamma=1$ must be treated specifically. In that case, we have a
linear equation of state
\cite{chandra72,yabushita1,yabushita2,aarelat1,aarelat2}
\begin{eqnarray}
P=\alpha\epsilon,
\label{npu1}
\end{eqnarray}
where we have defined 
\begin{eqnarray}
\alpha=\frac{K}{c^2}.
\label{alpha}
\end{eqnarray}
This linear equation of state describes pressureless matter
($\alpha=0$), radiation ($\alpha=1/3$) and stiff matter
($\alpha=1$). 
The nonrelativistic limit corresponds to $\alpha\rightarrow 0$. Using
$\epsilon\sim\rho c^2$, we recover the isothermal equation of 
state (\ref{isog1}).

\subsubsection{Determination of $\rho_m$, $P(\rho_m)$ and $u(\rho_m)$}

The rest-mass density is
determined by Eq. (\ref{gesu1})  with the equation of state
(\ref{pu1}). We have
\begin{equation}
\label{pu1b}
\ln\rho_m=\int \frac{d\epsilon}{K\left (\frac{\epsilon}{c^2}\right )^{\gamma}+\epsilon}.
\end{equation}
The integral can be calculated analytically yielding
\begin{eqnarray}
\rho_m c^2=\frac{\epsilon}{\left\lbrack 1+\frac{K}{c^2}\left (\frac{\epsilon}{c^2}\right )^{\gamma-1}\right\rbrack^{1/(\gamma-1)}}.
\label{pu1c}
\end{eqnarray}
We have determined the constant of integration so that
$\epsilon\sim \rho_m c^2$ in the nonrelativistic limit.  Eq. (\ref{pu1c}) can be inverted to give
\begin{eqnarray}
\epsilon=\frac{\rho_m c^2}{\left (1-\frac{K\rho_m^{\gamma-1}}{c^2}\right )^{\frac{1}{\gamma-1}}}.
\label{pu1ci}
\end{eqnarray}
Substituting this result into Eq. (\ref{pu1}), we obtain
\begin{eqnarray}
P=\frac{K\rho_m^{\gamma}}{\left (1-\frac{K\rho_m^{\gamma-1}}{c^2}\right )^{\frac{\gamma}{\gamma-1}}}.
\label{pu1cib}
\end{eqnarray}
The internal energy is given by Eqs. (\ref{gesu2}) and (\ref{pu1ci})
giving
\begin{eqnarray}
u=\frac{\rho_m c^2}{\left (1-\frac{K\rho_m^{\gamma-1}}{c^2}\right )^{\frac{1}{\gamma-1}}}-\rho_m c^2.
\label{pu1di}
\end{eqnarray}
These 
results are consistent with those obtained in Appendix B.3 of
\cite{partially}. In the nonrelativistic limit, using $\epsilon\sim \rho_m
c^2\lbrack
1+\frac{K}{(\gamma-1)c^2}\rho_m^{\gamma-1}\rbrack$, we recover Eqs. (\ref{lac2})
and (\ref{lac3}) [recalling Eq. (\ref{addq})].

(i) For $\gamma=-1$ (Chaplygin gas), we obtain 
\begin{eqnarray}
\epsilon=\sqrt{(\rho_m c^2)^2-Kc^2},
\label{cgpu1c}
\end{eqnarray}
\begin{eqnarray}
P=\frac{Kc^2}{\sqrt{(\rho_m c^2)^2-Kc^2}},
\label{cgpu1ca}
\end{eqnarray}
\begin{eqnarray}
u=\sqrt{(\rho_m c^2)^2-Kc^2}-\rho_m c^2.
\label{cgpu1cb}
\end{eqnarray}

(ii) For $\gamma=2$ (BEC), we obtain 
\begin{eqnarray}
\epsilon=\frac{\rho_m c^2}{1-\frac{K\rho_m}{c^2}},
\label{becpu1c}
\end{eqnarray}
\begin{eqnarray}
P=\frac{K\rho_m^2}{\left (1-\frac{K\rho_m}{c^2}\right )^2},
\label{becpu1ca}
\end{eqnarray}
\begin{eqnarray}
u=\frac{K\rho_m^2}{1-\frac{K\rho_m}{c^2}}.
\label{becpu1cb}
\end{eqnarray}
For $K>0$ there is a maximum density $(\rho_m)_{\rm
max}=c^2/K$. The equation of state (\ref{becpu1ca})
was first obtained in \cite{partially}.

(iii) For $\gamma=0$ ($\Lambda$CDM model), we obtain
\begin{eqnarray}
\epsilon=\rho_m c^2-K,
\label{lcdmpu1c}
\end{eqnarray}
\begin{eqnarray}
P=K,
\end{eqnarray}
\begin{eqnarray}
u=-K.
\label{lcdmpu1cb}
\end{eqnarray}

(iv) For $\gamma=3$ (superfluid), we obtain 
\begin{eqnarray}
\epsilon=\frac{\rho_m c^2}{\left (1-\frac{K\rho_m^2}{c^2}\right )^{1/2}},
\label{becpu1csuper}
\end{eqnarray}
\begin{eqnarray}
P=\frac{K\rho_m^3}{\left (1-\frac{K\rho_m^2}{c^2}\right )^{3/2}},
\label{becpu1casuper}
\end{eqnarray}
\begin{eqnarray}
u=\frac{\rho_m c^2}{\left (1-\frac{K\rho_m^2}{c^2}\right )^{1/2}}-\rho_m c^2.
\label{becpu1cbsuper}
\end{eqnarray}
For $K>0$ there is a maximum density $(\rho_m)_{\rm
max}=c/\sqrt{K}$.

(v) For $\gamma=1$, Eq. (\ref{gesu1})  can be integrated into
\begin{eqnarray}
\rho_m=\left \lbrack\frac{\alpha\epsilon}{{\cal K}(\alpha)}\right \rbrack^{\frac{1}{1+\alpha}},
\label{upu1c}
\end{eqnarray}
where ${\cal K}(\alpha)$ is a constant that depends on $\alpha$. In the
nonrelativistic limit $\alpha\rightarrow 0$, the condition $\epsilon\sim\rho_m
c^2$ implies ${\cal K}(\alpha)\rightarrow \alpha c^2=K$. Combining Eq.
(\ref{upu1c}) with Eq. (\ref{npu1}), we obtain
\begin{eqnarray}
P={\cal K}(\alpha) \rho_m^{1+\alpha}.
\label{alpha1}
\end{eqnarray}
This is the equation of state of a polytrope of type II (see Sec. \ref{sec_pd})
with a
polytropic index $\Gamma=1+\alpha$ (i.e. $n=1/\alpha$) and a polytropic
constant
${\cal K}(\alpha)$.\footnote{The linear equation of
state (\ref{npu1})
with $\alpha=\Gamma-1$ corresponds to the ultrarelativistic limit of the
equation of state (\ref{pd5}) associated with a polytrope of type II with index
$\Gamma$. Indeed, for a polytrope $P=K\rho_m^{\Gamma}$, Eq.  (\ref{pd5}) yields
$P\sim (\Gamma-1)\epsilon$ in the ultrarelativistic limit. The index
$\Gamma=4/3$ corresponds to $\alpha=1/3$ (radiation) and the index  $\Gamma=2$
corresponds to $\alpha=1$ (stiff matter).} In the nonrelativistic limit
$\alpha\rightarrow 0$, we
obtain an isothermal equation of 
state $P=K\rho_m$ with a ``temperature'' $K$.  The internal energy (\ref{gesu2})
is  given by
\begin{eqnarray}
u=\rho_m c^2\left\lbrack \frac{{\cal K}(\alpha)}{\alpha c^2}\rho_m^{\alpha}-1\right\rbrack.
\label{npu1d}
\end{eqnarray}
It is similar to a Tsallis free energy density  $-{\cal K}
s_{q}$ (where $s_q=-\frac{1}{q-1}\rho_m^q$) of index $q=1+\alpha$ with a
``polytropic'' temperature
${\cal K}(\alpha)$. In the nonrelativistic
limit
$\alpha\rightarrow 0$ (i.e. $q\rightarrow 1$), Eq. (\ref{npu1d})
reduces to $u=K\rho_m\ln\rho_m$ (up
to an additive constant) and we recover Eq. (\ref{isog2}) [recalling Eq.
(\ref{addq})]. It is similar to the Boltzmann free energy density $-K s_{B}$
(where $s_B=-\rho_m\ln\rho_m$) with the
temperature $K$.  In the present context, the Tsallis
entropy arises from relativistic effects ($\alpha\neq 0\Rightarrow q\neq
1$).

\subsubsection{Determination
of $\rho$, $P(\rho)$ and $V_{\rm tot}(\rho)$}

The pseudo rest-mass density and the SF potential are determined by Eqs. (\ref{gesu6}) and (\ref{gesu7}) with the equation of state (\ref{pu1}). We have
\begin{eqnarray}
\ln\rho=\int \frac{1-\frac{K\gamma}{c^2} \left (\frac{\epsilon}{c^2}\right )^{\gamma-1}}{\epsilon+K \left (\frac{\epsilon}{c^2}\right )^{\gamma}}\,d\epsilon.
\label{pu1cq}
\end{eqnarray}
The integral can be calculated analytically yielding
\begin{eqnarray}
\rho c^2=\epsilon \left\lbrack 1+\frac{K}{c^2}\left (\frac{\epsilon}{c^2}\right )^{\gamma-1}\right\rbrack^{(1+\gamma)/(1-\gamma)}.
\label{pu2}
\end{eqnarray}
We have determined the constant of integration so that
$\epsilon\sim \rho c^2$ in the nonrelativistic limit.  The SF potential is given
by
\begin{equation}
\label{pu4}
V_{\rm tot}=\frac{1}{2}\left \lbrack \epsilon-K  \left (\frac{\epsilon}{c^2}\right )^{\gamma}\right\rbrack.
\end{equation}
Eqs. (\ref{pu1}), (\ref{pu2}) and (\ref{pu4})  define $P(\rho)$ and $V_{\rm tot}(\rho)$ in parametric form with parameter $\epsilon$.
In the nonrelativistic limit,
using $\epsilon\sim \rho c^2\lbrack
1-\frac{K(1+\gamma)}{(1-\gamma)c^2}\rho^{\gamma-1}\rbrack$ and Eq. (\ref{db10}),
we recover Eqs. (\ref{lac2}) and (\ref{lac3}).

(i) For $\gamma=-1$ (Chaplygin gas), we obtain
\begin{eqnarray}
\epsilon=\rho c^2,
\label{pu3}
\end{eqnarray}
\begin{eqnarray}
P=\frac{K}{\rho},
\label{pu3b}
\end{eqnarray}
\begin{equation}
\label{pu6b}
V_{\rm tot}(\rho)=\frac{1}{2}\left (\rho c^2-\frac{K}{\rho}\right ).
\end{equation}
Expression (\ref{pu6b}) of the SF potential was first given in \cite{btv}. We
note that the energy
density $\epsilon$ coincides with the
pseudo rest-mass energy density $\rho c^2$ [see Eq. (\ref{pu3})]. As a result,
$P(\rho)$ is a
Chaplygin equation of state of type III (see Sec.
\ref{sec_pt}). Therefore, the models I and III coincide in that case.

(ii) For $\gamma=2$ (BEC),  the energy density is determined by a cubic equation
\begin{equation}
\label{pu6c}
\rho c^2=\frac{\epsilon}{\left (1+\frac{K\epsilon}{c^4}\right )^3}.
\end{equation}
The solution $\epsilon(\rho)$ can be obtained by standard means. The total
potential $V_{\rm tot}(\rho)$ is then given by
\begin{equation}
\label{pu4b}
V_{\rm tot}=\frac{1}{2}\left \lbrack \epsilon-K  \left
(\frac{\epsilon}{c^2}\right )^{2}\right\rbrack
\end{equation}
with $\epsilon$ replaced by $\epsilon(\rho)$. Eqs. (\ref{pu6c})
and (\ref{pu4b}) also determine  $V_{\rm tot}(\rho)$ in parametric form.

(iii) For $\gamma=0$ ($\Lambda$CDM model), we obtain
\begin{eqnarray}
\epsilon=\rho c^2-K,
\label{lcdmpu3}
\end{eqnarray}
\begin{eqnarray}
P=K,
\end{eqnarray}
\begin{equation}
\label{lcdmpu6b}
V_{\rm tot}(\rho)=\frac{1}{2}\rho c^2-K.
\end{equation}
Comparing these results with Eqs. (\ref{lcdmpu1c})-(\ref{lcdmpu1cb}), we note
that $\rho=\rho_m$ and $V=u=-K$. The potential $V$ is constant.

(iv) For $\gamma=3$ (superfluid), the total
potential $V_{\rm tot}(\rho)$ is given in parametric form by
\begin{equation}
\label{pu6csuper}
\rho c^2=\frac{\epsilon}{\left \lbrack 1+\frac{K}{c^2}\left
(\frac{\epsilon}{c^2}\right )^2\right \rbrack^2},
\end{equation}
\begin{equation}
\label{pu4bsuper}
V_{\rm tot}=\frac{1}{2}\left \lbrack \epsilon-K  \left
(\frac{\epsilon}{c^2}\right )^{3}\right\rbrack.
\end{equation}
It is not possible to obtain more explicit expressions.

(v) For $\gamma=1$, Eq. (\ref{gesu6}) can be integrated into
\begin{eqnarray}
\rho=\left \lbrack\frac{\alpha\epsilon}{{\cal K}(\alpha)}\right\rbrack^{{(1-\alpha)/}{(1+\alpha)}},
\label{iu1}
\end{eqnarray}
where ${\cal K}(\alpha)$ is a constant that depends on $\alpha$.
In the nonrelativistic limit
$\alpha\rightarrow 0$, the condition $\epsilon\sim\rho c^2$ implies ${\cal
K}(\alpha)\rightarrow \alpha c^2=K$. Combining Eq. (\ref{iu1}) with Eq.
(\ref{npu1}), we obtain
\begin{eqnarray}
P={\cal K}(\alpha) \rho^{(1+\alpha)/(1-\alpha)}.
\label{iu2}
\end{eqnarray}
This is the equation of state of a polytrope of type III (see Sec.
\ref{sec_pt}) with a
polytropic index $\Gamma=(1+\alpha)/(1-\alpha)$  (i.e. $n=(1-\alpha)/(2\alpha)$)
and a polytropic constant ${\cal
K}(\alpha)$.\footnote{The linear equation of
state (\ref{npu1})
with $\alpha=(\Gamma-1)/(\Gamma+1)$ corresponds to the ultrarelativistic limit
of the
equation of state (\ref{pt4b}) associated with a polytrope of type III with
index
$\Gamma$. Indeed, for a polytrope $P=K\rho^{\Gamma}$, Eq.  (\ref{pt4b})
yields
$P\sim [(\Gamma-1)/(\Gamma+1)]\epsilon$ in the ultrarelativistic limit. The
index $\Gamma=2$ corresponds to $\alpha=1/3$ (radiation) and the index 
$\Gamma=\infty$ corresponds to $\alpha=1$ (stiff matter).}  In the
nonrelativistic limit
$\alpha\rightarrow 0$, we obtain an isothermal equation of state $P=K\rho$ 
with a ``temperature'' $K$.  The SF
potential  (\ref{gesu7}) is given by
\begin{eqnarray}
V_{\rm tot}=\frac{1-\alpha}{2\alpha}{\cal K}(\alpha)\rho^{(1+\alpha)/(1-\alpha)}.
\label{iu3}
\end{eqnarray}
This is a power-law potential.\footnote{Equation (\ref{iu3}) is
similar to a
Tsallis free
energy density of index $q=(1+\alpha)/(1-\alpha)$. Comments similar to 
those following Eq. (\ref{npu1d}) appy to the present situation. }
In the nonrelativistic limit $\alpha\rightarrow 0$, Eq. (\ref{iu3}) reduces to
$V_{\rm
tot}=K\rho\ln\rho$ (up to an additive constant) and we recover Eq.
(\ref{isog2}). For $\alpha=1$ (stiff matter) we obtain $V_{\rm tot}=0$,
corresponding to a free massless SF satisfying $\square\varphi=0$.

\subsubsection{Lagrangian ${\cal L}(X)$}

The Lagrangian ${\cal L}(X)$ is determined by Eq. (\ref{gesu10}) or Eq. (\ref{gesu11}) with the equation of state (\ref{pu1}). We have
\begin{equation}
\label{ln1}
\ln X=2\int \frac{dP}{\left (\frac{P}{K}\right )^{1/\gamma}c^2+P}
\end{equation}
or
\begin{equation}
\label{ln2}
\ln X=2\int \frac{\frac{K\gamma}{c^2}\left (\frac{\epsilon}{c^2}\right )^{\gamma-1}}{\epsilon+K\left (\frac{\epsilon}{c^2}\right )^{\gamma}}\, d\epsilon.
\end{equation}
The integrals can be calculated analytically yielding 
\begin{equation}
\label{ln3}
{\cal L}(X)=P=K\left\lbrace \frac{c^2}{-K} \left\lbrack 1-
\left (\frac{2X}{c^2}\right )^{\frac{\gamma-1}{2\gamma}}\right\rbrack\right\rbrace^{\frac{\gamma}{\gamma-1}}.
\end{equation}
We have determined the
constant of integration so that, in the
nonrelativistic limit, $X\sim c^2/2$ (see Sec. \ref{sec_nr}). In the
nonrelativistic limit, 
using Eq. (\ref{nr5}), we recover Eq. (\ref{lac7}). The
Lagrangian (\ref{ln3}) was first obtained in
\cite{bentoGCG} in relation to the GCG by using the procedure of \cite{btv}
that we have followed.

(i) For $\gamma=-1$ (Chaplygin gas), we obtain
\begin{equation}
\label{cgln3}
{\cal L}(X)=P=K\sqrt{\frac{c^2}{-K} \left (1-\frac{2X}{c^2}\right )}.
\end{equation}
The Lagrangian of the Chaplygin gas ($K<0$) is of the Born-Infeld type. Indeed, setting $K=-A$ so that $P=-Ac^2/\epsilon$ we get the Born-Infeld Lagrangian
\begin{equation}
\label{bi8}
{\cal L}_{\rm BI}=-(Ac^2)^{1/2}\sqrt{1-\frac{1}{c^2}\partial_{\mu}\theta
\partial^{\mu}\theta}.
\end{equation}
In the nonrelativistic limit, using Eq. (\ref{nr5}), it reduces to
\begin{equation}
\label{bi7}
{\cal L}_{\rm NR}=-(2A)^{1/2}\sqrt{
\dot\theta+\frac{1}{2}(\nabla\theta)^2},
\end{equation}
corresponding to Eq. (\ref{cl6}). Using Eq. (\ref{k2invb}) or Eq.
(\ref{ak8}), we
obtain the equation of 
motion 
\begin{equation}
\label{bbb}
D_\mu\left\lbrack \frac{\partial^\mu\theta}{\sqrt{|1-\frac{1}{c^2}\partial_\nu\theta\partial^\nu\theta |}}\right\rbrack=0.
\end{equation}
In the nonrelativistic limit, it reduces to Eq. (\ref{cl8}). The Born-Infeld
Lagrangian (\ref{bi8})  was obtained by
\cite{jp,jackiw} in
two different manners: (i) starting
from a heuristic relativistic Lagrangian
\begin{equation}
\label{heurchap}
L=-\int\left\lbrack \rho\dot\theta+\rho
c^2\sqrt{1+\frac{A}{\rho^2c^2}}\sqrt{1+\frac{(\nabla\theta)^2}{c^2}}
\right\rbrack\, d{\bf r}
\end{equation}
which generalizes the nonrelativistic Lagrangian from Eqs. 
(\ref{lcb2}) and (\ref{cl2}), writing the equations of motion, and
eliminating $\rho$ with the aid of the Bernoulli equation; (ii) for a
$d$-brane moving in a $(d+1,1)$
space-time. In the second case, it can be obtained from the Nambu-Goto action
in the
Cartesian
parametrization. The Born-Infeld Lagrangian (\ref{bi8}) was
also obtained in \cite{btv} for a complex SF in the TF regime by developing the
procedure that we have followed. It can also be directly
obtained from the k-essence formalism applied to the Chaplygin equation of
state (see Appendix \ref{sec_lrsf}).

(ii) For $\gamma=2$ (BEC),  we obtain
\begin{equation}
\label{becln3}
{\cal L}(X)=P=K\left\lbrace \frac{c^2}{-K} \left\lbrack 1-
\left (\frac{2X}{c^2}\right )^{1/4}\right\rbrack\right\rbrace^{2}.
\end{equation}

(iii) For $\gamma=0$ ($\Lambda$CDM
model) the k-essence Lagrangian is ill-defined.

(iv) For $\gamma=3$ (superfluid),  we obtain
\begin{equation}
\label{becln3super}
{\cal L}(X)=P=K\left\lbrace \frac{c^2}{-K} \left\lbrack 1-
\left (\frac{2X}{c^2}\right )^{1/3}\right\rbrack\right\rbrace^{3/2}.
\end{equation}

(v) For $\gamma=1$,  Eq. (\ref{gesu10}) or Eq. (\ref{gesu11}) can be easily
integrated yielding 
\begin{equation}
\label{ln5}
{\cal L}(X)=P={\cal A}(\alpha)  \left (
\frac{2X}{c^2}\right )^{\frac{\alpha+1}{2\alpha}},
\end{equation}
where ${\cal
A}(\alpha)$ is a constant that depends on $\alpha$. The Lagrangian is a pure
power-law. It
was first given in \cite{gm}.
Using Eq. (\ref{k2invb}) or Eq.
(\ref{ak8}), we obtain 
the equation of motion
\begin{equation}
\label{bbb2}
D_\mu\left\lbrack \left (\frac{1}{c^2}\partial_\nu\theta\partial^\nu\theta\right )^{(1-\alpha)/2\alpha}\partial^\mu\theta\right\rbrack=0.
\end{equation}
In the nonrelativistic limit corresponding to $\alpha\rightarrow 0$, using Eq.
(\ref{nr5}), we recover Eqs. (\ref{isog5}) and (\ref{isog6}) with
${\cal A}(\alpha)\rightarrow K\rho_*$. In the case $\alpha=1$,
we obtain ${\cal L}\propto X$ and $\square\theta=0$ (free massless
SF).

\subsection{Polytropic equation of state  of type II}
\label{sec_pd}

The polytropic equation of state of type II writes \cite{tooper2}
\begin{eqnarray}
P=K\rho_m^{\gamma}.
\label{pd1}
\end{eqnarray}
Using Eq. (\ref{mtd6}), the internal energy is
\begin{eqnarray}
u=\frac{K}{\gamma-1}\rho_m^{\gamma}.
\label{wpd3}
\end{eqnarray}
It is similar to the Tsallis free energy. In the nonrelativistic limit, using
$\rho_m\sim \rho$, we recover
Eqs. (\ref{lac2}) and (\ref{lac3}) [recalling  Eq. (\ref{addq})].
Actually, Eqs. (\ref{pd1}) and (\ref{wpd3}) coincide with Eqs.
(\ref{lac2}) and (\ref{lac3}) with $\rho_m$ in place of $\rho$. We note that
$P=(\gamma-1)u$.

(i) For $\gamma=-1$ (Chaplygin gas), we obtain
\begin{eqnarray}
P=\frac{K}{\rho_m},\qquad u=-\frac{K}{2\rho_m}.
\label{cgpd1}
\end{eqnarray}

(ii) For $\gamma=2$ (BEC), we obtain
\begin{eqnarray}
P=K \rho_m^2,\qquad u=K\rho_m^{2}.
\label{becpd1}
\end{eqnarray}

(iii) For $\gamma=0$ ($\Lambda$CDM model), we obtain
\begin{eqnarray}
P=K,\qquad u=-K.
\end{eqnarray}

(iv) For $\gamma=3$ (superfluid), we obtain
\begin{eqnarray}
P=K \rho_m^3,\qquad u=\frac{1}{2}K\rho_m^{3}.
\label{becpd1super}
\end{eqnarray}

(v) The case $\gamma=1$, corresponding to an isothermal equation of state
\begin{eqnarray}
P=K\rho_m,
\label{pd1un}
\end{eqnarray}
must be treated specifically. Using Eq. (\ref{mtd6}), the internal energy is
\begin{eqnarray}
u=K\rho_m\left\lbrack \ln\left(\frac{\rho_m}{\rho_*}\right )-1\right\rbrack.
\label{wpd2}
\end{eqnarray}
It is similar to the Boltzmann free energy. In the nonrelativistic limit, using
$\rho_m\sim \rho$, we recover
Eqs. (\ref{isog1}) and (\ref{isog2}) [recalling  Eq. (\ref{addq})].
Actually, Eqs. (\ref{pd1un}) and (\ref{wpd2}) coincide with Eqs.
(\ref{isog1}) and (\ref{isog2})  with $\rho_m$ in place
of $\rho$.

\subsubsection{Determination of $\epsilon$ and $P(\epsilon)$}

The energy density is determined by Eq. (\ref{gesd1}) with Eq. (\ref{wpd3}). We obtain
\begin{eqnarray}
\epsilon=\rho_m
c^2+\frac{K}{\gamma-1}\rho_m^{\gamma}.
\label{pd3}
\end{eqnarray}
The  pressure is determined by Eq. (\ref{gesd2}) with Eq. (\ref{wpd3}). This returns Eq. (\ref{pd1}). Eliminating $\rho_m$ between Eq. (\ref{pd1}) and Eq.  (\ref{pd3}) we obtain  $P(\epsilon)$ under the inverse form  $\epsilon(P)$ as
\begin{eqnarray}
\epsilon=\left (\frac{P}{K}\right
)^{1/\gamma}c^2+\frac{P}{\gamma-1}.
\label{pd5}
\end{eqnarray}
In the nonrelativistic limit, using $\epsilon\sim
\rho c^2$, we recover Eq.
(\ref{lac2}).

(i) For $\gamma=-1$ (Chaplygin gas), we obtain
\begin{eqnarray}
\epsilon=\rho_m
c^2-\frac{K}{2\rho_m},
\label{cgpd3}
\end{eqnarray}
\begin{eqnarray}
\epsilon=\frac{Kc^2}{P}-\frac{P}{2},
\label{cgpd3a}
\end{eqnarray}
\begin{eqnarray}
\rho_m c^2=\frac{\epsilon\pm\sqrt{\epsilon^2+2Kc^2}}{2},
\label{cgpd3b}
\end{eqnarray}
\begin{eqnarray}
P=-\epsilon\pm\sqrt{\epsilon^2+2Kc^2}.
\label{cgpd3c}
\end{eqnarray}
We note that the Chaplygin gas of type II is different from
the Chaplygin gas of type I.

(ii) For $\gamma=2$ (BEC), we obtain
\begin{eqnarray}
\epsilon=\rho_m
c^2+K\rho_m^2,
\label{becpd3}
\end{eqnarray}
\begin{eqnarray}
\epsilon=\sqrt{\frac{P}{K}}c^2+P,
\label{becpd3a}
\end{eqnarray}
\begin{eqnarray}
\rho_m=\frac{-c^2\pm\sqrt{c^4+4K\epsilon}}{2K},
\label{becpd3b}
\end{eqnarray}
\begin{eqnarray}
P=\frac{1}{4K}\left\lbrack -c^2\pm\sqrt{c^4+4K\epsilon}\right\rbrack^2.
\label{becpd3c}
\end{eqnarray}
The equation of state
(\ref{becpd3c}) was first obtained in \cite{partially,stiff}.

(iii) For $\gamma=0$ ($\Lambda$CDM model), we obtain
\begin{eqnarray}
\epsilon=\rho_m c^2-K,
\label{deuxL1}
\end{eqnarray}
\begin{eqnarray}
P=K.
\label{deuxL2}
\end{eqnarray}

(iv) For $\gamma=3$ (superfluid), we obtain
\begin{eqnarray}
\epsilon=\rho_m
c^2+\frac{1}{2}K\rho_m^3,
\label{becpd3super}
\end{eqnarray}
\begin{eqnarray}
\epsilon=\left (\frac{P}{K}\right )^{1/3}c^2+\frac{1}{2}P.
\label{becpd3asuper}
\end{eqnarray}
This is a third degree equation which can be solved by standard
means to obtain $P(\epsilon)$.

(v) For $\gamma=1$, the energy density is determined by Eq. (\ref{gesd1}) with
Eq. (\ref{wpd2}). We obtain
\begin{eqnarray}
\epsilon=\rho_m c^2+K\rho_m\left\lbrack \ln\left(\frac{\rho_m}{\rho_*}\right )-1\right\rbrack.
\label{pd2}
\end{eqnarray}
The  pressure is determined by Eq. (\ref{gesd2}) with Eq. (\ref{wpd2}). This returns Eq. (\ref{pd1un}). Eliminating $\rho_m$ between Eq. (\ref{pd1un}) and Eqs. (\ref{pd2}) we obtain  $P(\epsilon)$ under the inverse form $\epsilon(P)$ as
\begin{eqnarray}
\epsilon=\frac{P}{K}c^2+P\left\lbrack \ln\left(\frac{P}{K\rho_*}\right )-1\right\rbrack.
\label{pd4}
\end{eqnarray}

{\it Remark:} For the index
$\gamma=1/2$, we can inverse Eq. (\ref{pd5}) to obtain  
\begin{eqnarray}
P=\frac{K^2}{c^2}\pm
K\sqrt{\frac{K^2}{c^4}+\frac{\epsilon}{c^2}}.
\label{becpd3cmn}
\end{eqnarray}
For the index $\gamma=3/2$, Eq. (\ref{pd3}) becomes 
\begin{equation}
\epsilon=\rho_m
c^2+2K\rho_m^{3/2}.
\label{becsoi2mni}
\end{equation}
This is a third degree equation for $\sqrt{\rho_m}$ which can
be solved by standard means. One can then obtain $P(\epsilon)$
explicitly.

\subsubsection{Determination of $\rho$, $P(\rho)$ and $V_{\rm tot}(\rho)$}

The pseudo rest-mass density and the SF 
potential are determined by Eqs. (\ref{gesd4b}) and
(\ref{gesd5})
with Eq. (\ref{wpd3}). We get
\begin{equation}
\rho=\frac{\rho_m}{1+\frac{\gamma}{\gamma-1}\frac{K}{c^2}\rho_m^{\gamma-1}},
\label{soi2}
\end{equation}
\begin{equation}
V_{\rm tot}=\frac{1}{2}\left\lbrack \rho_m c^2-K\frac{\gamma-2}{\gamma-1}\rho_m^{\gamma}\right\rbrack.
\label{pd9}
\end{equation}
Equations (\ref{pd1}), (\ref{soi2}) and (\ref{pd9}) determine  $P(\rho)$ and
$V_{\rm
tot}(\rho)$ in parametric form with parameter $\rho_m$. In the nonrelativistic
limit,
using $\rho_m\sim \rho \lbrack
1+\frac{\gamma}{\gamma-1}\frac{K}{c^2}\rho^{\gamma-1}\rbrack$ and Eq.
(\ref{db10}), we recover Eqs. (\ref{lac2}) and (\ref{lac3}).

(i) For $\gamma=-1$ (Chaplygin gas), $\rho_m$ is determined by a cubic equation
\begin{equation}
2\rho_m^3-2\rho\rho_m^2-\frac{K\rho}{c^2}=0,
\label{cgsoi2}
\end{equation}
which can be solved by standard means. The SF potential is
given by
\begin{equation}
V_{\rm tot}=\frac{1}{2}\left (\rho_m c^2+\frac{3K}{2\rho_m}\right ).
\label{cgsoi2b}
\end{equation}
One can then obtain $P(\rho)$ and $V_{\rm tot}(\rho)$ explicitly.

(ii) For $\gamma=2$ (BEC), we obtain
\begin{equation}
\rho=\frac{\rho_m}{1+\frac{2K}{c^2}\rho_m},
\label{becsoi2}
\end{equation}
\begin{equation}
\rho_m=\frac{\rho}{1-\frac{2K}{c^2}\rho},
\label{becsoi2b}
\end{equation}
\begin{equation}
P=\frac{K\rho^2}{\left (1-\frac{2K}{c^2}\rho\right )^2},
\label{becsoi2c}
\end{equation}
\begin{equation}
V_{\rm tot}=\frac{1}{2}\rho c^2+\frac{K\rho^2}{1-\frac{2K\rho}{c^2}}.
\label{becsoi2d}
\end{equation}

(iii) For $\gamma=0$ ($\Lambda$CDM model), we obtain 
\begin{eqnarray}
\rho=\rho_m,\qquad P=K,
\label{deuxL3}
\end{eqnarray}
\begin{eqnarray}
V_{\rm tot}=\frac{1}{2}\rho c^2-K.
\label{deuxL4}
\end{eqnarray}

(iv) For $\gamma=3$ (superfluid), we obtain
\begin{equation}
\rho=\frac{\rho_m}{1+\frac{3K}{2c^2}\rho_m^2},
\label{becsoi2super}
\end{equation}
\begin{equation}
\rho_m=\frac{c^2}{3K\rho}\left ( 1\pm \sqrt{1-\frac{6K}{c^2}\rho^2}\right ),
\label{becsoi2bsuper}
\end{equation}
\begin{equation}
P=K\left (\frac{c^2}{3K\rho}\right )^3\left ( 1\pm
\sqrt{1-\frac{6K}{c^2}\rho^2}\right )^3,
\label{becsoi2csuper}
\end{equation}
\begin{eqnarray}
V_{\rm tot}&=&\frac{c^4}{6K\rho}\left ( 1\pm
\sqrt{1-\frac{6K}{c^2}\rho^2}\right )\nonumber\\
&\times&\left\lbrack
\frac{4}{3}-\frac{c^2}{9K\rho^2}\left (1\pm
\sqrt{1-\frac{6K}{c^2}\rho^2}\right )\right\rbrack.
\label{becsoi2dsuper}
\end{eqnarray}

(v) For $\gamma=1$, the pseudo rest-mass density and the SF potential are 
determined by Eqs. (\ref{gesd4b}) and (\ref{gesd5}) with Eq.
(\ref{wpd2}). We get
\begin{equation}
\rho=\frac{\rho_m}{1+\frac{K}{c^2}\ln\left(\frac{\rho_m}{\rho_*}\right )},
\label{soi1}
\end{equation}
\begin{equation}
V_{\rm tot}=\frac{1}{2}\left\lbrack
\rho_m c^2+K\rho_m \ln\left(\frac{\rho_m}{\rho_*}\right )-2K\rho_m\right\rbrack.
\label{pd8}
\end{equation}
Equations (\ref{pd1un}), (\ref{soi1}) and (\ref{pd8}) determine  $P(\rho)$ and
$V_{\rm
tot}(\rho)$ in parametric form with parameter $\rho_m$. In the nonrelativistic
limit, using $\rho_m\simeq \rho\lbrack
1+({K}/{c^2})\ln({\rho}/{\rho_*} )\rbrack$ and Eq. (\ref{db10}), we recover Eq.
(\ref{isog2}).

{\it Remark:} For the index $\gamma=1/2$, Eq. (\ref{soi2}) can be written as
\begin{equation}
\rho_m^{3/2}-\rho\rho_m^{1/2}+\frac{K\rho}{c^2}=0.
\label{becsoi2mn}
\end{equation}
This is a third degree equation for $\sqrt{\rho_m}$ which can
be solved by standard means. One can then obtain $P(\rho)$ and $V_{\rm
tot}(\rho)$ explicitly. For the index $\gamma=3/2$ we find that
\begin{equation}
P=K\left (\frac{3K}{2c^2}\rho\pm\sqrt{\frac{9K^2}{4c^4}\rho^2+\rho}\right )^3,
\label{becsoi2mnf}
\end{equation}
\begin{eqnarray}
V_{\rm tot}&=&\frac{1}{2}\left
(\frac{3K}{2c^2}\rho\pm\sqrt{\frac{9K^2}{4c^4}\rho^2+\rho}\right
)^{2}c^2\nonumber\\
&\times& \left \lbrack 1+\frac{K}{2c^2}\rho\left
(\frac{3K}{c^2}\pm\sqrt{\frac{9K^2}{c^4}+\frac{4}{\rho}}\right )\right\rbrack.
\label{becsoi2mnff}
\end{eqnarray}

\subsubsection{Lagrangian ${\cal L}(X)$}

The Lagrangian ${\cal L}(X)$ is determined by Eqs. (\ref{gesd8}), (\ref{pd1})
and (\ref{wpd3}). We get
\begin{equation}
X=\frac{1}{2c^2}\left\lbrack c^2+\frac{K\gamma}{\gamma-1}\rho_m^{\gamma-1}\right\rbrack^2.
\label{soi4}
\end{equation}
This equation can be inverted to give 
\begin{equation}
\rho_m^{\gamma-1}=-\frac{\gamma-1}{\gamma}\frac{c^2}{K}\left\lbrack  1-\left (\frac{2X}{c^2}\right )^{1/2}\right\rbrack.
\label{pd9e}
\end{equation}
We then obtain
\begin{equation}
{\cal L}(X)=P=K\left \lbrace -\frac{\gamma-1}{\gamma}\frac{c^2}{K}\left\lbrack  1- \left (\frac{2X}{c^2}\right )^{1/2}\right\rbrack\right\rbrace^{\frac{\gamma}{\gamma-1}}.
\label{pd9g}
\end{equation}
In the nonrelativistic limit, using Eq. (\ref{nr5}), we recover Eq.
(\ref{lac7}).

(i) For $\gamma=-1$ (Chaplygin gas), we obtain
\begin{equation}
{\cal L}(X)=P=K\sqrt{\frac{2c^2}{-K}\left\lbrack  1- \left (\frac{2X}{c^2}\right )^{1/2}\right\rbrack}.
\label{cgpd9g}
\end{equation}

(ii) For $\gamma=2$ (BEC), we obtain
\begin{equation}
{\cal L}(X)=P=K\left \lbrace \frac{c^2}{-2K}\left\lbrack  1- \left (\frac{2X}{c^2}\right )^{1/2}\right\rbrack\right\rbrace^{2}.
\label{becpd9g}
\end{equation}

(iii) For $\gamma=0$ ($\Lambda$CDM model), the k-essence Lagrangian is
ill-defined.

(iv) For $\gamma=3$ (superfluid), we obtain
\begin{equation}
{\cal L}(X)=P=K\left \lbrace \frac{2c^2}{-3K}\left\lbrack  1- \left
(\frac{2X}{c^2}\right )^{1/2}\right\rbrack\right\rbrace^{3/2}.
\label{becpd9gsuper}
\end{equation}

(v) For $\gamma=1$, the Lagrangian ${\cal L}(X)$ is determined by Eqs.
(\ref{gesd8}), (\ref{pd1un}) and (\ref{wpd2}). This yields
\begin{equation}
X=\frac{1}{2c^2}\left\lbrack c^2+K\ln\left (\frac{\rho_m}{\rho_*}\right )\right\rbrack^2.
\label{soi3}
\end{equation}
This equation can be inverted to give 
\begin{equation}
\rho_m=\rho_* e^{-\frac{c^2}{K}\left\lbrack 1-\left (\frac{2X}{c^2}\right
)^{1/2}\right\rbrack }.
\label{pd9d}
\end{equation}
We then obtain
\begin{equation}
{\cal L}(X)=P=K\rho_* e^{-\frac{c^2}{K}\left\lbrack 1-\left
(\frac{2X}{c^2}\right )^{1/2}\right\rbrack }.
\label{pd9f}
\end{equation}
In the nonrelativistic limit, using Eq. (\ref{nr5}), we recover Eq.
(\ref{isog5}).

\subsection{Polytropic equation of state of type III}
\label{sec_pt}

The polytropic equation of state of type III writes
\cite{abrilas,csf} 
\begin{eqnarray}
P=K\rho^{\gamma}.
\label{pt1}
\end{eqnarray}
Using Eq. (\ref{etoilebis}), the  SF potential is
\begin{eqnarray}
V_{\rm tot}(\rho)=\frac{1}{2}\rho c^2+\frac{K}{\gamma-1}\rho^{\gamma}.
\label{pt2b}
\end{eqnarray}
It is similar to the Tsallis free energy. In the nonrelativistic limit, we
recover Eqs.
(\ref{lac2}) and (\ref{lac3}) [recalling Eq. (\ref{db10})]. Actually, Eqs.
(\ref{pt1})
and (\ref{pt2b}) coincide with Eqs. (\ref{lac2}) and (\ref{lac3}). The SF
potential $V(\rho)$ corresponds to a pure power-law. We note that
$P=(\gamma-1)V$.

(i) For $\gamma=-1$ (Chaplygin gas),  we obtain
\begin{eqnarray}
P=\frac{K}{\rho},\qquad V_{\rm tot}=\frac{1}{2}\rho c^2-\frac{K}{2\rho},
\label{cgpt1}
\end{eqnarray}
as found in \cite{abrilas,csf}. We recover the potential from Eq. (\ref{pu6b})
for the reason explained in Sec. 
\ref{sec_bel}.

(ii) For $\gamma=2$ (BEC), we obtain
\begin{eqnarray}
P=K\rho^2,\qquad V_{\rm tot}=\frac{1}{2}\rho c^2+K\rho^{2}.
\label{becpt1}
\end{eqnarray}
The SF potential $V(\rho)=K\rho^{2}$ from Eq. (\ref{becpt1}) with $K={2\pi
a_s\hbar^2}/{m^3}$  corresponds to the
standard $|\varphi|^4$ potential of a BEC
\cite{colpi,abrilas,csf}.

(iii) For $\gamma=0$ ($\Lambda$CDM model), we obtain 
\begin{eqnarray}
P=K, \qquad V_{\rm tot}=\frac{1}{2}\rho c^2-K.
\label{troisL1}
\end{eqnarray}
The SF potential  $V(\rho)=-K$ from Eq. (\ref{troisL1}) is constant. Using
$K=-\rho_\Lambda c^2$, we see that $V(\rho)=\rho_\Lambda
c^2$  is equal to the cosmological density \cite{csf}.

(iv) For $\gamma=3$ (superfluid), we obtain
\begin{eqnarray}
P=K\rho^3,\qquad V_{\rm tot}=\frac{1}{2}\rho c^2+\frac{1}{2}K\rho^{3}.
\label{becpt1super}
\end{eqnarray}

(v) The case $\gamma=1$, corresponding to a linear equation of state
\begin{eqnarray}
P=K\rho,
\label{pt1un}
\end{eqnarray}
must be treated specifically. Using Eq. (\ref{etoilebis}), the SF potential
is
\begin{eqnarray}
V_{\rm tot}(\rho)=\frac{1}{2}\rho c^2+K\rho \left\lbrack \ln\left (\frac{\rho}{\rho_*}\right )-1\right\rbrack.
\label{pt2}
\end{eqnarray}
It is similar to the Boltzmann free energy. In the nonrelativistic
limit, we recover  Eqs. (\ref{isog1}) and (\ref{isog2})
[recalling Eq. (\ref{db10})]. Actually, Eqs. (\ref{pt1un}) and (\ref{pt2})
coincide with Eqs. (\ref{isog1}) and (\ref{isog2}).
The potential (\ref{pt2}) was first obtained in
\cite{tunnel,csf}.

\subsubsection{Determination of $\epsilon$ and $P(\epsilon)$}
\label{sec_bel}

The energy density is determined by Eq. (\ref{gest1}) with Eq. (\ref{pt2b}). This yields
\begin{eqnarray}
\epsilon=\rho c^2+\frac{\gamma+1}{\gamma-1}K\rho^{\gamma}.
\label{pt3b}
\end{eqnarray}
The pressure is determined by Eq. (\ref{gest2}) with Eq.  (\ref{pt2b}). This returns Eq. (\ref{pt1}). Eliminating $\rho$ between Eqs. (\ref{pt1}) and  (\ref{pt3b}), we obtain $P(\epsilon)$ under the inverse form $\epsilon(P)$ as
\begin{eqnarray}
\epsilon=\left (\frac{P}{K}\right
)^{1/\gamma}c^2+\frac{\gamma+1}{\gamma-1}P.
\label{pt4b}
\end{eqnarray}
In the nonrelativistic
limit, using $\epsilon\sim \rho c^2$, we recover Eq. (\ref{lac2}).

(i) For $\gamma=-1$ (Chaplygin gas), we obtain
\begin{eqnarray}
\epsilon=\rho c^2,
\label{cgpt3b}
\end{eqnarray}
\begin{eqnarray}
P=\frac{Kc^2}{\epsilon}.
\label{cgpt3bn}
\end{eqnarray}
This returns  the Chaplygin gas of type I (see Sec. \ref{sec_pu}). Therefore,
the Chaplygin gas models
of type I and III coincide.

(ii) For $\gamma=2$ (BEC), we obtain
\begin{eqnarray}
\epsilon=\rho c^2+3K\rho^2,
\label{becpt3b1}
\end{eqnarray}
\begin{eqnarray}
\epsilon=\sqrt{\frac{P}{K}}c^2+3P,
\label{becpt3b2}
\end{eqnarray}
\begin{eqnarray}
\rho=\frac{-c^2\pm\sqrt{c^4+12K\epsilon}}{6K},
\label{becpt3bn}
\end{eqnarray}
\begin{eqnarray}
P=\frac{1}{36K}\left\lbrack -c^2\pm\sqrt{c^4+12K\epsilon}\right\rbrack^2.
\label{becpt3b}
\end{eqnarray}
This equation of state was first obtained in
\cite{colpi} (see also \cite{abrilph,abrilas,csf}).

(iii) For $\gamma=0$ ($\Lambda$CDM model), we obtain 
\begin{eqnarray}
P=K, \qquad \epsilon=\rho c^2-K.
\label{troisL2}
\end{eqnarray}

(iv) For $\gamma=3$ (superfluid), we obtain
\begin{eqnarray}
\epsilon=\rho c^2+2K\rho^3,
\label{becpt3bsuper}
\end{eqnarray}
\begin{eqnarray}
\epsilon=\left (\frac{P}{K}\right
)^{1/3}c^2+2P.
\label{becpt3bsuperb}
\end{eqnarray}
This is a third degree equation which can be solved by
standard
means to obtain $P(\epsilon)$.

(v) For $\gamma=1$, the energy density is determined by Eq. (\ref{gest1}) with
Eq. (\ref{pt2}). This yields
\begin{eqnarray}
\epsilon=\rho c^2+2K\rho \ln\left (\frac{\rho}{\rho_*}\right )-K\rho.
\label{pt3}
\end{eqnarray}
The pressure is determined by Eq. (\ref{gest2}) with Eq. (\ref{pt2}). This returns Eq. (\ref{pt1un}). Eliminating $\rho$ between Eqs. (\ref{pt1un}) and (\ref{pt3}), we obtain $P(\epsilon)$ under the inverse form $\epsilon(P)$ as
\begin{eqnarray}
\epsilon=\frac{P}{K}c^2+2P \ln\left (\frac{P}{K\rho_*}\right )-P.
\label{pt4}
\end{eqnarray}
In the nonrelativistic limit, using $\epsilon\sim \rho c^2$, we
recover Eq.
(\ref{isog1}).

{\it Remark:} For the index
$\gamma=1/2$,  we can inverse Eq. (\ref{pt4b}) to obtain 
\begin{eqnarray}
P=\frac{3K^2}{2c^2}\pm K\sqrt{\frac{9K^2}{4c^4}+\frac{\epsilon}{c^2}}.
\label{becpt3bmn}
\end{eqnarray}
For the index $\gamma=3/2$, Eq. (\ref{pt3b}) becomes 
\begin{equation}
\epsilon=\rho
c^2+5K\rho^{3/2}.
\label{becsoi2mnij}
\end{equation}
This is a third degree equation for $\sqrt{\rho_m}$ which can
be solved by standard means. One can then obtain $P(\epsilon)$
explicitly.

\subsubsection{Determination of $\rho_m$, $P(\rho_m)$ and  $u(\rho_m)$}

The rest-mass density and the internal energy are determined by Eqs.
(\ref{gest4a}) and (\ref{gest6}) with
Eq. (\ref{pt2b}). We get
\begin{equation}
\rho_m=\rho\sqrt{1+\frac{2\gamma}{\gamma-1}\frac{K}{c^2}\rho^{\gamma-1}},
\label{theo2}
\end{equation}
\begin{equation}
u=\rho c^2+\frac{\gamma+1}{\gamma-1}K\rho^{\gamma}-\rho_m(\rho)c^2.
\label{gpd9}
\end{equation}
Eqs. (\ref{pt1}), (\ref{theo2}) and (\ref{gpd9})  define  $P(\rho_m)$ and
$u(\rho_m)$ 
in parametric form with parameter $\rho$. In the nonrelativistic limit,
using $\rho_m\sim \rho \lbrack
1+\frac{\gamma}{\gamma-1}\frac{K}{c^2}\rho^{\gamma-1}\rbrack$, we recover Eqs.
(\ref{lac2}) and (\ref{lac3}) [recalling Eq. (\ref{addq})].

(i) For $\gamma=-1$ (Chaplygin gas), we obtain
\begin{eqnarray}
\rho_m c^2=\sqrt{(\rho c^2)^2+Kc^2},
\label{cgpu1dn}
\end{eqnarray}
\begin{eqnarray}
\rho c^2=\sqrt{(\rho_m c^2)^2-Kc^2},
\label{cgpu1cn}
\end{eqnarray}
\begin{eqnarray}
P=\frac{Kc^2}{\sqrt{(\rho_m c^2)^2-Kc^2}},
\label{cgpu1can}
\end{eqnarray}
\begin{eqnarray}
u=\sqrt{(\rho_m c^2)^2-Kc^2}-\rho_m c^2.
\label{cgpu1cbn}
\end{eqnarray}

(ii) For $\gamma=2$ (BEC), $\rho$ is determined by a cubic equation
\begin{eqnarray}
\frac{4K}{c^2}\rho^3+\rho^2-\rho_m^2=0, 
\label{becpu1cn}
\end{eqnarray}
which can be solved by standard means. The internal energy
is given by
\begin{eqnarray}
u=\rho c^2+3K\rho^2-\rho_m(\rho)c^2.
\end{eqnarray}
One can then obtain  $P(\rho_m)$ and  $u(\rho_m)$ explicitly.

(iii) For $\gamma=0$ ($\Lambda$CDM model), we obtain 
\begin{eqnarray}
P=K, \qquad \rho_m=\rho, \qquad u=-K.
\label{troisL3}
\end{eqnarray}
We note that the rest-mass density coincides with the pseudo rest-mass density
($\rho_m=\rho$).

(iv) For $\gamma=3$ (superfluid), we obtain
\begin{eqnarray}
\rho_m=\rho\sqrt{1+\frac{3K}{c^2}\rho^2},
\label{cgpu1dnsuper}
\end{eqnarray}
\begin{eqnarray}
\rho=\left (
-\frac{c^2}{6K}\pm\frac{c^2}{6K}\sqrt{1+\frac{12K}{c^2}\rho_m^2}\right )^{
1/2},
\label{cgpu1cnsuper}
\end{eqnarray}
\begin{eqnarray}
P=K\left (
-\frac{c^2}{6K}\pm\frac{c^2}{6K}\sqrt{1+\frac{12K}{c^2}\rho_m^2}\right )^{
3/2},
\label{cgpu1cansuper}
\end{eqnarray}
\begin{eqnarray}
u=c^2\left (
-\frac{c^2}{6K}\pm\frac{c^2}{6K}\sqrt{1+\frac{12K}{c^2}\rho_m^2}\right )^{
1/2}\nonumber\\
\times\left
(\frac{2}{3}\pm\frac{1}{3}\sqrt{1+\frac{12K}{c^2}\rho_m^2}\right )-\rho_m c^2.
\label{cgpu1cbnsuper}
\end{eqnarray}

(v) For $\gamma=1$ the rest-mass density and the internal energy are determined
by Eqs.
(\ref{gest4a}) and (\ref{gest6})
with Eq. (\ref{pt2}). This gives 
\begin{equation}
\rho_m=\rho\sqrt{1+\frac{2K}{c^2}\ln\left (\frac{\rho}{\rho_*}\right )},
\label{theo1}
\end{equation}
\begin{equation}
u=\rho c^2+2K\rho \ln\left(\frac{\rho}{\rho_*}\right )-K\rho-\rho_m(\rho)c^2.
\label{gpd8}
\end{equation}
Equations (\ref{pt1un}), (\ref{theo1}) and (\ref{gpd8})  determine   $P(\rho_m)$
and
$u(\rho_m)$ in parametric form with parameter $\rho$.  In the nonrelativistic
limit, using $\rho_m\sim \rho\lbrack
1+\frac{K}{c^2}\ln\left ({\rho}/{\rho_*}\right )\rbrack$, we recover Eqs.
(\ref{isog1}) and (\ref{isog2}) [recalling Eq. (\ref{addq})].

\subsubsection{Lagrangian ${\cal L}(X)$}

The Lagrangian ${\cal L}(X)$ is determined by Eq. (\ref{gest7})  with Eq. (\ref{pt2b}). We get
\begin{equation}
X=\frac{1}{2}c^2+\frac{K\gamma}{\gamma-1}\rho^{\gamma-1}.
\label{qpd9c}
\end{equation}
This relation can be reversed to give 
\begin{equation}
\rho^{\gamma-1}=-\frac{\gamma-1}{2\gamma}\frac{c^2}{K}\left (1-\frac{2X}{c^2}\right ).
\label{qpd9e}
\end{equation}
We then obtain
\begin{equation}
{\cal L}(X)=P=K\left\lbrack -\frac{\gamma-1}{2\gamma}\frac{c^2}{K}\left (1-\frac{2X}{c^2}\right )\right\rbrack^{\frac{\gamma}{\gamma-1}}.
\label{qpd9g}
\end{equation}
In the nonrelativistic limit, using Eq. (\ref{nr5}), we recover Eq.
(\ref{lac7}). Actually, Eq. (\ref{qpd9g}) coincides with Eq. 
(\ref{lac7}) with $c^2/2-X$ in place of $x$. Interestingly,
the Lagrangian (\ref{qpd9g}) corresponds to the Lagrangian  introduced
heuristically in \cite{banerjee} in relation to the GCG
[see their Eq. (33)]. 
In \cite{banerjee} it was obtained from a heuristic relativistic Lagrangian
\begin{equation}
L=-\int\left\lbrack \rho\dot\theta+\rho
c^2\sqrt{1+\frac{2K}{\gamma-1}\frac{1}{\rho^{1-\gamma}c^2}}\sqrt{1+\frac{
(\nabla\theta)^2 } { c^2 } }
\right\rbrack\, d{\bf r}
\end{equation}
which generalizes the Lagrangian from Eq.
(\ref{heurchap}). Our approach provides therefore a justification of the
Lagrangian (\ref{qpd9g})
from a more rigorous relativistic theory. We note that this Lagrangian differs
from the Lagrangian (\ref{ln3}) introduced in
\cite{bentoGCG} except for the particular index $\gamma=-1$ corresponding to the
Chaplygin gas (see below). It is also different from the Lagrangian
(\ref{pd9g}) even for $\gamma=-1$. This is an effect
of the inequivalence between the equations of state of types I, II and III.

(i) For $\gamma=-1$ (Chaplygin gas), we obtain
\begin{equation}
\label{cgln3n}
{\cal L}(X)=P=K\sqrt{\frac{c^2}{-K} \left (1-\frac{2X}{c^2}\right )},
\end{equation}
like in Eq. (\ref{cgln3}).

(ii) For $\gamma=2$ (BEC), we obtain
\begin{equation}
{\cal L}(X)=P=K\left\lbrack \frac{c^2}{-4K}\left (1-\frac{2X}{c^2}\right )\right\rbrack^{2}.
\label{qpd9h}
\end{equation}

(iii) For $\gamma=0$ ($\Lambda$CDM model), the k-essence Lagrangian is
ill-defined.

(iv) For $\gamma=3$ (superfluid), we obtain
\begin{equation}
{\cal L}(X)=P=K\left\lbrack \frac{c^2}{-3K}\left (1-\frac{2X}{c^2}\right
)\right\rbrack^{3/2}.
\label{qpd9hsuper}
\end{equation}

(v) For $\gamma=1$, the Lagrangian ${\cal L}(X)$ is determined by Eq.
(\ref{gest7})  with Eq. (\ref{pt2}). We get
\begin{equation}
X=\frac{1}{2}c^2+K\ln\left (\frac{\rho}{\rho_*}\right ).
\label{qpd9b}
\end{equation}
This relation can be reversed to give 
\begin{equation}
\rho=\rho_* e^{-\frac{c^2}{2K}\left (1-\frac{2X}{c^2}\right )}.
\label{qpd9d}
\end{equation}
We then obtain
\begin{equation}
{\cal L}(X)=P=K\rho_* e^{-\frac{c^2}{2K}\left (1-\frac{2X}{c^2}\right )}.
\label{qpd9f}
\end{equation}
In the nonrelativistic limit, using Eq. (\ref{nr5}), we recover Eq.
(\ref{isog5}). Actually,  Eq. (\ref{qpd9f}) coincides with Eq. 
(\ref{isog5}) with $c^2/2-X$ in place of $x$.

\section{Logotropes}
\label{sec_logo}

In this section, we apply the general results of Sec. \ref{sec_ges} to the case 
of a logotropic equation of state.

\subsection{Logotropic equation of state of type I}
\label{sec_lu}

The logotropic equation of state of type I writes
\cite{epjp}
\begin{eqnarray}
P=A\ln\left (\frac{\epsilon}{\epsilon_*}\right ).
\label{lu1}
\end{eqnarray}

\subsubsection{Determination of $\rho_m$, $P(\rho_m)$ and $u(\rho_m)$}

The rest-mass density and the internal energy are determined by 
Eqs. (\ref{gesu1}) and (\ref{gesu2}) with the equation of state (\ref{lu1})
yielding
\begin{equation}
\label{lu1b}
\ln\rho_m=\int\frac{d\epsilon}{A\ln\left (\frac{\epsilon}{\epsilon_*}\right )+\epsilon},
\end{equation}
\begin{equation}
\label{lu1bn}
u=\epsilon-\rho_m(\epsilon)c^2.
\end{equation}
Equations (\ref{lu1})-(\ref{lu1bn}) determine $P(\rho_m)$ and $u(\rho_m)$ in
parametric form with parameter
$\epsilon$. Unfortunately, the integral in Eq. (\ref{lu1b}) cannot be calculated
analytically.

\subsubsection{Determination of $\rho$, $P(\rho)$ and $V_{\rm tot}(\rho)$}

The pseudo rest-mass density and the SF potential are determined by Eqs.
(\ref{gesu6}) and (\ref{gesu7}) with the equation of state (\ref{lu1}) yielding
\begin{eqnarray}
\ln\rho=\int \frac{1-\frac{A}{\epsilon}}{\epsilon+A\ln\left
(\frac{\epsilon}{\epsilon_*}\right )}\, d\epsilon,
\label{lu2}
\end{eqnarray}
\begin{equation}
\label{lu3}
V_{\rm tot}=\frac{1}{2}\left \lbrack\epsilon-A\ln\left
(\frac{\epsilon}{\epsilon_*}\right )\right \rbrack.
\end{equation}
They can also be determined by Eqs. (\ref{gesu8}) and (\ref{gesu9})  with Eq. (\ref{lu1}) yielding
\begin{eqnarray}
\ln\rho=\int \frac{\frac{\epsilon_*}{A}e^{P/A}-1}{\epsilon_* e^{P/A}+P}\, dP,
\label{lu2g}
\end{eqnarray}
\begin{equation}
\label{lu3g}
V_{\rm tot}=\frac{1}{2}\left (\epsilon_* e^{P/A}-P\right ).
\end{equation}
Equations (\ref{lu1}) and  (\ref{lu2})-(\ref{lu3g}) determine $P(\rho)$, or
$\rho(P)$, and $V_{\rm tot}(\rho)$ in
parametric form with parameter $\epsilon$ or $P$. Unfortunately, the integrals
in Eqs. (\ref{lu2}) and (\ref{lu2g})  cannot be calculated analytically.

\subsubsection{Lagrangian ${\cal L}(X)$}

The Lagrangian ${\cal L}(X)$ is determined by Eq. (\ref{gesu10}) 
or Eq. (\ref{gesu11}) with Eq. (\ref{lu1}) yielding
\begin{equation}
\label{lu4}
\ln X=2\int \frac{dP}{\epsilon_* e^{P/A}+P}
\end{equation}
or
\begin{equation}
\label{lu4g}
\ln X=2\int \frac{\frac{A}{\epsilon}}{\epsilon+A \ln\left
(\frac{\epsilon}{\epsilon_*}\right )}\, d\epsilon.
\end{equation}
These equations determine $P(X)$, thus ${\cal L}(X)$.
Unfortunately, the
integrals in Eqs. (\ref{lu4}) and (\ref{lu4g})  cannot be calculated
analytically.

\subsection{Logotropic equation of state of type II}
\label{sec_ld}

The logotropic equation of state of type II writes \cite{epjp} 
\begin{eqnarray}
P=A\ln\left (\frac{\rho_m}{\rho_*}\right ).
\label{ld1}
\end{eqnarray}
Using Eq. (\ref{mtd6}), the internal energy is
\begin{eqnarray}
u=-A\ln\left (\frac{\rho_m}{\rho_*}\right )-A.
\label{ld2}
\end{eqnarray}
In the nonrelativistic
limit, using $\rho_m\sim \rho$, we recover
Eqs. (\ref{lal1}) and (\ref{lal2}) [recalling  Eq. (\ref{addq})].
Actually, Eqs. (\ref{ld1}) and (\ref{ld2}) coincide with Eqs.
(\ref{lal1}) and (\ref{lal2}) with $\rho_m$ in place of $\rho$.

\subsubsection{Determination of $\epsilon$ and $P(\epsilon)$}

The energy density is determined by Eq. (\ref{gesd1})  with Eq. (\ref{ld2}). We obtain
\begin{eqnarray}
\epsilon=\rho_m c^2-A\ln\left (\frac{\rho_m}{\rho_*}\right )-A.
\label{ld3}
\end{eqnarray}
The pressure is determined by Eq. (\ref{gesd2}) with Eq. (\ref{ld2}). This returns Eq. (\ref{ld1}).
Eliminating $\rho_m$ between Eqs. (\ref{ld1}) and (\ref{ld3}) we obtain  $P(\epsilon)$ under the inverse form $\epsilon(P)$ as
\begin{eqnarray}
\epsilon=\rho_* c^2 e^{P/A}-P-A.
\label{ld4}
\end{eqnarray}
In the nonrelativistic limit, using $\epsilon\sim
\rho c^2$, we recover Eq. (\ref{lal1}).

\subsubsection{Determination of $\rho$, $P(\rho)$ and $V_{\rm tot}(\rho)$}

The pseudo rest-mass density and the SF potential are determined by Eqs.
(\ref{gesd4b}) and (\ref{gesd5}) with Eq. (\ref{ld2}). We get
\begin{eqnarray}
\rho=\frac{\rho_m^2}{\rho_m-\frac{A}{c^2}},
\label{ld6wa}
\end{eqnarray}
\begin{eqnarray}
V_{\rm tot}=\frac{1}{2}\left\lbrack \rho_m c^2-2A\ln\left
(\frac{\rho_m}{\rho_*}\right )-A\right\rbrack.
\label{ld8w}
\end{eqnarray}
Equation (\ref{ld6wa}) can be inverted to give
\begin{eqnarray}
\rho_m=\frac{\rho\pm\sqrt{\rho^2-\frac{4A\rho}{c^2}}}{2}.
\label{ld6wan}
\end{eqnarray}
Combined with Eqs. (\ref{ld1}) and
(\ref{ld8w}) we explicitly obtain $P(\rho)$ and $V_{\rm tot}(\rho)$ under
the form
\begin{eqnarray}
P=A\ln\left
(\frac{\rho\pm\sqrt{\rho^2-\frac{4A\rho}{c^2}}}{2\rho_*}\right ),
\label{ld1exp}
\end{eqnarray}
\begin{equation}
V_{\rm tot}=
\frac{\rho\pm\sqrt{\rho^2-\frac{4A\rho}{c^2}}}{4} c^2-A\ln\left
(\frac{\rho\pm\sqrt{\rho^2-\frac{4A\rho}{c^2}}}{2\rho_*}\right
)-\frac{A}{2}.
\label{ld8wexp}
\end{equation}
In the
nonrelativistic limit, using $\rho_m\sim \rho (1-A/\rho c^2)$ and Eq.
(\ref{db10}), we recover Eqs. (\ref{lal1}) and (\ref{lal2}).

\subsubsection{Lagrangian ${\cal L}(X)$}

The Lagrangian ${\cal L}(X)$ is determined by Eqs. (\ref{gesd8}), (\ref{ld1})
and (\ref{ld2}). We get
\begin{eqnarray}
X=\frac{c^2}{2}\left (1-\frac{A}{\rho_m c^2}\right )^2.
\label{ld9b}
\end{eqnarray}
This relation can be
inverted to give
\begin{eqnarray}
\rho_m=\frac{A/c^2}{1-\left (\frac{2X}{c^2}\right )^{1/2}}.
\label{ld9b2}
\end{eqnarray}
We then obtain
\begin{equation}
\label{ld9}
{\cal L}(X)=P=-A\ln \left \lbrack \frac{\rho_*
c^2}{A}\left (1-\sqrt{\frac{2X}{c^2}}\right )\right\rbrack.
\end{equation}
In the nonrelativistic limit, using Eq. (\ref{nr5}), we recover Eq.
(\ref{lal6}).

{\it Remark:} Starting from Eq. (\ref{pd9g}), taking the limit
$\gamma\rightarrow 0$, $K\rightarrow +\infty$ with $K\gamma=A$ constant, and
proceeding as in Eq. (\ref{lal7}), we obtain Eq. (\ref{ld9}) up to an
additional constant. More generally, we can recover in the same manner the
other equations of this section.

\subsection{Logotropic equation of state of type III}
\label{sec_lt}

The logotropic equation of state of type III writes \cite{logosf} 
\begin{eqnarray}
P=A\ln\left (\frac{\rho}{\rho_*}\right ).
\label{lt1}
\end{eqnarray}
Using Eq. (\ref{etoilebis}), the SF potential is 
\begin{eqnarray}
V_{\rm tot}(\rho)=\frac{1}{2}\rho c^2-A\ln \left (\frac{\rho}{\rho_*}\right )-A.
\label{lt2}
\end{eqnarray}
In the nonrelativistic limit, we recover  Eqs. (\ref{lal1}) and
(\ref{lal2}) [recalling Eq. (\ref{db10})]. Actually, Eqs.
(\ref{lt1})
and (\ref{lt2}) coincide with Eqs. (\ref{lal1}) and (\ref{lal2}). The SF
potential $V(\rho)$ is logarithmic.

\subsubsection{Determination of $\epsilon$ and $P(\epsilon)$}

The energy density is determined by Eqs. (\ref{gest1}) with Eq. (\ref{lt2}).
This yields
\begin{eqnarray}
\epsilon=\rho c^2-A\ln \left (\frac{\rho}{\rho_*}\right )-2A.
\label{lt4}
\end{eqnarray}
The  pressure is determined by Eq.  (\ref{gest2}) with Eq. (\ref{lt2}). This returns Eq. (\ref{lt1}).
Eliminating $\rho$ between Eqs. (\ref{lt1}) and (\ref{lt4}), we obtain $P(\epsilon)$ under the inverse form $\epsilon(P)$ as
\begin{eqnarray}
\epsilon=\rho_* c^2 e^{P/A}-P-2A.
\label{lt3}
\end{eqnarray}
In the nonrelativistic limit, using $\epsilon\sim
\rho c^2$, we recover Eq. (\ref{lal1}).

\subsubsection{Determination of $\rho_m$, $P(\rho_m)$ and  $u(\rho_m)$}

The rest-mass density and the internal energy are determined by Eqs.
(\ref{gest4a}) and (\ref{gest6}) with Eq. (\ref{lt2}). We get
\begin{equation}
\label{lest4wb}
\rho_m=\sqrt{\rho\left (\rho-\frac{2A}{c^2}\right )},
\end{equation}
\begin{equation}
\label{lest6w}
u=\rho c^2-A\ln \left (\frac{\rho}{\rho_*}\right )-2A-\rho_m(\rho)c^2.
\end{equation}
Equation (\ref{lest4wb}) can be inverted to
give
\begin{eqnarray}
\rho=\frac{A}{c^2}+\sqrt{\frac{A^2}{c^4}+\rho_m^2}.
\label{ld6wab}
\end{eqnarray}
Combined with Eqs. (\ref{lt1}) and (\ref{lest6w}) we explicitly obtain 
$P(\rho_m)$ and $u(\rho_m)$ under the form
\begin{eqnarray}
P=A\ln\left
(\frac{A}{\rho_*c^2}+\sqrt{\frac{A^2}{\rho_*^2c^4}+\frac{\rho_m^2}{\rho_*^2}}
\right ),
\label{lt1exp}
\end{eqnarray}
\begin{equation}
\label{lest6wexp}
u=\sqrt{A^2+\rho_m^2c^4}-A\ln\left
(\frac{A}{\rho_*c^2}+\sqrt{\frac{A^2}{\rho_*^2c^4}+\frac{\rho_m^2}{\rho_*^2}}
\right )-A-\rho_mc^2.
\end{equation}
In the nonrelativistic
limit, using $\rho_m\simeq \rho-A/c^2$, we recover Eqs.
(\ref{lal1}) and (\ref{lal2}) [recalling Eq. (\ref{addq})].

\subsubsection{Lagrangian ${\cal L}(X)$}

The Lagrangian ${\cal L}(X)$ is determined by Eq. (\ref{gest7})   with Eq. (\ref{lt2}). We get
\begin{equation}
\label{pt5b}
X=\frac{1}{2}c^2-\frac{A}{\rho}.
\end{equation}
This equation can be reversed to give
\begin{equation}
\label{pt5b2}
\rho=\frac{A}{\frac{1}{2}c^2-X}.
\end{equation}
We then obtain
\begin{equation}
\label{lt8}
{\cal L}(X)=P=-A\ln \left \lbrack \frac{\rho_*
c^2}{2A}\left (1-\frac{2X}{c^2}\right )\right\rbrack.
\end{equation}
In the nonrelativistic limit, using Eq. (\ref{nr5}), we recover Eq.
(\ref{lal6}). Actually, Eq. (\ref{lt8}) coincides with Eq. 
(\ref{lal6}) with $c^2/2-X$ in place of $x$. Using Eq. (\ref{k2invb}) or
Eq. (\ref{ak8}), we obtain 
the equation of motion
\begin{equation}
\label{bbbq}
D_\mu\left\lbrack
\frac{\partial^\mu\theta}{1-\frac{1}{c^2}
\partial_\nu\theta\partial^\nu\theta}\right\rbrack=0.
\end{equation}

{\it Remark:} Starting from Eq. (\ref{qpd9g}), taking the limit
$\gamma\rightarrow 0$, $K\rightarrow +\infty$ with $K\gamma=A$ constant, and
proceeding as in Eq. (\ref{lal7}), we obtain Eq. (\ref{lt8}) up to an
additional constant. More generally, we can recover in the same manner the
other equations of this section.

\section{Conclusion}

In this paper, we have showed that the equation of state of a relativistic
barotropic fluid could be specified in different manners depending
on whether the pressure $P$ is expressed in terms of the energy density
$\epsilon$ (model I), the rest-mass density $\rho_m$ (model II), or the pseudo
rest-mass density $\rho$ (model III). In model II, specifying the equation of
state $P(\rho_m)$ is equivalent to specifying the internal energy $u(\rho_m)$.
In model III, specifying the equation of state $P(\rho)$ is equivalent to
specifying the potential $V(\rho)$ of the complex SF to which the fluid is
associated in the TF limit. In the nonrelativistic limit, these three
formulations coincide. 

We have shown how these different models are connected to each
other. We have established general equations allowing us to determine
[$\epsilon$, $P(\epsilon)$], [$\rho_m$, $P(\rho_m)$, $u(\rho_m)$] and [$\rho$,
$P(\rho)$ $V(\rho)$] once an equation of state is specified under the form I, II
or III. 

In model III, we have determined the hydrodynamic representation of a complex SF
with a potential $V(|\varphi|^2)$ and the form of its Lagrangian. In the
TF approximation, we can use the Bernoulli equation to obtain a reduced
Lagrangian of the form ${\cal L}(X)$ with
$X=\frac{1}{2}\partial_{\mu}\theta\partial^{\mu}\theta$, where $\theta$ is the
phase of the  SF. This is a k-essence Lagrangian whose expression is
determined by the potential of the complex SF. We have established general
equations allowing us to obtain ${\cal L}(X)$ once an equation of state is
specified under the form I, II or III. 

For illustration, we have applied our formalism to polytropic, isothermal and
logotropic
equations of state of type I, II and III that  have been proposed as UDM models.
We have recovered previously obtained results, and we have derived new results.
For example, we have established the general analytical expression of the
k-essence Lagrangian of polytropic and isothermal equations of state of type I,
II and III. For $\gamma=-1$ (Chaplygin gas), the models of type I and III are
equivalent and return the Born-Infeld action, while the model of type II leads
to a different action. We have also established the
general analytical expression of the k-essence Lagrangian associated with a
logotrope of type
II and III (the k-essence Lagrangian associated with a logotrope of type I
cannot be obtained analytically). 

In a future contribution \cite{prepmixed}, we will apply our general formalism
to more
complicated equations of state which can be viewed as a superposition of
polytropic, isothermal (linear) and logotropic equations of state.

The mixed equation of state of type I generically writes
\begin{equation}
\label{c1}
P=K\left (\frac{\epsilon}{c^2}\right
)^{\gamma}+\alpha\epsilon-\epsilon_{\Lambda}+A\ln\left
(\frac{\epsilon}{\epsilon_P}\right ),
\end{equation}
where we can add several polytropic terms with different indices $\gamma$.
More specifically, we can consider generalized polytropic models of type I of
the form 
\begin{equation}
\label{c2}
P=-(\alpha+1)\epsilon \left
(\frac{\epsilon}{\epsilon_P}\right
)^{1/|n_e|}+\alpha\epsilon-(\alpha+1)\epsilon \left
(\frac{\epsilon_\Lambda}{\epsilon}\right
)^{1/|n_l|},
\end{equation}
or
\begin{equation}
\label{c3}
P=-(\alpha+1)\frac{\epsilon^2}{\epsilon_P}
+\alpha\epsilon-(\alpha+1)\epsilon_\Lambda.
\end{equation}
Polytropic and isothermal (linear) equations of state of type I have been
studied in the
context of relativistic stars
\cite{tooper1,chavharko,partially,chandra72,yabushita1,yabushita2,aarelat1,
aarelat2} and cosmology
\cite{kmp,gkmp,bentoGCG,cosmopoly1,cosmopoly2,cosmopoly3}. Mixed models
of type I of the form of Eqs. (\ref{c1})-(\ref{c3}) have been introduced
and studied in cosmology in Refs.
\cite{cosmopoly1,cosmopoly2,cosmopoly3,aip,jgravity,universe,iop}. The
equation of state (\ref{c3}) describes the early inflation, the intermediate
decelerating expansion, and the late accelerating expansion of the universe in a
unified manner.

The mixed equation of state of type II generically writes
\begin{equation}
\label{c4}
P=K\rho_m^{\gamma}+\alpha\rho_m c^2-\rho_{\Lambda}c^2+A\ln\left
(\frac{\rho_m}{\rho_P}\right ),
\end{equation}
where we can add several polytropic terms with different indices $\gamma$. It is
associated with an internal
energy of the form
\begin{eqnarray}
\label{c5}
u=\frac{K}{\gamma-1}\rho_m^{\gamma}+\alpha\rho_m
c^2\left\lbrack\ln\left
(\frac{\rho_m}{\rho_*}\right
)-1\right\rbrack\nonumber\\
+\rho_{\Lambda}c^2-A\ln\left
(\frac{\rho_m}{\rho_P}\right )-A.
\end{eqnarray}
Polytropic, isothermal (linear) and logotropic equations of state of type
II have been
studied in the
context of relativistic stars \cite{tooper2,partially} and cosmology
\cite{stiff,epjp}. It is often assumed that DM is
pressureless ($P=0$) so that $\alpha=0$. However, a nonvanishing value
of $\alpha$ can account for thermal effects as in \cite{clm1,clm2,modeldm}. In
that case $\alpha c^2=k_B T/m$. Mixed models
of type II of the form of Eqs.  (\ref{c4}) and (\ref{c5}) have been
introduced and studied in Refs. \cite{stiff,epjp}.

The mixed equation of state of type III generically writes
\begin{equation}
\label{c6}
P=K\rho^{\gamma}+\alpha\rho c^2-\rho_{\Lambda}c^2+A\ln\left
(\frac{\rho}{\rho_P}\right ),
\end{equation}
where we can add several polytropic terms with different indices $\gamma$. It is
associated with a complex SF
potential of the form
\begin{eqnarray}
\label{c7}
V_{\rm tot}=\frac{1}{2}\rho c^2+\frac{K}{\gamma-1}\rho^{\gamma}+\alpha\rho
c^2\left\lbrack\ln\left
(\frac{\rho}{\rho_*}\right
)-1\right\rbrack\nonumber\\
+\rho_{\Lambda}c^2-A\ln\left
(\frac{\rho}{\rho_P}\right )-A,
\end{eqnarray}
where we recall that $\rho=(m/\hbar)^2|\varphi|^2$. Polytropic, isothermal
(linear)
and logotropic equations of state of type III have been studied in the
context of relativistic stars \cite{colpi,chavharko,partially} and cosmology
\cite{shapiro,abrilas,csf,graal}. Mixed models
of type III of the form of Eqs.  (\ref{c6}) and (\ref{c7}) have been
introduced and studied in Refs. \cite{graal,csf}.

\appendix

\section{General identities for a nonrelativistic cold gas}
\label{sec_gicg}

The first principle of
thermodynamics for a nonrelativistic gas can be written as
\begin{equation}
\label{fo1}
d\left (\frac{u}{\rho}\right )=-P d\left (\frac{1}{\rho}\right )+T d\left
(\frac{s}{\rho}\right ),
\end{equation}
where $u$ is the density of internal energy, $s$ the density of entropy,
$\rho=nm$
 the mass density, $P$ the pressure, and $T$ the temperature. For a cold ($T=0$)
or isentropic ($s/\rho_m={\rm cst}$) gas, it reduces to
\begin{equation}
\label{fo2}
d\left (\frac{u}{\rho}\right )=-P d\left (\frac{1}{\rho}\right
)=\frac{P}{\rho^2}d\rho.
\end{equation}
Introducing the enthalpy per particle
\begin{equation}
\label{fo3}
h=\frac{P+u}{\rho},
\end{equation}
we get
\begin{equation}
\label{fo4}
du=h d\rho\qquad {\rm and}\qquad dh=\frac{dP}{\rho}.
\end{equation}
For a barotropic gas for which $P=P(\rho)$, the foregoing
equations can be
written as
\begin{equation}
\label{fo5}
P(\rho)=-\frac{d(u/\rho)}{d(1/\rho)}=\rho^2\left \lbrack
\frac{u(\rho)}{\rho}\right \rbrack'=\rho u'(\rho)-u(\rho),
\end{equation}
\begin{equation}
\label{fo5b}
c_s^2=P'(\rho)=\rho u''(\rho),\qquad h(\rho)=\frac{P(\rho)+u(\rho)}{\rho},
\end{equation}
\begin{equation}
\label{fo6}
\qquad h(\rho)=u'(\rho),\qquad h'(\rho)=\frac{P'(\rho)}{\rho},
\end{equation}
where $c_s^2=P'(\rho)$ is the squared speed of sound. Eq. (\ref{fo5}) determines the equation of state $P(\rho)$ as a function of the internal energy $u(\rho)$.  Inversely, the internal energy is determined by the equation of state according to the relation
\begin{eqnarray}
u(\rho)=\rho\int\frac{P(\rho)}{\rho^2}\, d\rho,
\label{fo6b}
\end{eqnarray}
which is the solution of the
differential equation
\begin{equation}
\label{fo5b2}
\rho \frac{du}{d\rho}-u(\rho)=P(\rho).
\end{equation}

Comparing Eq. (\ref{fo6b}) with Eq. (\ref{etoile}), we see that
the
potential $V(\rho)$ represents the density of internal energy
\begin{equation}
\label{fo7}
u(\rho)=V(\rho).
\end{equation}
We then have
\begin{equation}
\label{nfo5}
P(\rho)=\rho^2\left \lbrack \frac{V(\rho)}{\rho}\right \rbrack'=\rho
V'(\rho)-V(\rho),
\end{equation}
\begin{equation}
\label{bfo5b}
c_s^2=P'(\rho)=\rho V''(\rho),\qquad h(\rho)=\frac{P(\rho)+V(\rho)}{\rho},
\end{equation}
\begin{equation}
\label{fo6bb}
\qquad h(\rho)=V'(\rho),\qquad h'(\rho)=\frac{P'(\rho)}{\rho},
\end{equation}
\begin{eqnarray}
V(\rho)=\rho\int\frac{P(\rho)}{\rho^2}\, d\rho.
\label{fo6c}
\end{eqnarray}

{\it Remark:} The first principle of thermodynamics can be written as
\begin{equation}
\label{ngr1}
du=Tds+\mu dn.
\end{equation}
This can be viewed as the variational principle ($\delta s/k_B-\beta\delta
u+\alpha\delta n=0$ with $\beta=1/k_B T$ and $\alpha=\mu/k_B T$) associated with
the maximization of the entropy density $s$ at fixed energy density $u$ and
particle density $n$ \cite{gr1}. Combined with the Gibbs-Duhem relation
\cite{gr1}
\begin{equation}
\label{ngr2}
s=\frac{u+P-\mu n}{T},
\end{equation}
we obtain Eq. (\ref{fo1}) and
\begin{equation}
\label{ngr2b}
sdT-dP+nd\mu=0.
\end{equation}
If $T={\rm cst}$, then $dP=nd\mu$. For $T=0$, the foregoing equations reduce to
\begin{equation}
\label{ngr3}
du=\mu dn,\qquad \mu=\frac{u+P}{n},\qquad dP=nd\mu,
\end{equation}
which are equivalent to Eqs. (\ref{fo3}) and (\ref{fo4}) with $\mu=mh$.
Therefore, the enthalpy $h({\bf r})$ is equal to the local chemical potential
$\mu({\bf r})$ by unit of mass: $h({\bf r})=\mu({\bf r})/m$.

\section{K-essence Lagrangian of a real SF}
\label{sec_lrsf}

\subsection{General results}
\label{sec_kess}

We consider a relativistic real SF $\varphi(x^\mu)=\varphi(x,y,z,t)$
characterized by the action
\begin{equation}
S=\int \mathcal{L} \sqrt{-g}\, d^4x,
\label{ak1}
\end{equation}
where $\mathcal{L}=\mathcal{L}(\varphi,\partial_\mu\varphi)$
is the Lagrangian density and $g={\rm det}(g_{\mu\nu})$ is the determinant of
the metric tensor. The Lagrangian of a relativistic real SF
$\varphi$ is usually written under the canonical form as
\begin{eqnarray}
{\cal
L}=X-V(\varphi),
\label{ak2}
\end{eqnarray}
where 
\begin{eqnarray}
X=\frac{1}{2}\partial_{\mu}\varphi\partial^{\mu}\varphi
\label{ak3}
\end{eqnarray}
is the kinetic energy and $V$ is the potential energy.\footnote{Note that $V$
represents here the total potential including the rest mass term. For brevity,
we write $V$ instead of $V_{\rm tot}$.} In that
case, all the physics of the problem is contained in the potential term.
However, some authors have proposed to take $V=0$ and modify the kinetic term.
This leads to a Lagrangian of the form
\begin{eqnarray}
{\cal
L}={\cal L}(X)
\label{ak4}
\end{eqnarray}
that  is called a k-essence Lagrangian \cite{ams}. In that case, the
physics
of the problem
is encapsulated in 
the noncanonical kinetic term ${\cal L}(X)$.\footnote{K-essence Lagrangians were
initially
introduced to describe inflation (k-inflation) \cite{adm,gm}. They were later
used
to described dark energy \cite{chiba,ams,ams2}. K-essence Lagrangians can also
be
obtained
from a canonical complex SF in the TF limit $\hbar\rightarrow 0$
\cite{btv,btvB2}.
In that case, the real SF $\varphi$ corresponds to the  action (phase) $\theta$
of
the complex SF (see Sec. \ref{sec_k}).} Eq. (\ref{ak4}) is a pure k-essence
Lagrangian. More general Lagrangians 
\begin{eqnarray}
{\cal
L}={\cal L}(X,\varphi)
\label{ak5}
\end{eqnarray}
can depend both on $X$ and $\varphi$ \cite{adm,gm}. The particular forms ${\cal
L}=V(\varphi)F(X)$ and ${\cal L}=F(X)-V(\varphi)$ have been  specifically
introduced
in Refs. \cite{adm} and \cite{mv,babichev,flh}  respectively.

The least action principle $\delta S=0$, which is equivalent to
the Euler-Lagrange equation
\begin{eqnarray}
\label{ak6}
D_{\mu}\left\lbrack \frac{\partial {\cal
L}}{\partial(\partial_\mu\varphi)}\right\rbrack-\frac{\partial {\cal
L}}{\partial\varphi}=0,
\end{eqnarray}
yields the equation of motion
\begin{equation}
\label{ak7}
D_\mu\left ( \frac{\partial {\cal L}}{\partial X}\partial^\mu\varphi\right )-\frac{\partial {\cal L}}{\partial\varphi}=0.
\end{equation}
For the Lagrangian (\ref{ak4}), it reduces to 
\begin{equation}
\label{ak8}
D_\mu\left\lbrack {\cal L}'(X)\partial^\mu\varphi\right\rbrack=0.
\end{equation}

For the Lagrangian (\ref{ak4}) the current is given by
\begin{eqnarray}
\label{j1b}
J^{\mu}=-\frac{\partial {\cal L}}{\partial
(\partial_\mu\varphi)},
\end{eqnarray}
yielding
\begin{equation}
\label{charge11}
J_{\mu}=-{\cal L}'(X)\partial_\mu\varphi.
\end{equation}
Equation (\ref{ak8}) can then be written as $D_{\mu}J^{\mu}=0$.
It can therefore be viewed as a continuity equation expressing the
local conservation of the charge (or the local conservation of the boson
number) given by $Q=\frac{e}{mc}\int J^0\sqrt{-g}\, d^3x$, i.e.,
\begin{equation}
\label{charge12}
Q=-\frac{e}{mc}\int {\cal L}'(X)\partial^0\varphi\sqrt{-g}\, d^3x.
\end{equation}
In the present context, the conservation of the charge (or boson
number) is related to the invariance of the Lagrangian
density under the constant shift $\varphi\rightarrow\varphi+{\rm cst}$ of the
SF (Noether theorem).\footnote{Note that the more general Lagrangian (\ref{ak5})
does not conserve the charge
(or the boson number).}
Introducing the quadrivelocity 
\begin{eqnarray}
u_{\mu}=-\frac{\partial_{\mu}\varphi}{\sqrt{2X}}c,
\label{charge13}
\end{eqnarray}
which satisfies by construction the identity $u_{\mu}u^{\mu}=c^2$,
we get
\begin{equation}
\label{charge15}
J_{\mu}={\cal L}'(X)\sqrt{2X}\frac{u_\mu}{c}.
\end{equation}
We can therefore rewrite the continuity equation as
\begin{equation}
\label{charge15b}
D_\mu\left\lbrack {\cal L}'(X)\sqrt{2X}u^\mu\right\rbrack=0.
\end{equation}
Comparing Eqs. (\ref{charge15}) and (\ref{charge15b}) with $J_{\mu}=\rho_m
u_{\mu}$ and $D_{\mu}(\rho_m u^{\mu})=0$, we find
that the
rest-mass density is given by
\begin{equation}
\label{charge16}
\rho_m={\cal L}'(X)\sqrt{2X}\frac{1}{c}.
\end{equation}

The energy-momentum tensor is given by Eq. (\ref{em1}). It satisfies the
conservation law $D_{\mu}T^{\mu\nu}=0$. For a real SF we have 
\begin{eqnarray}
\label{ak9b}
T_{\mu}^{\nu}=\frac{\partial {\cal L}}{\partial
(\partial_\nu\varphi)}\partial_\mu\varphi-g_{\mu}^{\nu}{\cal L}.
\end{eqnarray}
The energy-momentum tensor associated with the Lagrangian
(\ref{ak5}) is
\begin{eqnarray}
T_{\mu\nu}=\frac{\partial {\cal L}}{\partial X}\partial_{\mu}\varphi\partial_{\nu}\varphi-g_{\mu\nu}{\cal L}.
\label{ak11}
\end{eqnarray}
Introducing the quadrivelocity from Eq. (\ref{charge13}) we get
\begin{eqnarray}
T_{\mu\nu}=2X \frac{\partial {\cal L}}{\partial X}
\frac{u_{\mu}u_{\nu}}{c^2}-g_{\mu\nu}{\cal L}.
\label{ak13b}
\end{eqnarray}
The   energy-momentum tensor (\ref{ak11}) can be written under the perfect fluid
form 
\begin{eqnarray}
T_{\mu\nu}=(\epsilon+P)\frac{u_{\mu}u_{\nu}}{c^2}-P g_{\mu\nu},
\label{ak14}
\end{eqnarray}
where $\epsilon$ is the energy density and $P$ is the pressure, provided that we
make the 
identifications 
\begin{eqnarray}
P={\cal L}\quad {\rm and}\quad \epsilon+P=2X\frac{\partial {\cal L}}{\partial X}.
\label{ak15}
\end{eqnarray}
As a result, the pressure and the energy density associated with
the Lagrangian (\ref{ak5}) are given by
\begin{eqnarray}
P={\cal L}(X,\varphi),
\label{ak17}
\end{eqnarray}
\begin{eqnarray}
\epsilon=2X\frac{\partial P}{\partial X}-P.
\label{ak16}
\end{eqnarray}
The Lagrangian plays the role of an
effective pressure. If the Lagrangian satisfies the condition $X\partial
P/\partial X\ll P$ for some range of $X$ and $\varphi$, then the equation of
state
is $P\simeq -\epsilon$ (vacuum energy) and we have an inflationary solution
\cite{mukhanov}. On the other hand, for the Lagrangian ${\cal
L}=V(\varphi)X$ corresponding to $P\propto X$ i.e. $X\partial
P/\partial X=P$  we obtain the stiff equation of state $P=\epsilon$.
In that case, the equation of motion (\ref{ak7}) becomes
$D_\mu(V(\varphi)\partial^{\mu}\varphi)-\frac{1}{2}
V'(\varphi)\partial_\mu\varphi\partial^\mu\varphi=0$. It reduces to
$\square\varphi=0$ when ${\cal
L}=A X$.

The equation of state parameter and the squared speed of sound are given by
\cite{gm}
\begin{eqnarray}
w=\frac{P}{\epsilon}=\frac{P}{2X\frac{\partial P}{\partial X}-P}
\label{ak18}
\end{eqnarray}
and
\begin{eqnarray}
c_s^2=\frac{\frac{\partial P}{\partial X}}{\frac{\partial \epsilon}{\partial X}}c^2=\frac{\frac{\partial P}{\partial X}}{\frac{\partial P}{\partial X}+2X\frac{\partial^2P}{\partial X^2}}c^2.\label{ak19}
\end{eqnarray}
We note that $c_s\simeq c$ if
$2X\partial^2P/\partial X^2\ll \partial P/\partial
X$. This is the case  in particular for the Lagrangian ${\cal
L}=V(\varphi)X$ discussed above for which $c_s=c$ exactly.

In the general case, we
have $P=P(X,\varphi)$ and $\epsilon=\epsilon(X,\varphi)$
so that the fluid is not necessarily barotropic. However, for a $k$-essence SF
described by a Lagrangian of the form of Eq. (\ref{ak4}), we have $P=P(X)$ and
$\epsilon=\epsilon(X)$ implying $P=P(\epsilon)$. In that case, the fluid is barotropic
and $c_s^2=P'(\epsilon)c^2$. On the other hand, using Eqs.
(\ref{mtd4a}), (\ref{charge16}), (\ref{ak17}) and 
(\ref{ak16}), we find that the enthalpy is
given by
\begin{equation}
h=c\sqrt{2X}.
\label{en3wq}
\end{equation}

{\it Remark:} For the Lagrangian  ${\cal
L}=V(\varphi)X^{(\alpha+1)/2\alpha}$, 
we obtain
the linear equation of state $P=\alpha\epsilon$. This includes stiff matter
($\alpha=1$; ${\cal
L}=V(\varphi)X$), radiation  ($\alpha=1/3$; ${\cal
L}=V(\varphi)X^{2}$) and a cosmological constant ($\alpha=-1$; ${\cal
L}=V(\varphi)$). The equation of motion (\ref{ak7}) becomes
\begin{eqnarray}
\frac{\alpha+1}{2\alpha}D_{\mu}\left
\lbrack V(\varphi)X^{\frac{1-\alpha}{2\alpha}}\partial^{\mu}\varphi\right
\rbrack-V'(\varphi)X^{\frac{\alpha+1}{2\alpha}}=0.
\end{eqnarray}
It reduces to
\begin{eqnarray}
D_{\mu}\left
(X^{\frac{1-\alpha}{2\alpha}}\partial^{\mu}\varphi\right
)=0
\end{eqnarray}
when  ${\cal
L}=A X^{(\alpha+1)/2\alpha}$. For the Lagrangian  ${\cal
L}=V(\varphi)\ln X$ (which can be viewed as a limit of the Lagrangian
$V(\varphi)X^{(\alpha+1)/2\alpha}$ for $\alpha\rightarrow -1$ and 
$V\rightarrow +\infty$ with $V(\alpha+1)$ finite), we obtain
$P=2V(\varphi)-\epsilon$ which reduces to the affine equation of state
$P=2A-\epsilon$ \cite{cosmopoly2,cosmopoly3} when $V(\varphi)=A$. In that case,
the equation of
motion (\ref{ak7}) becomes
\begin{eqnarray}
D_{\mu}\left
\lbrack
\frac{V(\varphi)}{X}\partial^{\mu}\varphi\right
\rbrack-V'(\varphi)\ln X=0.
\end{eqnarray}
It reduces to
\begin{eqnarray}
D_{\mu}\left
(\frac{1}{X}\partial^{\mu}\varphi\right
)=0
\end{eqnarray}
when  ${\cal
L}=A \ln X$.

\subsection{Canonical SF}
\label{sec_canosf}

The Lagrangian of a canonical  SF is 
\begin{eqnarray}
{\cal L}=\frac{1}{2}g^{\mu\nu}\partial_{\mu}\varphi\partial_{\nu}
\varphi-V(\varphi),
\label{ak24}
\end{eqnarray}
where the first term is the kinetic energy and the second term is minus the
potential energy. It is of the form ${\cal L}=X-V(\varphi)$. The least
action principle
$\delta S=0$, which is equivalent
to the Euler-Lagrange equation (\ref{ak6}), leads 
to the KG equation 
\begin{equation}
\label{ak25}
\square\varphi+\frac{dV}{d\varphi}=0,
\end{equation}
where $\square=D_\mu\partial^{\mu}$ is the d'Alembertian. A canonical real SF
does not conserve the charge.

The energy-momentum tensor (\ref{ak9b}) associated with the canonical Lagrangian
(\ref{ak24}) is
\begin{eqnarray}
T_{\mu\nu}=\partial_{\mu}\varphi\partial_{\nu}\varphi-g_{\mu\nu}{\cal L}.
\label{ak27}
\end{eqnarray}
Repeating the procedure of Appendix \ref{sec_kess} we find that  the energy
density and the pressure are given 
by 
\begin{eqnarray}
\epsilon=\frac{1}{2}\partial_{\mu}\varphi\partial^{\mu}
\varphi+V(\varphi),
\label{ak28}
\end{eqnarray}
\begin{eqnarray}
P=\frac{1}{2}\partial_{\mu}\varphi\partial^{\mu}
\varphi-V(\varphi).
\label{ak29}
\end{eqnarray}
Since $\epsilon=X+V(\varphi)$ and $P=X-V(\varphi)$, we find that
$P=\epsilon-2V(\varphi)$,
$w=[X-V(\varphi)]/[X+V(\varphi)]$ and $c_s=c$. For a
canonical SF, the speed of sound is equal to the speed of light.

When $X\gg V$, we obtain $\epsilon=X$ and $P=X$ leading to the equation of state
$P=\epsilon$ corresponding to stiff matter. This is  the so-called kination
regime \cite{kination}. This regime is achieved in particular when $V=0$. In
that
case, the Lagrangian ${\cal L}=X$ describes a noninteracting massless
SF and the KG equation reduces to $\square\varphi=0$. When
$X\ll V$, we obtain $\epsilon=V$ and $P=-V$ leading to the equation of state
$P=-\epsilon$ 
corresponding to the vacuum energy. This regime is achieved in
particular when $\varphi=\varphi_0$ is constant ($X=0$) and lies at the bottom
of the
potential ($V'(\varphi_0)=0$). In cosmology, this equation of state
leads to a de Sitter era where $\epsilon=V(\varphi_0)$ is constant and the scale
factor increases exponentially rapidly with time as $a\propto {\rm exp}[(8\pi
G\epsilon/3c^2)^{1/2}t]$.
The condition $X\ll V$ corresponds to the slow-roll regime \cite{mukhanov}.

When $V(\varphi)=V_0$ is constant, the Lagrangian   ${\cal L}=X-V_0$ describes a
massless SF in the presence of a cosmological constant ($\epsilon_\Lambda=V_0$).
In that case,
$\epsilon=X+V_0$ and $P=X-V_0$ leading to the 
affine equation of state $P=\epsilon-2V_0=\epsilon-2\epsilon_\Lambda$
\cite{cosmopoly2}. In
cosmology,  when $V_0>0$, this equation of state generically
leads to a stiff matter era followed by a de Sitter era (or a de Sitter era
alone when $X=0$, i.e., $\varphi={\rm cst}$). When $V_0=0$  we recover the
Lagrangian ${\cal L}=X$ a free massless
SF. In that case, $\epsilon=X$ and $P=X$ leading to the stiff equation of state
$P=\epsilon$. In
cosmology, we have a pure stiff matter era.

{\it Remark:} The Lagrangian of a 
particle of mass $m$ and position $q(t)$ in Newtonian mechanics is $L=(1/2)m
\dot q^2-V(q)$. Its impulse is  $p=\partial L/\partial \dot q=m\dot q$ and its
energy is $E=p\dot q-L=\dot q\partial L/\partial \dot q-L=(1/2)m\dot q^2+V(q)$,
i.e., $E=p^2/(2m)+V(q)$. Its equation of motion is given by the
Euler-Lagrange equation $(d/dt)(\partial L/\partial\dot q)-\partial
L/\partial q=0$ yielding $m\ddot q=-V'(q)$. The Lagrangian equations of a
canonical SF are similar
to the Lagrangian equations of a nonrelativistic particle in which the SF
$\varphi(x^\mu)$ plays the role of $q(t)$ and the SF potential $V(\varphi)$ the
role of $V(q)$.

\subsection{Tachyonic SF}
\label{sec_tach}

The Lagrangian of a tachyonic SF is
\begin{eqnarray}
{\cal
L}=-V(\varphi)\sqrt{1-\frac{1}{c^2}g^{\mu\nu}\partial_{\mu}\varphi\partial_{\nu}
\varphi}.
\label{ak35}
\end{eqnarray}
This corresponds to the
Born-Infeld Lagrangian (\ref{biintro}) 
multiplied by $V(\varphi)$. It is of the form ${\cal
L}=-V(\varphi)\sqrt{1-2X/c^2}$. This Lagrangian was introduced
by Sen \cite{sen1,sen2,sen3} in the context of string theory and $d$-branes and
further discussed in
\cite{gibbons,frolov,felder,paddytachyon,pc,bagla,gkmp,feinstein}. The relation
to k-essence fields was made in
\cite{frolov,paddytachyon,feinstein}. The least action principle $\delta S=0$,
which is equivalent to the Euler-Lagrange equation (\ref{ak6}), leads to the
equation of motion 
\begin{equation}
\label{ak36}
D_{\mu}\left\lbrack
\frac{V(\varphi)/c^2}{\sqrt{1-\frac{1}{c^2}\partial_{\mu}\varphi\partial^{\mu}
\varphi}}\partial^{\mu}\varphi\right\rbrack+V'(\varphi)\sqrt{1-\frac{1}{c^2}
\partial_{\mu}\varphi\partial^{\mu}
\varphi}=0
\end{equation}
or, equivalently, 
\begin{equation}
\label{ak36autre}
D_{\mu}\partial^{\mu}\varphi+\frac{D_{\mu}\partial_{\nu}\varphi}{1-\partial_{\mu
} \varphi
\partial^{\mu}\varphi}\partial^{\mu}\varphi\partial^{\nu}\varphi+(\ln V)'c^2=0.
\end{equation}
A real tachyonic SF
does not conserve the charge.

The energy-momentum tensor (\ref{ak9b}) associated with the tachyonic Lagrangian
(\ref{ak35}) is
\begin{equation}
\label{ak37}
T_{\mu\nu}=\frac{V(\varphi)/c^2}{\sqrt{1-\frac{1}{c^2}\partial_{\mu}
\varphi\partial^ {\mu}
\varphi}}\partial_{\mu}\varphi\partial_{\nu}\varphi-g_{\mu\nu}{\cal L}.
\end{equation}
Repeating the procedure of Appendix \ref{sec_kess} we find that  the energy 
density and the pressure are given by 
\begin{eqnarray}
\epsilon=\frac{V(\varphi)}{\sqrt{1-\frac{1}{c^2}\partial_{\mu}\varphi\partial^{
\mu}\varphi}},
\label{ak38}
\end{eqnarray}
\begin{eqnarray}
P=-V(\varphi)\sqrt{1-\frac{1}{c^2}\partial_{\mu}\varphi\partial^{\mu}\varphi}.
\label{ak39}
\end{eqnarray}
Since $\epsilon=V(\varphi)/\sqrt{1-2X/c^2}$ and $P=-V(\varphi)\sqrt{1-2X/c^2}$,
we find
that $P=-V(\varphi)^2/\epsilon$, $w=-(1-2X/c^2)$ and
$c_s^2=(1-2X/c^2)c^2=-wc^2$.

When $V(\varphi)=V_0$ is constant, the Lagrangian (\ref{ak35}) reduces to the
Born-Infeld Lagrangian (\ref{biintro}) and we obtain 
$\epsilon=V_0/\sqrt{1-2X/c^2}$
and
$P=-V_0\sqrt{1-2X/c^2}$ leading to the Chaplygin equation of state
$P=-V^2_0/\epsilon$ (inversely, the Chaplygin equation of state
$P=-Ac^2/\epsilon$ leads to the Born-Infeld Lagrangian (\ref{biintro})
corresponding to Eq.  (\ref{ak35}) with $V(\varphi)=V_0$
constant). Therefore, the Chaplygin gas can be considered as the
simplest   tachyon model where the tachyon field is associated
with a purely kinetic Lagrangian. The relation between the
tachyonic Lagrangian with a constant
potential (reducing to the Born-Infeld Lagrangian) and the  Chaplygin gas
\cite{kmp} was first made by \cite{frolov}. The fact that the Chaplygin gas
is associated with the Born-Infeld Lagrangian was understood by
\cite{hoppe,jp,jackiw,btv}. The relation between the Chaplygin gas, the
Born-Infeld Lagrangian, k-essence Lagrangians and tachyon fields were further
discussed in \cite{btvB1,btvB2,gkmp,gorini5,gkmsf}.

{\it Remark:} The Lagrangian of  a particle of mass $m$ and position $q(t)$ in
special relativity is $L=-mc^2\sqrt{1-\dot q^2/c^2}$.  Its impulse is $p=m\dot
q/\sqrt{1-\dot q^2/c^2}$ and its energy is $E=mc^2/\sqrt{1-\dot q^2/c^2}$, i.e.,
 $E^2=p^2c^2+m^2c^4$. Its equation of motion is given by the
Euler-Lagrange equation $(d/dt)(\partial L/\partial\dot q)-\partial
L/\partial q=0$ yielding $(d/dt)(m\dot q/\sqrt{1-{\dot
q}^2/c^2})+m'(q)c^2\sqrt{1-{\dot q}^2/c^2}=0$. The Lagrangian equations of a
tachyonic SF are similar to
the Lagrangian equations of a relativistic particle in which the SF
$\varphi(x^\mu)$ plays the role of $q(t)$ and the SF potential $V(\varphi)$ the
role of the mass $m$ that would depend on $q$ in the 
general case.

\subsection{Nonrelativistic limit}
\label{sec_kessnr}

The action of a nonrelativistic real SF
is
\begin{equation}
S=\int \mathcal{L}\, d^3x dt,
\label{no1}
\end{equation}
where $\mathcal{L}=\mathcal{L}(\varphi,\dot\varphi,\nabla\varphi)$
is the Lagrangian density. We consider a pure k-essence Lagrangian of the
form 
\begin{eqnarray}
{\cal
L}={\cal L}(x),
\label{no3}
\end{eqnarray}
where 
\begin{eqnarray}
x=\dot\varphi+\frac{1}{2}(\nabla\varphi)^2.
\label{no2}
\end{eqnarray}
More general k-essence Lagrangians  
\begin{eqnarray}
{\cal
L}={\cal L}(x,\varphi)
\label{no4}
\end{eqnarray}
can depend both on $x$ and $\varphi$.
The least action principle $\delta S=0$, which 
is equivalent to the Euler-Lagrange equation
\begin{eqnarray}
\label{no5}
\frac{\partial}{\partial t}\left (\frac{\partial
{\cal L}}{\partial\dot\varphi}\right)+\nabla\cdot \left (\frac{\partial
{\cal L}}{\partial\nabla\varphi}\right)-\frac{\partial {\cal L}}{\partial \varphi}=0,
\end{eqnarray}
yields the equation of motion
\begin{eqnarray}
\label{no6}
\frac{\partial}{\partial t}\left ( \frac{\partial {\cal L}}{\partial x}\right )+\nabla\cdot \left (  \frac{\partial {\cal L}}{\partial x}\nabla\varphi\right )-\frac{\partial {\cal L}}{\partial\varphi}=0.
\end{eqnarray}
For the Lagrangian (\ref{no3}), it reduces to
\begin{eqnarray}
\label{no7}
\frac{\partial}{\partial t}\left \lbrack{\cal L}'(x)\right\rbrack+\nabla\cdot \left \lbrack {\cal L}'(x)\nabla\varphi\right\rbrack=0.
\end{eqnarray}

For the Lagrangian (\ref{no3}) the current is given
by 
\begin{eqnarray}
\label{j1c}
J^{\mu}=-\frac{\partial {\cal L}}{\partial
(\partial_\mu\varphi)}.
\end{eqnarray}
It determines the mass density  
\begin{eqnarray}
\label{j2b}
\rho\equiv -\frac{\partial\cal L}{\partial \dot\varphi}=-{\cal L}'(x)
\end{eqnarray}
and the mass flux 
\begin{eqnarray}
\label{j3b}
{\bf J}\equiv -\frac{\partial\cal L}{\partial (\nabla\varphi)}=-{\cal
L}'(x)\nabla\varphi.
\end{eqnarray}
Equation (\ref{no7}) can be written as
\begin{eqnarray}
\label{j4}
\frac{\partial\rho}{\partial t}+\nabla\cdot {\bf J}=0.
\end{eqnarray}
It expresses the local conservation of mass. This conservation
law is associated with the invariance of the Lagrangian
density under the transformation $\varphi\rightarrow \varphi+{\rm cst}$ (Noether
theorem). Introducing the velocity 
\begin{eqnarray}
{\bf u}=\nabla\varphi,
\label{no13}
\end{eqnarray}
we get
\begin{eqnarray}
{\bf J}=\rho {\bf u},
\end{eqnarray}
and the continuity equation
\begin{eqnarray}
\label{no21}
\frac{\partial\rho}{\partial t}+\nabla\cdot \left (\rho{\bf u}\right )=0.
\end{eqnarray}

The energy-momentum tensor is given by
\begin{eqnarray}
\label{no8}
T_{\mu}^{\nu}=\partial_\mu\varphi \frac{\partial {\cal L}}{\partial (\partial_\nu\varphi)}-{\cal L}\delta_{\mu}^{\nu}.
\end{eqnarray}
The local conservation of energy and impulse can be
written as
\begin{eqnarray}
\frac{\partial T_{00}}{\partial t}-\partial_i T_{0i}=0,
\end{eqnarray}
\begin{eqnarray}
-\frac{\partial T_{i0}}{\partial t}+\partial_j
T_{ij}=0.
\end{eqnarray}
For the
Lagrangian (\ref{no4}) we obtain the energy density
\begin{eqnarray}
\label{no9}
T_{00}\equiv {\dot\varphi}\frac{\partial {\cal
L}}{\partial\dot\varphi}-{\cal L}=\frac{\partial {\cal L}}{\partial
x}\dot\varphi-{\cal L}(x),
\end{eqnarray}
the momentum density
\begin{eqnarray}
\label{no10}
-T_{i0}\equiv -\partial_i\varphi\frac{\partial {\cal
L}}{\partial\dot\varphi}=-\frac{\partial {\cal L}}{\partial
x}\partial_i\varphi,
\end{eqnarray}
the energy flux
\begin{eqnarray}
\label{no11}
-T_{0i}\equiv \dot\varphi\frac{\partial {\cal
L}}{\partial(\partial_i\varphi)}=\frac{\partial {\cal L}}{\partial
x}\dot\varphi\partial_i\varphi,
\end{eqnarray}
and the momentum fluxes (stress tensor)
\begin{equation}
\label{no12}
T_{ij}\equiv -\partial_i\varphi\frac{\partial {\cal
L}}{\partial(\partial_j\varphi)}+{\cal L}\delta_{ij}=-\frac{\partial {\cal
L}}{\partial x} \partial_i\varphi
\partial_j\varphi+{\cal L}\delta_{ij}.
\end{equation}
Introducing the velocity from Eq. (\ref{no13}), we get
\begin{eqnarray}
T_{ij}=-\frac{\partial {\cal L}}{\partial x} u_i u_j+{\cal L}\delta_{ij}.
\label{no14}
\end{eqnarray}
The  energy-momentum tensor $T_{ij}$ can be written under the perfect fluid form
\begin{eqnarray}
T_{ij}=\rho u_{i}u_{j}+P \delta_{ij}
\label{no15}
\end{eqnarray}
provided that we make the identifications
\begin{eqnarray}
P={\cal L}(x,\varphi)
\label{no16a}
\end{eqnarray}
and
\begin{eqnarray}
 \rho=-\frac{\partial {\cal L}}{\partial x}=-\frac{\partial P}{\partial x}.
\label{no16b}
\end{eqnarray}
The equation of state parameter and the squared speed of sound
are given by
\begin{eqnarray}
\label{gf1}
w=\frac{P}{\rho c^2}=-\frac{P}{\frac{\partial P}{\partial x}c^2},
\end{eqnarray}
\begin{eqnarray}
\label{gf1b}
c_s^2=\frac{\frac{\partial P}{\partial x}}{\frac{\partial\rho}{\partial
x}}=-\frac{\frac{\partial P}{\partial x}}{\frac{\partial^2 P}{\partial x^2}}.
\end{eqnarray}
Using Eqs. (\ref{no16a}) and (\ref{no16b}), we can rewrite Eq. (\ref{no6}) and
Eqs.
(\ref{no9})-(\ref{no11}) as
\begin{eqnarray}
\label{no20}
\frac{\partial\rho}{\partial t}+\nabla\cdot \left (\rho{\bf u}\right
)+\frac{\partial {\cal L}}{\partial\varphi}=0,
\end{eqnarray}
\begin{eqnarray}
\label{no17}
T_{00}=-\rho\dot\varphi-{\cal L}(x),
\end{eqnarray}
\begin{eqnarray}
\label{no18}
-T_{i0}=\rho u_i,
\end{eqnarray}
\begin{eqnarray}
\label{no19}
-T_{0i}=-\rho \dot\varphi u_i.
\end{eqnarray}
For the Lagrangian (\ref{no3}), Eq. (\ref{no20}) reduces to the continuity
equation (\ref{no21}). In that case, the momentum density is equal to the
mass flux: $-T_{i0}=J_i$. On the other hand,
using Eq. (\ref{no2}), the energy density can be
written as
\begin{eqnarray}
T_{00}=\frac{1}{2}\rho {\bf u}^2+V(\rho),
\end{eqnarray}
where we have defined the potential $V(\rho)$ by the Legendre
transform $V(\rho)=-\rho x-{\cal L}(x)$.
Using Eq. (\ref{no16b}), we get $V'(\rho)=-x$. Then, using Eq. (\ref{no16a}),
we obtain $P(\rho)=\rho V'(\rho)-V(\rho)$ returning Eqs. (\ref{mad11b}) and
(\ref{gc4z}). Therefore, $V(\rho)$ coincides with the potential introduced in
Sec.
\ref{sec_nra}. Similarly, the energy flux can be written as
\begin{eqnarray}
-T_{0i}=\rho\left\lbrack \frac{1}{2} {\bf u}^2+V'(\rho)\right\rbrack u_i.
\end{eqnarray}
These results are consistent with the results obtained in Appendix
\ref{sec_stress} when $\hbar=0$.

{\it Remark:} Eqs. (\ref{no16a}) and (\ref{no16b}) are the counterparts of Eqs.
(\ref{ak17}) and (\ref{ak16}) in the relativistic case. Indeed, using Eqs.
(\ref{nr5}) and $\epsilon\sim \rho c^2$ valid in the nonrelativistic limit, 
Eqs. (\ref{ak17}) and (\ref{ak16}) imply $P={\cal L}(x,\varphi)$ and 
\begin{eqnarray}
\label{no22}
\rho \sim \frac{\epsilon}{c^2}\sim -\frac{\partial
P}{\partial x}\sim  - \frac{\partial {\cal L}}{\partial x},
\end{eqnarray}
returning Eqs. (\ref{no16a}) and (\ref{no16b}). 

\subsection{Cosmological evolution}
\label{sec_cos}

We now consider a spatially homogeneous SF described by a k-essence Lagrangian
in an expanding universe. In that case\footnote{In this Appendix and in
Appendices \ref{sec_csfapp} and \ref{sec_tsf}, $t$ stands for $ct$.}
\begin{equation}
\label{cos1c}
X=\frac{1}{2}\dot\varphi^2.
\end{equation}
On the other hand, the
energy-momentum tensor is diagonal $T_\mu^\nu={\rm
diag}(\epsilon,-P,-P,-P)$.\footnote{In the homogeneous case,
using Eq. (\ref{ak14}) with $u^0=c$ and $u^i=0$ we get $T_0^{0}=\epsilon$
and $T_{i}^i=-P$.} The energy
density and the pressure of
the  SF are given by
\begin{equation}
\label{cos1e}
\epsilon=T_0^0=\frac{\partial {\cal L}}{\partial\dot\varphi}\dot\varphi-{\cal
L},\qquad P=-T_i^i={\cal L},
\end{equation}
returning Eqs. (\ref{ak16}) and  (\ref{ak17}).

For a Lagrangian density of the form ${\cal L}={\cal L}(X,\varphi)$,
the equation of motion (\ref{ak7}) of the SF becomes
\begin{equation}
\label{cos1}
\left (\frac{\partial{\cal L}}{\partial X}+2X\frac{\partial^2{\cal L}}{\partial
X^2}\right )\ddot\varphi+3H\frac{\partial{\cal L}}{\partial
X}\dot\varphi+2X\frac{\partial^2{\cal L}}{\partial X\partial
\varphi}-\frac{\partial{\cal L}}{\partial \varphi}=0.
\end{equation}
This equation is
equivalent to the energy conservation equation (\ref{econs}). Indeed, taking the
time derivative of $\epsilon$ from Eq. (\ref{ak16}) and substituting the
result into Eq. (\ref{econs}) we get 
\begin{eqnarray}
\label{cos1b}
\left (\frac{\partial{\cal L}}{\partial X}+2X\frac{\partial^2{\cal L}}{\partial
X^2}\right )\dot X&+&\left (2X\frac{\partial^2{\cal L}}{\partial X\partial
\varphi}-\frac{\partial{\cal L}}{\partial \varphi}\right
)\dot\varphi\nonumber\\
&+&6HX\frac{\partial{\cal L}}{\partial
X}=0.
\end{eqnarray}
Recalling Eq. (\ref{cos1c}), we obtain Eq. (\ref{cos1}). We can check that this
equation returns Eqs.
(\ref{ak30}) and (\ref{ak30lt}) for a canonical and a tachyonic SF respectively.

For a Lagrangian density of the form ${\cal L}=V(\varphi)F(X)$, Eq. (\ref{cos1})
reduces to
\begin{equation}
\label{cos1d}
(F_X+2XF_{XX})\ddot\varphi+3HF_X\dot\varphi+(2XF_X-F)\frac{V'}{V}=0.
\end{equation}
In the particular case ${\cal
L}=V(\varphi)X$, corresponding to a stiff equation of state [see the comment 
after
Eq. (\ref{ak17})], we get
\begin{equation}
\label{cos1db}
\ddot\varphi+3H\dot\varphi+X\frac{V'}{V}=0.
\end{equation}

For a Lagrangian density of the form ${\cal L}=F(X)-V(\varphi)$, Eq.
(\ref{cos1})
reduces to
\begin{equation}
\label{cos1dn}
(F_X+2XF_{XX})\ddot\varphi+3HF_X\dot\varphi+V'(\varphi)=0.
\end{equation}

For a pure k-essence Lagrangian ${\cal L}={\cal L}(X)$, Eq. (\ref{cos1})
reduces to
\begin{equation}
\label{cos2}
({\cal L}'+2X{\cal L}'')\ddot\varphi+3H {\cal L}'\dot\varphi=0.
\end{equation}
We also have [see Eq. (\ref{cos1b})]
\begin{equation}
\label{cos5}
({\cal L}'+2X{\cal L}'')\frac{dX}{da}+\frac{6}{a} X {\cal L}'=0.
\end{equation}
This equation integrates to give
\begin{equation}
\label{cos6}
\sqrt{X}{\cal L}'(X)=\frac{k}{a^3}.
\end{equation}
Using Eq. (\ref{charge16}), we see that this equation is equivalent to the
conservation of the rest-mass: $\rho_m\propto a^{-3}$. Eq. (\ref{cos6}) was
first obtained by Chimento \cite{chimento} and Scherrer \cite{scherrer} but they
did not realize the relation with the rest-mass density. Our approach provides
therefore a physical interpretation of their result.

\section{Equation of state of type I}
\label{sec_mtu}

In this Appendix, we consider a barotropic fluid described by an equation of
state of type I where the pressure $P=P(\epsilon)$ is specified as a function of
the energy density. We show that, in a cosmological context, it is possible to
associate to this fluid a real SF with a potential $V(\varphi)$ which is fully
determined by the equation of state. As an illustration, we determine the real
SF potential associated
with a polytropic equation of state of type I.

\subsection{Friedmann equations}
\label{sec_angr}

If we consider an expanding homogeneous background and adopt the
Friedmann-Lema\^itre-Robertson-Walker (FLRW) metric, the Einstein field
equations 
reduce to the Friedmann equations 
\begin{equation}
\label{ak22}
H^2=\frac{8\pi
G}{3c^2}\epsilon,
\end{equation}
\begin{equation}
\label{ak22b}
2\dot H+3H^2=-\frac{8\pi
G}{c^2} P,
\end{equation}
where $H=\dot a/a$ is  the Hubble
parameter and $a(t)$ is the scale factor. To obtain Eq. (\ref{ak22}), we have
assumed that the universe is  flat ($k=0$) in agreement with the inflation
paradigm \cite{guthinflation} and the observations of the cosmic microwave
background (CMB) \cite{planck2014,planck2016}. On the other hand, we have set
the cosmological constant   to zero ($\Lambda=0$) since dark energy can be taken
into account in the equation of state $P(\epsilon)$ or in the SF potential
$V(\varphi)$ (quintessence). Eq. (\ref{ak22b}) can also be written as
\begin{equation}
\label{ak22c}
\frac{\ddot a}{a}=-\frac{4\pi G}{3c^2}(3P+\epsilon),
\end{equation}
showing that the expansion of the universe is decelerating when $P>-\epsilon/3$
and accelerating when $P<-\epsilon/3$.

Using Eqs. (\ref{ak22}) and (\ref{ak22b}), we obtain the energy conservation
equation
\begin{equation}
\label{econs}
\frac{d\epsilon}{dt}+3H\left
(\epsilon+P\right )=0.
\end{equation}
This equation can be directly obtained from the conservation of the
energy-momentum tensor $D_\mu T^{\mu\nu}=0$ which results from the Bianchi
identities. The energy density
increases with the scale factor when $P>-\epsilon$ and decreases with the scale
factor when $P<-\epsilon$. The latter case corresponds to a phantom universe.

For a given equation of state $P(\epsilon)$ we can solve Eq. (\ref{econs}) to get 
\begin{equation}
\label{econsint}
\ln a=-\frac{1}{3}\int \frac{d\epsilon}{\epsilon+P(\epsilon)}.
\end{equation}
This equation determines $\epsilon(a)$. We can then solve the Friedmann equation
(\ref{ak22}) to obtain the temporal evolution 
of the scale factor $a(t)$.

A polytropic equation of state of type I is defined by
\begin{equation}
\label{mar1}
P=K\left (\frac{\epsilon}{c^2}\right )^{\gamma}\quad {\rm with}\quad \gamma=1+1/n.
\end{equation}
Assuming $1+(K/c^2)(\epsilon/c^2)^{1/n}\ge 0$, i.e., $P\ge -\epsilon$
corresponding to a nonphantom universe,\footnote{The
case of a phantom universe is treated in \cite{cosmopoly3}.}
the energy conservation equation (\ref{econs}) can be
integrated into \cite{cosmopoly1,cosmopoly2}
\begin{equation}
\label{mar2}
\epsilon=\frac{\rho_* c^2}{\left\lbrack (a/a_*)^{3/n}\mp 1\right\rbrack^n},
\end{equation}
where $\rho_*=(c^2/|K|)^n$ and $a_*$ is a constant of integration. The upper sign corresponds to $K>0$ and the lower sign to $K<0$.

The case $\gamma=1$, corresponding to a linear equation of state 
\begin{equation}
\label{mar3}
P=\alpha\epsilon
\end{equation}
with $\alpha=K/c^2$, must be treated specifically. In that case, the solution of Eq. (\ref{econs}) can be written as
\begin{equation}
\label{mar4}
\epsilon=\frac{\rho_* c^2}{\left ({a}/{a_*}\right )^{3(1+\alpha)}},
\end{equation}
where $\rho_*a_*^{3(1+\alpha)}$ is a constant of integration.

{\it Remark:} Unfortunately, for the logotropic equation of state of type I [see
Eq. (\ref{lu1})], the energy conservation equation 
\begin{equation}
\label{econsintlog}
\ln a=-\frac{1}{3}\int \frac{d\epsilon}{\epsilon+A\ln(\epsilon/\epsilon_*)}
\end{equation}
cannot be
integrated explicitly.

\subsection{Canonical SF}
\label{sec_csfapp}

We consider a spatially homogeneous canonical SF 
in an expanding universe with a Lagrangian 
\begin{equation}
\label{ak30lag}
L=\frac{1}{2}\dot\varphi^2-V(\varphi).
\end{equation}
It evolves according to the KG
equation [see Eq. (\ref{cos1})]
\begin{equation}
\label{ak30}
\ddot \varphi+3H\dot\varphi+\frac{dV}{d\varphi}=0
\end{equation}
coupled to the Friedmann equation (\ref{ak22}).
The  SF tends to
run down the potential towards lower energies while experiencing a Hubble
friction. The energy-momentum tensor is diagonal $T_\mu^\nu={\rm
diag}(\epsilon,-P,-P,-P)$. The energy
density and the pressure of
the  SF are given by [see Eq. (\ref{cos1e})]
\begin{equation}
\label{ak31}
\epsilon=\frac{1}{2}\dot\varphi^2+V(\varphi),
\end{equation}
\begin{equation}
\label{ak32}
P=\frac{1}{2}\dot\varphi^2-V(\varphi).
\end{equation}
We note that, here,  $V$ represents the total SF potential including the
rest-mass term. When the kinetic 
term dominates we obtain the stiff equation of state $P=\epsilon$. When the
potential term dominates, we obtain the equation of state
$P=-\epsilon$ corresponding to the vacuum energy. We can easily check that the
KG equation (\ref{ak30}) with Eqs.
(\ref{ak31}) and (\ref{ak32}) implies the energy conservation
equation (\ref{econs}) (see Appendix \ref{sec_ecesf}).
Inversely, the the energy conservation
equation (\ref{econs})  with Eqs.
(\ref{ak31}) and (\ref{ak32})  implies the KG equation (\ref{ak30}).
The equation of
state parameter $w=P/\epsilon$ is given by
\begin{equation}
\label{ak34}
w=\frac{\frac{1}{2}\dot\varphi^2-V(\varphi)}{\frac{1}{2}\dot\varphi^2+V(\varphi)
}.
\end{equation}
It satisfies $-1\le w\le 1$. The speed
of sound is equal to the speed of light ($c_s=c$).

Using standard techniques \cite{em,paddytachyon,cst,bamba}, we can obtain the SF
potential as follows
\cite{cosmopoly2}. From
Eqs. (\ref{ak31}) and (\ref{ak32}), we get
\begin{eqnarray}
\label{mtu8}
\dot\varphi^2=(w+1)\epsilon.
\end{eqnarray}
Then, using
$\dot\varphi=(d\varphi/da)Ha$ and the
Friedmann equation (\ref{ak22}), we find that the relation between the
 SF and the
scale factor is given by 
\begin{eqnarray}
\label{mtu9}
\frac{d\varphi}{da}=\left (\frac{3c^4}{8\pi G}\right
)^{1/2}\frac{\sqrt{1+w}}{a}.
\end{eqnarray}
We note that $\varphi$ is a monotonic function of $a$. We have selected the
solution where $\varphi$ increases with $a$. On the other hand, according to
Eqs. (\ref{ak31}) and (\ref{ak32}), we have
\begin{eqnarray}
\label{mtu10}
V=\frac{1}{2}(1-w)\epsilon.
\end{eqnarray}
Therefore, the
potential $V(\varphi)$ of
the
canonical SF is determined in parametric form by the equations
\begin{equation}
\label{mtu11}
\varphi(a)=\left (\frac{3c^4}{8\pi G}\right )^{1/2}\int
\sqrt{1+w(a)}\, \frac{da}{a},
\end{equation}
\begin{equation}
\label{mtu12}
V(a)=\frac{1}{2}\left\lbrack 1-w(a)\right\rbrack
\epsilon(a).
\end{equation}
We note that $\varphi$ is defined up to an additive constant.

The canonical SF potential corresponding to a polytropic equation of state
of type I [see  Eq. (\ref{ak22})]  
has been determined in Sec. 8.1. of \cite{cosmopoly2}.  It is given by
\begin{eqnarray}
V=\frac{1}{2}\rho_* c^2 \frac{\cosh^2\psi+1}{\cosh^{\frac{2\gamma}{\gamma-1}}\psi}\qquad (K<0),
\label{sfr2}
\end{eqnarray}
\begin{eqnarray}
V=\frac{1}{2}\rho_* c^2 \frac{\sinh^2\psi-1}{\sinh^{\frac{2\gamma}{\gamma-1}}\psi}\qquad (K>0),
\label{sfr1}
\end{eqnarray}
where
\begin{eqnarray}
\psi=\left (\frac{8\pi G}{3c^4}\right )^{1/2}\frac{3}{2}(\gamma-1)\varphi.
\label{sfr4}
\end{eqnarray}
The relation between the scale factor and the SF is
\begin{eqnarray}
\frac{a}{a_*}=\sinh^{\frac{2}{3(\gamma-1)}}\psi \qquad (K<0),
\label{sfr2b}
\end{eqnarray}
\begin{eqnarray}
\frac{a}{a_*}=\cosh^{\frac{2}{3(\gamma-1)}}\psi \qquad (K>0),
\label{sfr1b}
\end{eqnarray}
These expressions are valid for $\psi\ge 0$.

(i) For $\gamma=-1$ (Chaplygin gas), we get
\begin{eqnarray}
V=\frac{1}{2}\rho_* c^2 \left (\cosh\psi+\frac{1}{\cosh\psi}\right )\qquad (K<0),
\label{sfr6}
\end{eqnarray}
\begin{eqnarray}
V=\frac{1}{2}\rho_* c^2 \left (\sinh\psi-\frac{1}{\sinh\psi}\right )\qquad (K>0),
\label{sfr5}
\end{eqnarray}
with $\rho_*=\sqrt{|K|/c^2}$. This SF potential was first
obtained in \cite{kmp}.

(ii) For $\gamma=2$ (BEC), we get
\begin{eqnarray}
V=\frac{1}{2}\rho_* c^2 \frac{\cosh^2\psi+1}{\cosh^{4}\psi}\qquad (K<0),
\label{sfr8}
\end{eqnarray}
\begin{eqnarray}
V=\frac{1}{2}\rho_* c^2 \frac{\sinh^2\psi-1}{\sinh^{4}\psi}\qquad (K>0),
\label{sfr7}
\end{eqnarray}
with $\rho_*=c^2/|K|$. This SF potential was first
obtained in \cite{cosmopoly2}.

(iii) For $\gamma=0$ ($\Lambda$CDM model), we get
\begin{eqnarray}
V=\frac{1}{2}\rho_* c^2 (\cosh^2\psi+1)\qquad (K<0),
\label{sfr8b}
\end{eqnarray}
\begin{eqnarray}
V=\frac{1}{2}\rho_* c^2 (\sinh^2\psi-1)\qquad (K>0),
\label{sfr7b}
\end{eqnarray}
with $\rho_*=|K|/c^2$.  This SF potential was first obtained in \cite{gkmp} and
rediscovered independently in \cite{cosmopoly2}.

(iv) For $\gamma=3$ (superfluid), we get
\begin{eqnarray}
V=\frac{1}{2}\rho_* c^2 \frac{\cosh^2\psi+1}{\cosh^{3}\psi}\qquad (K<0),
\label{sfr8c}
\end{eqnarray}
\begin{eqnarray}
V=\frac{1}{2}\rho_* c^2 \frac{\sinh^2\psi-1}{\sinh^{3}\psi}\qquad (K>0),
\label{sfr7c}
\end{eqnarray}
with $\rho_*=\sqrt{c^2/|K|}$.

(v) For $\gamma=1$, we get
\begin{eqnarray}
V(\varphi)=\frac{1}{2}\rho_* c^2 (1-\alpha)e^{-3\sqrt{\alpha+1}\left (\frac{8\pi
G}{3c^4}\right )^{1/2}\varphi}.
\label{sfr9}
\end{eqnarray}
This exponential potential was obtained in \cite{paddytachyon} but
it appeared in earlier works on inflation and quintessence
\cite{ratra,wett1,fj2,halliwell,lm}. In
that case, the relation between the scale factor and the SF is
\begin{eqnarray}
\varphi(a)=\left (\frac{3c^4}{8\pi G}\right )^{1/2}\sqrt{1+\alpha}\,
\ln\left (\frac{a}{a_*}\right ).
\label{sfr9b}
\end{eqnarray}
For $\alpha=1$ (stiff
matter), we find that $V(\varphi)=0$. On the other hand, for $\alpha=-1$ (vacuum
energy), we find that $\dot\varphi=0$ so that $\varphi=\varphi_0$ is constant.
This is
possible according to the equation of motion (\ref{ak30}) provided
that $V'(\varphi_0)=0$. Therefore, $\varphi_0$ must be at the minimum of the
potential $V(\varphi)$. In that case, $\epsilon=V(\varphi_0)=V_0$ and
$P=-V(\varphi_0)=-V_0$, yielding $P=-\epsilon$. Note that $V(\varphi)$ is not
necessarily constant but it must
have a minimum $V_0>0$.

\subsection{Tachyonic SF}
\label{sec_tsf}

We consider a spatially homogeneous tachyonic SF
\cite{gibbons,frolov,felder,paddytachyon,pc,bagla,gkmp,feinstein} in an
expanding
universe with a Lagrangian 
\begin{equation}
\label{ak30lt}
L=-V(\varphi)\sqrt{1-\dot\varphi^2}.
\end{equation}
It evolves according to the
equation [see Eq. (\ref{cos1})]
\begin{equation}
\label{ak40}
\frac{\ddot
\varphi}{1-\dot\varphi^2}+3H\dot\varphi+\frac{1}{V}\frac{dV}{d\varphi}=0
\end{equation}
coupled to the Friedmann equation (\ref{ak22}). The  SF tends to
run down the potential towards lower energies while experiencing a Hubble
friction. The energy-momentum tensor is diagonal $T_\mu^\nu={\rm
diag}(\epsilon,-P,-P,-P)$.  The density and the pressure of
the  SF are given by [see Eq. (\ref{cos1e})]
\begin{eqnarray}
\epsilon=\frac{V(\varphi)}{\sqrt{1-\dot\varphi^2}},
\label{ak41}
\end{eqnarray}
\begin{eqnarray}
P=-V(\varphi)\sqrt{1-\dot\varphi^2}.
\label{ak42}
\end{eqnarray}
We can easily check that the equation of motion (\ref{ak40}) with Eqs.
(\ref{ak41}) and (\ref{ak42}) implies the energy conservation
equation (\ref{econs}) (see Appendix
\ref{sec_ecesf}). Inversely, the energy conservation equation
(\ref{econs}) with Eqs.
(\ref{ak41}) and (\ref{ak42}) implies the 
equation of
motion (\ref{ak40}).
The equation of state parameter $w=P/\epsilon$ is given by
\begin{eqnarray}
w=\dot\varphi^2-1.
\label{ak44}
\end{eqnarray}
It satisfies $-1\le w\le 0$. The squared speed
of sound is given by $c_s^2/c^2=1-{\dot \varphi}^2=-w$. It satisfies $0\le w\le
1$.

Using standard techniques \cite{em,paddytachyon,cst,bamba}, we can obtain the SF
potential as follows
\cite{cosmopoly2}. From
Eqs. (\ref{ak41}) and (\ref{ak42}), we obtain
\begin{equation}
\label{tachyon3}
\dot\varphi^2=1+w.
\end{equation}
Using $\dot\varphi=(d\varphi/da) H a$, and the
Friedmann equation (\ref{ak22}), we get 
\begin{equation}
\label{tachyon4}
\frac{d\varphi}{da}=\left (\frac{3c^4}{8\pi G}\right
)^{1/2}\frac{\sqrt{1+w}}{\sqrt{\epsilon} a}.
\end{equation}
We note that $\varphi$ is a monotonic function of $a$. We have selected the
solution where $\varphi$ increases with $a$. On the other hand, from Eqs.
(\ref{ak41}) and (\ref{ak42}), we have
\begin{equation}
\label{tachyon8}
V^2=-w\epsilon^2.
\end{equation}
Therefore, the
potential $V(\varphi)$ of
the tachyonic SF is determined in parametric form by the equations
\begin{equation}
\label{tachyon4b}
\varphi(a)=\left (\frac{3c^4}{8\pi G}\right
)^{1/2}\int \frac{\sqrt{1+w(a)}}{\sqrt{\epsilon(a)}}\, \frac{da}{a},
\end{equation}
\begin{equation}
\label{tachyon8b}
V(a)=\sqrt{-w(a)}\epsilon(a).
\end{equation}

The tachyonic SF potential corresponding to a polytropic 
equation of state of type I [see  Eq. (\ref{ak22})]  has been determined in
Sec. 8.2. of \cite{cosmopoly2}.
It is defined only for $K<0$.  It is given in parametric
form by 
\begin{equation}
\label{tachyon8bp}
V=\frac{\rho_*c^2}{(x^2+1)^{\frac{1+\gamma}{2(\gamma-1)}}},
\end{equation}
\begin{eqnarray}
\label{psix}
\psi=\int (x^2+1)^{\frac{2-\gamma}{2(\gamma-1)}}\, dx,
\end{eqnarray}
where we have introduced the variable
\begin{eqnarray}
\psi=\sqrt{\rho_* c^2}\left (\frac{8\pi G}{3c^4}\right
)^{1/2}\frac{3}{2}(\gamma-1)\varphi.
\label{sfr4bb}
\end{eqnarray}
The relation between the scale factor and the SF is given by Eq. (\ref{psix})
with
\begin{eqnarray}
x=\left (\frac{a}{a_*}\right )^{\frac{3}{2}(\gamma-1)}.
\end{eqnarray}
The integral in Eq. (\ref{psix}) can be expressed in terms of hypergeometric
functions. Simple analytical expressions can be
obtained in special
cases.

(i) For $\gamma=-1$ and $K<0$ (Chaplygin gas), we find that
$V(\varphi)=\rho_* c^2$ with $\rho_*=\sqrt{|K|/c^2}$. In
that
case, the potential  
is constant \cite{gkmp}. This leads to the Born-Infeld Lagrangian (see
Appendix
\ref{sec_tach}).

(ii) For $\gamma=2$ and $K<0$ (BEC), we get 
\begin{equation}
\label{tachyon12}
V(\psi)= \frac{\rho_* c^2}{(\psi^2+1)^{3/2}}
\end{equation}
with $\rho_*=c^2/|K|$. We have $a/a_*=\psi^{2/3}$ with $\psi\ge 0$. This
potential was first obtained in \cite{cosmopoly2}.

(iii) For $\gamma=0$ and $K<0$ ($\Lambda$CDM model), we get
\begin{equation}
\label{tachyon17}
V(\psi)=\frac{\rho_{*} c^2}{\cos\psi}
\end{equation}
with $\rho_*=|K|/c^2$. We have $a/a_*=1/\tan(\psi)^{2/3}$ with $0\le
\psi\le\pi/2$.
This potential was first obtained in \cite{gkmp} and
rediscovered independently in \cite{cosmopoly2}.

(iv) For $\gamma=3$ and $K<0$ (superfluid), it is not possible to obtain
explicit expressions.

(v) For $\gamma=1$ and $-1<\alpha<0$, we get
\begin{equation}
\label{tachyon27}
V(\varphi)=\frac{\sqrt{-\alpha}}{1+\alpha}\frac{c^4}{6\pi G}\frac{1}{\varphi^2}.
\end{equation}
This inverse square law potential was first obtained in
\cite{paddytachyon,feinstein}. In
that case, the relation
between the scale factor and the SF is 
\begin{equation}
\label{tachyon28}
\varphi=\frac{2}{3}\frac{1}{\sqrt{\rho_* c^2}} \left (\frac{3c^4}{8\pi G}\right
)^{1/2}\frac{1}{\sqrt{1+\alpha}}\left (\frac{a}{a_*}\right )^{3(1+\alpha)/2}.
\end{equation}
For $\alpha=-1$ (vacuum
energy), we find that $\dot\varphi=0$ so that $\varphi=\varphi_0$ is constant.
This is
possible according to the equation of motion (\ref{ak40}) provided
that $V'(\varphi_0)=0$. Therefore, $\varphi_0$ must be at the minimum of the
potential $V(\varphi)$. In that case, $\epsilon=V(\varphi_0)=V_0$ and
$P=-V(\varphi_0)=-V_0$, yielding $P=-\epsilon$. Note that $V(\varphi)$ is not
necessarily constant but it must
have a minimum $V_0>0$.

{\it Remark:} For $\gamma=1/2$ and $K<0$, we get
\begin{equation}
\label{tachyon24}
V(\psi)=\frac{\rho_* c^2}{(1-\psi^2)^{3/2}}
\end{equation}
with $\rho_*=(|K|/c^2)^2$. We have  $(a/a_*)^{3/4}=\sqrt{1-\psi^2}/\psi$ with
$0\le \psi\le 1$. This
potential was first obtained in \cite{cosmopoly2}.

\section{Equation of state of type II}
\label{sec_mtd}

In this Appendix, we consider a barotropic fluid 
described by an equation of state of type II where the pressure $P=P(\rho_m)$ is
specified as a function of the rest-mass density. After recalling general
results, we apply this equation of state to a cosmological context. 

\subsection{General results}

The first principle of
thermodynamics for a relativistic gas can be
written as
\begin{equation}
\label{mtd1}
d\left (\frac{\epsilon}{\rho_m}\right )=-Pd\left
(\frac{1}{\rho_m}\right )+Td\left (\frac{s}{\rho_m}\right ),
\end{equation}
where
\begin{eqnarray}
\label{mtd2}
\epsilon=\rho_m c^2+u(\rho_m)
\end{eqnarray}
is the energy density including the rest-mass energy density $\rho_m c^2$ (where
$\rho_m=n m$ 
is the rest-mass density) and the internal energy density $u(\rho_m)$, $s$ is
the density of entropy, $P$ is the pressure, and $T$ is the temperature. We
assume that
$Td(s/\rho_m)=0$. This corresponds
to cold ($T=0$) or isentropic ($s/\rho_m={\rm cst}$) gases. In that
case, Eq.
(\ref{mtd1}) reduces to
\begin{equation}
\label{mtd3}
d\left (\frac{\epsilon}{\rho_m}\right )=-Pd\left
(\frac{1}{\rho_m}\right )=\frac{P}{\rho_m^2}\, d\rho_m.
\end{equation}
This equation can be rewritten as
\begin{equation}
\label{mtd4}
\frac{d\epsilon}{d\rho_m}=\frac{P+\epsilon}{\rho_m}
\end{equation}
where the term in the right hand side is the  enthalpy $h$. We have
\begin{equation}
\label{mtd4a}
h=\frac{P+\epsilon}{\rho_m},
\qquad h=\frac{d\epsilon}{d\rho_m},\qquad dh=\frac{dP}{\rho_m}. 
\end{equation}
Eq. (\ref{mtd3}) can be
integrated
into
\begin{eqnarray}
\label{mtd5}
\epsilon=\rho_m c^2+\rho_m \int \frac{P(\rho_m)}{\rho_m^2}\, d\rho_m
\end{eqnarray}
establishing that
\begin{eqnarray}
\label{mtd6}
u(\rho_m)=\rho_m \int \frac{P(\rho_m)}{\rho_m^2}\, d\rho_m.
\end{eqnarray}
This equation determines the internal energy as a function of the equation of
state $P(\rho_m)$. Inversely, the equation of state is determined by the
internal energy $u(\rho_m)$ from the relation
\begin{equation}
\label{mtd7}
P(\rho_m)=-\frac{d(u/\rho_m)}{d(1/\rho_m)}=\rho_m^2\left\lbrack
\frac{u(\rho_m)}{\rho_m}\right\rbrack'=\rho_m u'(\rho_m)-u(\rho_m).
\end{equation}
We note that
\begin{eqnarray}
\label{mtd8}
P'(\rho_m)=\rho_m u''(\rho_m).
\end{eqnarray}
The squared speed of sound is
\begin{eqnarray}
\label{mtd8x}
c_s^2=P'(\epsilon)c^2=\frac{\rho_m
\epsilon''(\rho_m)}{\epsilon'(\rho_m)}c^2=\frac{\rho_m u''(\rho_m)c^2}{
c^2+u'(\rho_m)}.
\end{eqnarray}

{\it Remark:} The first principle of thermodynamics can be
written as \cite{gr1}
\begin{equation}
\label{gr1}
d\epsilon=Tds+\mu dn.
\end{equation}
Combined with the Gibbs-Duhem relation \cite{gr1}
\begin{equation}
\label{gr2}
s=\frac{\epsilon+P-\mu n}{T},
\end{equation}
we obtain Eq. (\ref{mtd1}) and
\begin{equation}
\label{gr2b}
sdT-dP+nd\mu=0.
\end{equation}
If $T={\rm cst}$, then $dP=nd\mu$. For $T=0$, the foregoing equations reduce to
\begin{equation}
\label{gr3}
d\epsilon=\mu dn,\qquad \mu=\frac{\epsilon+P}{n},\qquad dP=nd\mu,
\end{equation}
which are equivalent to Eq. (\ref{mtd4a}) with $\mu=mh$.

\subsection{Cosmology}

Let us apply these equations in a cosmological context, namely for a homogeneous
fluid in an expanding background. 
Combining the energy conservation equation (\ref{econs}) 
with Eq. (\ref{mtd4}), we obtain
\begin{eqnarray}
\frac{d\rho_m}{dt}+3H\rho_m=0.
\label{mtd10}
\end{eqnarray}
This equation expresses the conservation of the particle number (or
rest-mass). It can be integrated into $\rho_m\propto a^{-3}$.
Inserting this relation into Eq. (\ref{mtd2}), we see that $\rho_m$ represents
DM while $u$ represents DE. Therefore, DM represents the rest-mass energy
density and DE represents the internal energy density. This decomposition
provides therefore a simple (and nice) interpretation of DM and DE in terms of a
single DF \cite{epjp,lettre}. Owing to this interpretation, we can write
\begin{eqnarray}
\rho_m c^2=\frac{\Omega_{\rm m,0}\epsilon_0}{a^3}
\label{mtd11}
\end{eqnarray}
and
\begin{eqnarray}
\label{mtd2b}
\epsilon=\frac{\Omega_{\rm m,0}\epsilon_0}{a^3}+u\left (\frac{\Omega_{\rm m,0}\epsilon_0}{c^2a^3} \right ),
\end{eqnarray}
where $\epsilon_0$ is the present energy density of the universe and
$\Omega_{\rm m,0}$ is the present proportion of DM.
For given $P(\rho_m)$ or
$u(\rho_m)$ we can get  $\epsilon(a)$ from Eq.  (\ref{mtd2b}). We can then solve
the Friedmann equation (\ref{ak22}) to obtain the temporal evolution of the
scale factor $a(t)$.

{\it Remark:} Eqs. (\ref{mtd2}) and (\ref{mtd7}) determine
the equation of state $P=P(\epsilon)$. As a result, we can obtain Eq.
(\ref{mtd11}) directly from Eqs.
(\ref{mtd2}), (\ref{mtd7}) and the energy conservation equation
(\ref{econs}) .
Indeed, combining these equations we obtain Eq. (\ref{mtd10}) which  integrates
to give Eq. (\ref{mtd11}).

\subsection{Two-fluid model}
\label{sec_twofluids}

In the model of type II, we have a single
dark fluid
with an equation of state $P=P(\rho_m)$. Still, the
energy density (\ref{mtd2}) is the sum of two terms, a rest-mass density term
$\rho_m$ which mimics DM and an internal energy term $u(\rho_m)$ which mimics
DE. It is interesting to consider a two-fluid model which leads to the
same results as the single dark fluid model, at least for what concerns the
evolution of the homogeneous background. In this two-fluid model, one fluid
corresponds to pressureless DM with an equation of state $P_{\rm m}=0$ and a
density $\rho_m c^2=\Omega_{\rm m,0}\epsilon_0/a^3$ determined by the energy
conservation equation for DM, and the other fluid corresponds to DE with an
equation of state $P_{\rm de}(\epsilon_{\rm de})$ and an energy density
$\epsilon_{\rm de}(a)$ determined by the energy
conservation equation for DE. We can obtain the equation of state of DE yielding
the same results as the one-fluid model by taking
\begin{eqnarray}
P_{\rm de}=P(\rho_m),\qquad \epsilon_{\rm de}=u(\rho_m).
\end{eqnarray}
In other words, the  equation of state $P_{\rm de}(\epsilon_{\rm de})$ of DE
 in the two-fluid model corresponds to the
relation $P(u)$ in the single fluid model. Explicit examples of the
correspondance between the one and two-fluid models are given below. We note
that
although
the one and two-fluid models are equivalent for the evolution  of the
homogeneous background, they may differ for what concerns the formation of the
large-scale structures of the Universe and for inhomogeneous systems in
general.

In the two-fluid model associated with the
Chaplygin gas of type I (or III), the DE has an equation of state 
\begin{eqnarray}
P_{\rm de}=\frac{2Kc^2\epsilon_{\rm de}}{\epsilon_{\rm
de}^2-Kc^2},
\label{twg1}
\end{eqnarray}
which is obtained by eliminating $\rho_m$ between Eqs. (\ref{cgpu1ca}) and
(\ref{cgpu1cb}), and by identifying $P(u)$ with $P_{\rm de}(\epsilon_{\rm
de})$.

In the two-fluid model associated with the BEC of type I, the DE
has an equation of state 
\begin{eqnarray}
P_{\rm de}=\frac{4K\epsilon_{\rm de}^2}{\left
\lbrack-\frac{K\epsilon_{\rm de}}{c^2}\pm\sqrt{\left
(\frac{K\epsilon_{\rm de}}{c^2}\right )^2+4K\epsilon_{\rm de}}\right \rbrack^2},
\label{twg2}
\end{eqnarray}
which is obtained by eliminating $\rho_m$ between Eqs. (\ref{becpu1ca}) and
(\ref{becpu1cb}), and by identifying $P(u)$ with $P_{\rm de}(\epsilon_{\rm
de})$.

In the two-fluid model associated with the $\Lambda$CDM model, the DE
has an equation of state 
\begin{eqnarray}
P_{\rm de}=-\epsilon_{\rm de}.
\label{twg3}
\end{eqnarray}

In the two-fluid model associated with a polytrope of type II, the DE
has an equation of state 
\begin{eqnarray}
P_{\rm de}=(\gamma-1)\epsilon_{\rm de},
\label{twg4}
\end{eqnarray}
which is obtained by eliminating $\rho_m$ between Eqs. (\ref{pd1}) and
(\ref{wpd3}), and by identifying $P(u)$ with $P_{\rm de}(\epsilon_{\rm
de})$.

In the two-fluid model associated with a logotrope of type II, the DE
has an equation of state \cite{jcap}
\begin{eqnarray}
P_{\rm de}=-\epsilon_{\rm de}-A,
\label{twg5}
\end{eqnarray}
which is obtained by eliminating $\rho_m$ between Eqs. (\ref{ld1}) and
(\ref{ld2}), and by identifying $P(u)$ with $P_{\rm de}(\epsilon_{\rm
de})$.

\section{Equation of state of type III}
\label{sec_mtt}

In this Appendix, we consider a barotropic 
fluid described by an equation of state of type III where the pressure
$P=P(\rho)$ is specified as a function of the pseudo rest-mass density.  As
explained in Sec. \ref{sec_rtf} this hydrodynamic description arises naturally
when
considering a complex SF with a potential $V(|\varphi |^2)$ in the TF
approximation. Here, we consider the case of a spatially
homogeneous complex SF
in an expanding background.

\subsection{General results}

Let us first establish general results that are valid beyond the TF
approximation.

We consider a spatially homogeneous complex SF in an expanding
universe with a Lagrangian
\begin{equation}
\label{ak30comp}
L=\frac{1}{2c^2}|\dot\varphi|^2-V_{\rm tot}(\varphi).
\end{equation}
Its cosmological evolution is governed by the
KGF equations
\begin{eqnarray}
\frac{1}{c^2}\frac{d^2\varphi}{dt^2}+\frac{3H}{c^2}\frac{d\varphi}{dt}
+2\frac{dV_{\rm tot}}{d|\varphi|^2}\varphi=0,
\label{hj1}
\end{eqnarray}
\begin{equation}
\label{hj2}
H^2=\frac{8\pi
G}{3c^2}\epsilon.
\end{equation}
The energy density
$\epsilon(t)$ and the pressure $P(t)$ of the SF are given by
\begin{equation}
\epsilon\equiv T_0^0=\frac{\partial
L}{\partial\dot\varphi}\dot\varphi+\frac{\partial
L}{\partial\dot\varphi^*}\dot\varphi^*-L,\qquad P\equiv -T_i^i={L},
\end{equation}
yielding
\begin{equation}
\epsilon=\frac{1}{2c^2}\left |\frac{d\varphi}{d
t}\right|^2+V_{\rm tot}(|\varphi|^2),
\label{hj3}
\end{equation}
\begin{equation}
P=\frac{1}{2c^2}\left |\frac{d\varphi}{d
t}\right|^2-V_{\rm tot}(|\varphi|^2).
\label{hj4}
\end{equation}
When the kinetic 
term dominates we obtain the stiff equation of state $P=\epsilon$. When the
potential term dominates, we obtain the equation of state
$P=-\epsilon$ corresponding to the vacuum energy.

In the following, we use the hydrodynamic representation of the SF (see
Sec. \ref{sec_rtf} and \cite{abrilas}). The  Lagrangian is 
\begin{equation}
\label{hak30comp}
L=\frac{1}{2c^2}\rho\dot\theta^2+\frac{\hbar^2}{8m^2\rho
c^2}\dot\rho^2-V_{\rm tot}(\rho).
\end{equation}
The energy density
$\epsilon(t)$ and the pressure $P(t)$ of the SF are given by 
\begin{equation}
\epsilon\equiv T_0^0=\frac{\partial
L}{\partial\dot\theta}\dot\theta+\frac{\partial
L}{\partial\dot\rho}\dot\rho-L,\qquad P\equiv -T_i^i={L},
\end{equation}
yielding
\begin{equation}
\epsilon=\frac{1}{2c^2}\rho\dot\theta^2+\frac{\hbar^2}{8m^2\rho
c^2}\dot\rho^2+V_{\rm tot}(\rho),
\label{hhj3}
\end{equation}
\begin{equation}
P=\frac{1}{2c^2}\rho\dot\theta^2+\frac{\hbar^2}{8m^2\rho
c^2}\dot\rho^2-V_{\rm tot}(\rho).
\label{hhj4}
\end{equation}

The equation $D_\nu T^{\mu\nu}=0$ leads to the energy conservation equation 
\begin{eqnarray}
\frac{d\epsilon}{dt}+3H(\epsilon+P)=0.
\label{hj5}
\end{eqnarray}
This equation can also be obtained from the KG equation (\ref{hj1}) with Eqs.
(\ref{hj5}) and (\ref{hj4}) (see Appendix
\ref{sec_ecesf}). Inversely, the energy conservation equation (\ref{hj5}) with
Eqs.
(\ref{hj5}) and (\ref{hj4}) 
implies the KG equation (\ref{hj1}).

The equation $D_\mu J^\mu=0$, which is equivalent to the continuity
equation (\ref{db8}), can be written as
\begin{eqnarray}
\label{hj6}
\frac{d}{dt}\left (E_{\rm tot} \rho a^3\right )=0,
\end{eqnarray}
where
\begin{eqnarray}
\label{hj7}
E_{\rm tot}=\hbar\omega=-m\dot\theta=-\dot S_{\rm
tot}
\end{eqnarray}
is the energy of the SF ($\omega=-\dot\Theta$ with $\Theta=m\theta/\hbar$ is its
pulsation). Eq. (\ref{hj6}) expresses the
conservation of the charge of the complex  SF (or equivalently the
conservation
of the boson number). It can be written as
\begin{eqnarray}
\label{hj8}
\rho E_{\rm tot}=\frac{Q m^2c^2}{a^3},
\end{eqnarray}
where $Q=Ne$ is a constant of integration representing the charge
 of the SF which is proportional to the boson number $N$
\cite{arbeycosmo,gh,shapiro,abrilph,abrilas,kasuya,boyle}.
Indeed, according to Eq. (\ref{charge3}), the charge of the SF is defined
by\footnote{We have taken $e=1$
so that the charge of the SF coincides with the boson number.}
\begin{eqnarray}
\label{hj9}
Q=\frac{1}{mc}\int J^0\sqrt{-g}\, d^3x,
\end{eqnarray}
where $J^0$ is the time component of the quadricurrent
$J^\mu=-\rho\partial^{\mu}\theta$ (see Sec. \ref{sec_db}). For a spatially
homogeneous SF in an expanding background, we have
\begin{eqnarray}
\label{hj10}
J^0=-\frac{1}{c}\rho\dot\theta=\frac{\hbar}{mc}\rho\omega=\frac{1}{mc}
\rho E_{\rm tot}
\end{eqnarray}
and $Q=J^0a^3/mc=\rho E_{\rm tot} a^3/m^2c^2$ yielding Eq. (\ref{hj8}).

The quantum Hamilton-Jacobi (or Bernoulli) equation (\ref{db9}) takes
the form
\begin{equation}
\label{hj11}
E_{\rm tot}^2=\hbar^2\frac{1}{\sqrt{\rho}}
\frac {
d^2\sqrt{\rho}}{dt^2}+3H\hbar^2\frac{1}{\sqrt{\rho}}\frac{
d\sqrt{\rho}}{dt}+2m^2c^2V'_{ \rm tot}(\rho).
\end{equation}

Finally, we have established in the general case (see Sec. \ref{sec_db})
that the
rest-mass
density is given by
\begin{eqnarray}
\rho_m=\frac{\rho}{c}  \sqrt{\partial_\mu\theta
\partial^\mu\theta}.
\label{hj12}
\end{eqnarray}
For a spatially homogeneous SF in an expanding background, we get
\begin{eqnarray}
\rho_m=-\frac{\rho}{c}
\partial_0\theta=-\frac{1}{c^2}\rho\dot\theta=\frac{\hbar}{mc^2}\rho\omega=\frac
{ 1 }
{mc^2}\rho E_ { \rm tot}.
\label{hj13}
\end{eqnarray}
Using Eq. (\ref{hj13}), Eqs. (\ref{hj6}) and (\ref{hj8}) can be rewritten
as
\begin{eqnarray}
\label{hj14}
\frac{d\rho_m}{dt}+3H\rho_m=0,
\end{eqnarray}
and
\begin{eqnarray}
\label{hj15}
\rho_m=\frac{Q m}{a^3}.
\end{eqnarray}
Equations (\ref{hj14}) and (\ref{hj15}) can also be obtained from the first law
of thermodynamics for a
cold fluid ($T=0$) in a homogeneous background (see Appendix \ref{sec_mtd}).
They
express the
conservation of
the particle number. Inversely,
Eq. (\ref{hj13}) can be directly obtained from Eq.  (\ref{hj8}) using
Eq. (\ref{hj15}). Comparing Eqs. (\ref{hj10})
and (\ref{hj13}), we note that
\begin{eqnarray}
\rho_m=\frac{J^0}{c}.
\label{hj16}
\end{eqnarray}
This relation is not generally valid (see Sec.
\ref{sec_db}). In the present case, it arises from the general identity 
$J^\mu=\rho_m u^\mu$  and the 
fact that $u^\mu=c\delta^{\mu}_0$ since the fluid
(SF) is static in the expanding background.

\subsection{TF approximation}

In the TF approximation ($\hbar\rightarrow 0$), the  Lagrangian
(\ref{hak30comp}) reduces to 
\begin{equation}
\label{hak30compft}
L=\frac{1}{2c^2}\rho\dot\theta^2-V_{\rm tot}(\rho),
\end{equation}
the energy density
$\epsilon(t)$ and the pressure $P(t)$ of the SF reduce to 
\begin{equation}
\epsilon=\frac{1}{2c^2}\rho\dot\theta^2+V_{\rm
tot}(\rho),
\label{hhj3tf}
\end{equation}
\begin{equation}
P=\frac{1}{2c^2}\rho\dot\theta^2-V_{\rm tot}(\rho),
\label{hhj4tf}
\end{equation}
and the quantum Hamilton-Jacobi (or
Bernoulli) equation (\ref{hj11}) reduces to\footnote{For a spatially
homogeneous SF, it is shown in Ref.
\cite{abrilas} that the TF approximation is equivalent to the fast oscillation
approximation $\omega\gg H$.}
\begin{eqnarray}
\label{hj17}
E_{\rm tot}^2=2m^2c^2V'_{\rm tot}(\rho).
\end{eqnarray}
Combining Eqs. (\ref{hj8}) and (\ref{hj17}), we obtain
\begin{eqnarray}
\label{hj18}
\frac{Q mc}{a^3}=\rho\sqrt{2 V'_{\rm tot}(\rho)}.
\end{eqnarray}
This equation determines the relation between the pseudo 
rest-mass density $\rho$ and the scale factor $a$. On the other hand, according
to Eqs.
(\ref{hj13}) and
(\ref{hj17}), the rest-mass densitity is given by
\begin{eqnarray}
\label{hj19}
\rho_m=\frac{\rho}{c}\sqrt{2 V'_{\rm tot}(\rho)}.
\end{eqnarray}
According to Eqs. (\ref{hj7}),
(\ref{hhj3tf}) and (\ref{hhj4tf}) the 
energy density and the pressure of the SF in the TF approximation are given by
\begin{eqnarray}
\epsilon=\rho V'_{\rm tot}(\rho)+V_{\rm
tot}(\rho),
\label{hj20}
\end{eqnarray}
\begin{eqnarray}
P=\rho V'_{\rm tot}(\rho)-V_{\rm
tot}(\rho).
\label{hj21}
\end{eqnarray}
Inversely, the SF potential is determined by the equation of state $P(\rho)$
according to
\begin{eqnarray}
V_{\rm tot}(\rho)=\rho\int\frac{P(\rho)}{\rho^2}\, d\rho.
\label{hj21b}
\end{eqnarray}
Equations (\ref{hj19})-(\ref{hj21b}) are always true  in the TF
approximation even for inhomogeneous systems (see Sec. \ref{sec_rtf}).For given
$P(\rho)$ or $V_{\rm tot}(\rho)$, we can obtain $\rho(a)$ from Eq.
(\ref{hj18}) and $\epsilon(a)$ from Eq. (\ref{hj20}). We can then solve the
Friedmann Eq. (\ref{hj2}) to obtain the temporal evolution of the scale factor
$a(t)$. Actually, since it is not always possible to invert Eq. (\ref{hj18}), we
can
proceed differently (see \cite{abrilas}). Taking the logarithmic derivative of
Eq. (\ref{hj18}),  we get
\begin{eqnarray}
\frac{\dot a}{a}=-\frac{1}{3}\frac{\dot \rho}{\rho}\left\lbrack 1+\frac{\rho V''_{\rm tot}(\rho)}{2V'_{\rm tot}(\rho)}\right\rbrack.
\label{hj22}
\end{eqnarray}
Then, using Eqs. (\ref{hj2}) and (\ref{hj20}), we obtain
\begin{eqnarray}
\frac{c^2}{24\pi G} \left (\frac{\dot \rho}{\rho}\right )^2=\frac{\rho V'_{\rm tot}(\rho)+V_{\rm
tot}(\rho)}{\left\lbrack 1+\frac{\rho V''_{\rm tot}(\rho)}{2V'_{\rm tot}(\rho)}\right\rbrack^2}.
\label{hj23}
\end{eqnarray}
For given $V_{\rm tot}(\rho)$ this is just a first order differential equation that can be solved by integration.

{\it Remark:} Eqs. (\ref{hj20}) and (\ref{hj21}) determine
the equation of state $P=P(\epsilon)$. As a result, we can obtain Eq.
(\ref{hj18}) directly from Eqs.
(\ref{hj20}), (\ref{hj21}) and the energy conservation equation (\ref{hj5}).
Indeed, combining these equations we obtain
\begin{eqnarray}
\left\lbrack 2V'_{\rm tot}(\rho)+\rho V''_{\rm tot}(\rho)\right\rbrack
\frac{d\rho}{dt}=-6H\rho V'_{\rm tot}(\rho).
\label{hj24}
\end{eqnarray}
leading to
\begin{eqnarray}
\int \frac{2V'_{\rm tot}(\rho)+\rho V''_{\rm
tot}(\rho)}{\rho
V'_{\rm tot}(\rho)}=-6\ln a.
\label{hj25}
\end{eqnarray}
Eq. (\ref{hj25}) integrates to give Eq. (\ref{hj18}).

\section{Analogies and differences between $u$ and $V$}
\label{sec_sum}

For a relativistic fluid of type II, we have established the identities (see
Appendix \ref{sec_mtd})
\begin{eqnarray}
\label{sum3}
\epsilon=\rho_m c^2+u(\rho_m),
\end{eqnarray}
\begin{eqnarray}
P=\rho_m u'(\rho_m)-u(\rho_m),
\label{sum4}
\end{eqnarray}
\begin{eqnarray}
\label{sum4b}
u(\rho_m)=\rho_m \int \frac{P(\rho_m)}{\rho_m^2}\, d\rho_m,
\end{eqnarray}
where $\rho_m$ is the rest-mass density and $u$ is the internal energy.

For a relativistic fluids of type III, we have established the identities (see
Sec. \ref{sec_rtf})
\begin{eqnarray}
\epsilon=\rho c^2+\rho V'(\rho)+V(\rho),
\label{sum5}
\end{eqnarray}
\begin{eqnarray}
P=\rho V'(\rho)-V(\rho),
\label{sum6}
\end{eqnarray}
\begin{eqnarray}
V(\rho)=\rho\int\frac{P(\rho)}{\rho^2}\, d\rho,
\label{sum6b}
\end{eqnarray}
where $\rho$ is the pseudo rest-mass density and $V$ is the potential of the complex SF.

We note that Eqs. (\ref{sum6}) and  (\ref{sum6b}) are identical to Eqs.
(\ref{sum4}) and (\ref{sum4b}) 
with $\rho$ instead of $\rho_m$ and $V$ instead of $u$. In general, the
variables $\rho$ and $V$ are different from the variables $\rho_m$ and
$u$. However, they coincide in the
nonrelativistic limit. For a nonrelativistic complex SF (BEC), Eqs.
(\ref{sum3})-(\ref{sum6b}) reduce to
\begin{eqnarray}
\epsilon\sim \rho c^2,
\label{sum1}
\end{eqnarray}
\begin{eqnarray}
P=\rho V'(\rho)-V(\rho),
\label{sum2}
\end{eqnarray}
\begin{eqnarray}
V(\rho)=\rho\int\frac{P(\rho)}{\rho^2}\, d\rho,
\label{sum2b}
\end{eqnarray}
where $\rho=\rho_m$ is the mass density and $V=u$ is the potential of the SF or
the internal energy of the corresponding barotropic fluid 
(see Appendix \ref{sec_gicg}).

\section{Energy conservation equation for a SF}
\label{sec_ecesf}

\subsection{Complex SF}

We consider a spatially homogeneous complex SF in an expanding background (see
Appendix \ref{sec_mtt}).  Taking the time derivative of the energy density given
by Eq. (\ref{hj3}), we get
\begin{equation}
\label{ecesf1}
\frac{d\epsilon}{dt}=\frac{1}{2c^2}\frac{d^2\varphi}{dt^2}\frac{d\varphi^*}{dt}
+V'_{\rm tot}(|\varphi|^2)\frac{d\varphi}{dt}\varphi^*+{\rm c.c.}
\end{equation}
Using the KG equation (\ref{hj1}), we obtain after simplification
\begin{equation}
\label{ecesf2}
\frac{d\epsilon}{dt}=-\frac{3H}{c^2}\left |\frac{d\varphi}{dt}\right |^2.
\end{equation}
From Eqs. (\ref{hj3}) and (\ref{hj4}) we have
\begin{equation}
\label{ecesf3}
\epsilon+P=\frac{1}{c^2}\left |\frac{d\varphi}{dt}\right |^2.
\end{equation}
Combining Eqs. (\ref{ecesf2}) and (\ref{ecesf3}), we obtain the energy
conservation equation (\ref{hj5}). Inversely,
from Eqs. (\ref{hj3}), (\ref{hj4})  and (\ref{hj5}), we can
directly derive the KG equation (\ref{hj1}).

\subsection{Real canonical SF}

We consider a spatially homogeneous real canonical SF in an expanding background
(see
Appendix \ref{sec_csfapp}).  Taking the time derivative of the energy
density given
by Eq. (\ref{ak31}), we get
\begin{equation}
\label{cecesf1}
\frac{d\epsilon}{dt}=\frac{d^2\varphi}{dt^2}\frac{d\varphi}{dt}
+V'(\varphi)\frac{d\varphi}{dt}.
\end{equation}
Using the KG equation (\ref{ak30}), we obtain after simplification
\begin{equation}
\label{cecesf2}
\frac{d\epsilon}{dt}=-3H{\dot\varphi}^2.
\end{equation}
From Eqs. (\ref{ak31}) and (\ref{ak32}) we have
\begin{equation}
\label{cecesf3}
\epsilon+P={\dot\varphi}^2.
\end{equation}
Combining Eqs. (\ref{cecesf2}) and (\ref{cecesf3}), we obtain the energy
conservation equation (\ref{econs}). Inversely,
from Eqs. (\ref{econs}), (\ref{ak31}) and (\ref{ak32}) 
we can directly derive the KG equation (\ref{ak30}).

\subsection{Real tachyonic SF}

We consider a spatially homogeneous real tachyonic SF in an expanding background
(see Appendix \ref{sec_tsf}).  Taking the time derivative of the
energy density given
by Eq. (\ref{ak41}), we get
\begin{equation}
\label{tecesf1}
\frac{d\epsilon}{dt}=\frac{V'(\varphi)}{\sqrt{1-{\dot\varphi}^2}}{\dot\varphi}
+\frac{V(\varphi)}{(1-{\dot\varphi}^{2})^{3/2}}{\dot\varphi}{\ddot\varphi}.
\end{equation}
Using the field equation (\ref{ak40}), we obtain after simplification
\begin{equation}
\label{tecesf2}
\frac{d\epsilon}{dt}=-3H\frac{V(\varphi)}{\sqrt{1-{\dot\varphi}^2}}{\dot
\varphi}^2.
\end{equation}
From Eqs. (\ref{ak41}) and (\ref{ak42}) we have
\begin{equation}
\label{tecesf3}
\epsilon+P=\frac{V(\varphi)}{\sqrt{1-{\dot\varphi}^2}}{\dot\varphi}^2.
\end{equation}
Combining Eqs. (\ref{tecesf2}) and (\ref{tecesf3}), we obtain the energy
conservation equation (\ref{econs}). Inversely, from Eqs. (\ref{econs}),
(\ref{ak41}) and (\ref{ak42}), we can directly derive the field
equation (\ref{ak40}).

\section{Some studies devoted to polytropic and logotropic equations of state of type I, II and III}
\label{sec_stu}

In this Appendix, we briefly mention studies devoted to polytropic and
logotropic
equations of state of type I, II and III in the context of stars, DM halos and
cosmology.

The study of nonrelativistic stars described by a polytropic 
equation of state dates back to the paper of Lane \cite{lane}. Isothermal stars
were first considered by Z\"ollner \cite{zollner}. A very complete study
of polytropic and isothermal stars is presented in the books of Emden
\cite{emden} and Chandrasekhar
\cite{chandrabook}. Nonrelativistic logotropic stars were studied by McLaughlin
and Pudritz \cite{pud}. The logotropic equation of state was applied to DM
halos by Chavanis \cite{logo,epjp,graal}.

General relativistic stars described by a polytropic equation of state of type I
were first considered by Tooper \cite{tooper1}. Polytropes of type I with index
$\gamma=2$ were specifically studied by Chavanis and Harko
\cite{chavharko,partially} in relation
to general relativistic BEC stars (however, this is not the correct equation of
state for these systems -- see below). General relativistic stars described by a
linear equation of state, extending the models of Newtonian isothermal stars,
were studied by Chandrasekhar \cite{chandra72} (see also
\cite{yabushita1,yabushita2,aarelat1,aarelat2} and references therein).
Cosmological models based on a
polytropic equation of state of type I with an arbitrary index $\gamma$ were
studied in
\cite{cosmopoly1,cosmopoly2,cosmopoly3}. The specific index $\gamma=-1$
corresponds to the Chaplygin gas \cite{kmp,gkmp} and the indices $-1\le
\gamma\le 0$ correspond to the GCG \cite{bentoGCG}.

General relativistic stars described by a polytropic equation of state of type
II were first considered by Tooper \cite{tooper2}. Polytropes of type II with
index $\gamma=2$ were specifically studied by Chavanis \cite{partially}
and Latifah {\it et al.} \cite{lsm} in relation
to general relativistic BEC stars (however, this is not the correct equation of
state for these systems -- see below). Cosmological models based on a polytropic
equation of state of type II with an arbitrary index $\gamma$ were studied
in \cite{stiff} (the index $\gamma=2$ of a BEC is specifically treated in the
main text of \cite{stiff} and the case of a  general index is treated in 
Appendix D of \cite{stiff}).  A cosmological model based on the logotropic
equation of state of type II was studied in \cite{epjp}.

General relativistic stars described by a polytropic equation of state of type
III were studied  by Colpi {\it et al.} \cite{colpi} and Chavanis and Harko
\cite{chavharko,partially} for the particular index
$\gamma=2$ corresponding to BECs. This is the hydrodynamic representation, valid
in the TF regime, of  a complex SF with a repulsive $|\varphi|^4$
self-interaction described by the KGE equations \cite{colpi}. Therefore, a
polytropic equation of state of type III with index $\gamma=2$, leading to
the equation of state (\ref{becpt3b}), is the correct equation of state of a
relativistic BEC with a quartic self-interaction in the TF regime. Cosmological
models based on
a polytropic equation of state of type III  with an arbitrary index $\gamma$
were studied in \cite{abrilas,csf} (the index $\gamma=2$ of a BEC is
specifically treated in the main text of  \cite{abrilas,csf} and the case
of an arbitrary index is treated in Appendix I of \cite{abrilas} and in
\cite{csf}). This is the hydrodynamic representation, valid in the TF regime or
in the fast oscillation regime, of  a complex SF with a $|\varphi|^4$ potential
described by the KGE equations \cite{shapiro}. A cosmological model based on the
logotropic equation of state of type III has been studied recently in
\cite{graal}.

\section{Conservation laws for a nonrelativistic SF}
\label{sec_stress}

In this Appendix, we establish the local conservation laws of mass, impulse and
energy for a nonrelativistic SF (see Sec. \ref{sec_nra}).

\subsection{Conservation laws in terms of hydrodynamic variables}
\label{sec_stressh}

The equation of continuity (\ref{mad6}) can be written as
\begin{eqnarray}
\label{stress2}
\frac{\partial\rho}{\partial t}+\nabla\cdot {\bf J}=0,
\end{eqnarray}
where $\rho$ is the mass density and 
\begin{eqnarray}
\label{stress1}
{\bf J}=\rho{\bf u}
\end{eqnarray}
is the density current. This equation expresses the local conservation of mass $M=\int \rho\, d{\bf r}$.

Using the continuity equation (\ref{mad6}), the quantum Euler equation 
(\ref{mad10}) can be rewritten as 
\begin{eqnarray}
\label{stress8}
\frac{\partial}{\partial t}(\rho {\bf u})+\nabla(\rho {\bf u}\otimes {\bf
u})+\nabla P+\frac{\rho}{m}\nabla Q={\bf 0}.
\end{eqnarray}
On the other hand, the quantum force can be written under the form (see Sec. 2.5
of \cite{chavtotal})
\begin{eqnarray}
\label{stress9}
-\frac{\rho}{m}\partial_i Q=-\partial_j P_{ij}^Q,
\end{eqnarray}
where  the anisotropic quantum pressure tensor $P_{ij}^Q$ is given by 
\begin{equation}
\label{stress10a}
P_{ij}^{Q}=
-\frac{\hbar^2}{4m^2}\rho\partial_i\partial_j\ln\rho=\frac{\hbar^2}{4m^2}
\left (\frac{1}{\rho}
\partial_i\rho\partial_j\rho-\partial_i\partial_j\rho\right )
\end{equation}
or, alternatively, by
\begin{equation}
\label{stress10b}
P_{ij}^{Q}=\frac{\hbar^2}{4m^2}\left (\frac{1}{\rho}\partial_i\rho\partial_j\rho-\delta_{ij}\Delta\rho\right ).
\end{equation}
Substituting Eq. (\ref{stress9}) into Eq. (\ref{stress8}),  we obtain
\begin{eqnarray}
\label{stress8b}
\frac{\partial}{\partial t}(\rho {\bf u})+\nabla(\rho {\bf u}\otimes {\bf u})
+\nabla P+\partial_j P^Q_{ij}={\bf 0}.
\end{eqnarray}
Introducing the momentum density
\begin{eqnarray}
\label{stress11}
-T_{i0}=\rho {\bf u},
\end{eqnarray}
we can rewrite Eq. (\ref{stress8b}) as
\begin{eqnarray}
\label{stress12}
-\frac{\partial T_{i0}}{\partial t}+\partial_j
T_{ij}=0,
\end{eqnarray}
where 
\begin{eqnarray}
\label{stress13}
T_{ij}=\rho u_i u_j+P\delta_{ij}+P_{ij}^Q
\end{eqnarray}
is the stress tensor. Using Eqs. (\ref{stress10a}) and (\ref{stress10b}), we get
\begin{eqnarray}
\label{stress14}
T_{ij}&=&\rho u_i u_j+P\delta_{ij}
-\frac{\hbar^2}{4m^2}\rho\partial_i\partial_j\ln\rho\nonumber\\
&=&\rho u_i u_j+P\delta_{ij}+
\frac{\hbar^2}{4m^2}
\left (\frac{1}{\rho}
\partial_i\rho\partial_j\rho-\partial_i\partial_j\rho\right )\nonumber\\
\end{eqnarray}
or, alternatively,
\begin{equation}
\label{stress15}
T_{ij}=\rho u_i u_j+\left \lbrack P(\rho)-\frac{\hbar^2}{4m^2}\Delta\rho\right
\rbrack\delta_{ij}
+\frac{\hbar^2}{4m^2}\frac{1}{\rho}\partial_i\rho\partial_j\rho.
\end{equation}
Eq. (\ref{stress12}) expresses the local conservation of the momentum ${\bf P}=\int \rho {\bf u}\, d{\bf r}$.

Introducing the energy density
\begin{eqnarray}
\label{stress3}
T_{00}=\rho e=\rho \frac{{\bf u}^2}{2}+\rho\frac{Q}{m}+V(\rho)
\end{eqnarray}
and combining the equation of continuity (\ref{mad6}) and the quantum Euler
equation (\ref{mad10}), we obtain (see Appendix E of
\cite{chavtotal}) 
\begin{eqnarray}
\label{stress4}
\frac{\partial}{\partial t}(\rho e)+\nabla\cdot (\rho e {\bf u})+\nabla\cdot (P{\bf u})+\nabla\cdot {\bf J}_Q=0,
\end{eqnarray}
where
\begin{eqnarray}
\label{stress5}
{\bf J}_Q=\frac{\hbar^2}{4m^2}\rho\frac{\partial \nabla\ln\rho}{\partial t}
\end{eqnarray}
is the quantum current. Introducing the energy current
\begin{eqnarray}
\label{stress6}
-T_{0i}&=&\rho e {\bf u}+P {\bf u}+{\bf J}_Q\nonumber\\
&=&\rho \left \lbrack \frac{{\bf
u}^2}{2}+\frac{Q}{m}+\frac{V(\rho)+P}{\rho}\right \rbrack {\bf
u}+{\bf J}_Q\nonumber\\
&=&\rho \left \lbrack \frac{{\bf u}^2}{2}+\frac{Q}{m}+V'(\rho)\right
\rbrack {\bf u}+{\bf J}_Q,
\end{eqnarray}
where we have used Eq. (\ref{mad11b}) to obtain the last equality, we can
rewrite Eq. (\ref{stress4}) as
\begin{eqnarray}
\label{stress7}
\frac{\partial T_{00}}{\partial t}-\partial_i T_{0i}=0.
\end{eqnarray}
This equation expresses the local conservation of energy $E=\int \rho e\, d{\bf
r}$. We also recall that $h(\rho)=V'(\rho)$ is the enthalpy.

For classical systems ($\hbar=0$), or for BECs in
the TF limit, the 
foregoing equations reduce to 
\begin{eqnarray}
\label{stress16}
T_{00}=\rho e=\rho \frac{{\bf u}^2}{2}+V(\rho),
\end{eqnarray}
\begin{eqnarray}
\label{stress17}
-T_{0i}=\rho \left \lbrack \frac{{\bf u}^2}{2}+V'(\rho)\right \rbrack {\bf u},
\end{eqnarray}
\begin{eqnarray}
\label{stress18}
-T_{i0}=\rho {\bf u},
\end{eqnarray}
\begin{eqnarray}
\label{stress19}
T_{ij}=\rho u_i u_j+P\delta_{ij}.
\end{eqnarray}

{\it Remark:} We note that $T_{0i}\neq T_{i0}$ because the theory is not Lorentz
invariant. By contrast, $T_{ij}=T_{ji}$ because the theory is invariant against
spatial rotations. We also note that the momentum density
is equal to the mass flux: 
\begin{eqnarray}
-T_{i0}=\rho {\bf u}={\bf J}.
\end{eqnarray}

\subsection{Conservation laws in terms of the wave function}
\label{sec_pisc}

Using Eqs.
(\ref{mad1})-(\ref{mad5}), the density current (\ref{stress1}) can be expressed in terms of the
wave function as
\begin{eqnarray}
\label{stress20}
{\bf J}=\frac{\hbar}{2im}\left (\psi^*\nabla\psi-\psi\nabla\psi^*\right ).
\end{eqnarray}
As a result, the equation of continuity (\ref{stress2}) takes the form
\begin{eqnarray}
\label{stress21}
\frac{\partial |\psi|^2}{\partial t}+\frac{\hbar}{2im}\nabla\cdot \left
(\psi^*\nabla\psi-\psi\nabla\psi^*\right )=0.
\end{eqnarray}

Similarly, the momentum density (\ref{stress11}) and the stress tensor
(\ref{stress15}) can be written in terms of the 
wave function  as  (see Appendix A of \cite{chavtotal})
\begin{eqnarray}
\label{stress23}
-T_{i0}={\bf J}=\frac{\hbar}{2im}\left (\psi^*\nabla\psi-\psi\nabla\psi^*\right
)
\end{eqnarray}
and
\begin{equation}
\label{stress29}
T_{ij}=\frac{\hbar^2}{m^2}{\rm Re}\left (\frac{\partial\psi}{\partial
x_i}\frac{\partial\psi^*}{\partial x_j}\right )
+\left \lbrack P(|\psi|^2)-\frac{\hbar^2}{4m^2}\Delta|\psi|^2\right
\rbrack\delta_{ij}
\end{equation}
or, alternatively,
\begin{eqnarray}
\label{stress30}
T_{ij}=\frac{\hbar^2}{2m^2}{\rm Re}\left (\frac{\partial\psi}{\partial
x_i}\frac{\partial\psi^*}{\partial
x_j}-\psi\frac{\partial^2\psi^*}{\partial x_i\partial x_j}\right )
+P\delta_{ij}.
\end{eqnarray}

Finally, the energy density (\ref{stress3}) and the energy current
(\ref{stress6}) can be written in terms of the 
wave function  as
\begin{eqnarray}
\label{stress22}
T_{00}=\frac{\hbar^2}{2m^2}|\nabla\psi|^2+V(|\psi|^2)
\end{eqnarray}
and
\begin{eqnarray}
\label{stress24}
-T_{0i}=\left \lbrack
\frac{\hbar^2}{2m^2}\frac{|\nabla\psi|^2}{|\psi|^2}+V'(|\psi|^2)\right
\rbrack {\bf J}+{\bf J}_Q
\end{eqnarray}
with
\begin{eqnarray}
\label{stress25}
{\bf J}_Q=\frac{\hbar^2}{4m^2}|\psi|^2\frac{\partial \nabla\ln|\psi|^2}{\partial t}.
\end{eqnarray}

\subsection{Conservation laws from the Lagrangian expressed in terms of the wave
function}
\label{sec_cpsi}

The current of a complex SF is given by 
\begin{eqnarray}
\label{j1d}
J^{\mu}=\frac{m}{i\hbar}\left \lbrack\psi\frac{\partial {\cal L}}{\partial
(\partial_\mu\psi)}-\psi^* \frac{\partial {\cal L}}{\partial
(\partial_\mu\psi^*)}\right\rbrack.
\end{eqnarray}
For the Lagrangian (\ref{lh1}) we obtain the mass density
\begin{eqnarray}
\label{j2}
J_{0}=|\psi|^2=\rho
\end{eqnarray}
and the mass flux 
\begin{eqnarray}
\label{j3}
{\bf J}=-\frac{i\hbar}{2m}(\psi^*\nabla\psi-\psi\nabla\psi^*).
\end{eqnarray}

The energy-momentum tensor of a complex SF is given by
\begin{eqnarray}
\label{em1b7}
T_{\mu}^{\nu}=\partial_\mu\psi\frac{\partial {\cal L}}{\partial
(\partial_\nu\psi)}+\partial_\mu\psi^* \frac{\partial {\cal L}}{\partial
(\partial_\nu\psi^*)}-{\cal L}\delta_{\mu}^{\nu}.
\end{eqnarray}
For the Lagrangian (\ref{lh1}) we obtain the energy density
\begin{eqnarray}
\label{em1b8}
T_{00}=\frac{\hbar^2}{2m^2}|\nabla\psi|^2+V(|\psi|^2),
\end{eqnarray}
the momentum density
\begin{eqnarray}
\label{em1b9}
-T_{i0}=-\frac{i\hbar}{2m}(\psi^*\nabla\psi-\psi\nabla\psi^*)=
{ \bf J},
\end{eqnarray}
the energy flux
\begin{eqnarray}
\label{em1b10}
-T_{0i}=-\frac{\hbar^2}{2m^2}(\dot\psi\nabla\psi^*+\dot\psi^*\nabla\psi),
\end{eqnarray}
and the momentum fluxes (stress tensor)
\begin{eqnarray}
\label{em1b11}
T_{ij}=\frac{\hbar^2}{m^2}{\rm Re} (\partial_i\psi \partial_j\psi^*)+{\cal
L}\delta_{ij}.
\end{eqnarray}
These are their general expressions.  If we use the GP
equation
(\ref{mfgp9new}), which is obtained after extremizing the action, we can rewrite
Eq.
(\ref{em1b10}) as Eq. (\ref{stress24}). Similarly,
if we use the expression (\ref{lh12b}) of the Lagrangian which also relies on
the GP equation, we can rewrite Eq. (\ref{em1b11}) as
Eq. (\ref{stress29}).
In this manner, we recover the
equations of Appendix \ref{sec_pisc} (up to terms that vanish by
integration).

{\it Remark:} The energy density is
\begin{eqnarray}
\label{em1b7b}
T_{00}&=&\dot\psi \frac{\partial{\cal L}}{\partial\dot\psi}+\dot\psi^*
\frac{\partial{\cal L}}{\partial\dot\psi^*}-{\cal L}\nonumber\\
&=&\pi\dot\psi + \pi^*\dot\psi^*-{\cal L}\nonumber\\
&=&\frac{i\hbar}{2m}\left (\psi^*\dot\psi-\psi\dot\psi^*\right )-{\cal L},
\end{eqnarray}
where $\pi=\partial {\cal L}/{\partial\dot\psi}=\frac{i\hbar}{2m}\psi^*$ is
the  conjugate momentum to $\psi$. This leads to Eqs. (\ref{lh3}) and
(\ref{lh4}). On the other hand, Eq. (\ref{lh4g}) can be
rewritten in the form of Hamilton equations
\begin{eqnarray}
\label{lh4ga}
\frac{\partial \psi}{\partial t}=\frac{1}{2}\frac{\delta H}{\delta \pi},\qquad
\frac{\partial \pi}{\partial t}=-\frac{1}{2}\frac{\delta H}{\delta \psi}.
\end{eqnarray}
They are equivalent to the GP equation (\ref{mfgp9new}) and its complex
conjugate.

\subsection{Conservation laws from the Lagrangian expressed in terms of
hydrodynamic variables}
\label{sec_chy}

The
current  of a complex SF in its hydrodynamic representation is given by
\begin{eqnarray}
\label{j1cn}
J^{\mu}=-m\frac{\partial {\cal L}}{\partial
(\partial_\mu S)}.
\end{eqnarray}
For the Lagrangian (\ref{lh8}) we obtain the mass density 
\begin{eqnarray}
\label{j2bn}
J_0=\rho
\end{eqnarray}
and the mass flux 
\begin{eqnarray}
\label{j3bn}
{\bf J}=\rho\frac{\nabla S}{m}=\rho {\bf u}.
\end{eqnarray}

The energy-momentum tensor of a complex SF in its hydrodynamic representation is
given by 
\begin{eqnarray}
\label{em1b13}
T_{\mu}^{\nu}=\partial_\mu\rho\frac{\partial {\cal L}}{\partial (\partial_\nu\rho)}+\partial_\mu S \frac{\partial {\cal L}}{\partial (\partial_\nu S)}-{\cal L}\delta_{\mu}^{\nu}.
\end{eqnarray}
For the Lagrangian (\ref{lh8}) we obtain the energy density
\begin{eqnarray}
\label{em1b14}
T_{00}&=&\frac{\rho}{2m^2}(\nabla
S)^2+\frac{\hbar^2}{8m^2}\frac{(\nabla\rho)^2}{\rho}+V(\rho)\nonumber\\
&=&\frac{1}{2}\rho {\bf u}^2+\rho \frac{Q}{m}+V(\rho),
\end{eqnarray}
the
momentum density
\begin{eqnarray}
\label{em1b16}
-T_{i0}=\frac{\rho}{m}\nabla S=\rho {\bf u}={\bf J},
\end{eqnarray}
the energy flux
\begin{eqnarray}
\label{em1b17}
-T_{0i}=-\frac{\partial\rho}{\partial
t}\frac{\hbar^2}{4m^2}\frac{\nabla\rho}{\rho}-\frac{\partial S}{\partial
t}\frac{\rho}{m^2}\nabla S,
\end{eqnarray}
and the momentum fluxes
\begin{eqnarray}
\label{em1b19}
T_{ij}&=&\frac{\hbar^2}{4m^2}\frac{1}{\rho}\partial_i\rho\partial_j\rho+\frac{
\rho}{m^2}\partial_iS\partial_jS+{\cal L}\delta_{ij}\nonumber\\
&=&\rho u_i
u_j+\frac{\hbar^2}{4m^2}\frac{1}{\rho}\partial_i\rho\partial_j\rho+{\cal
L}\delta_{ij}.
\end{eqnarray}
These are their general expressions. If we use the quantum
Hamilton-Jacobi (or Bernoulli) equation
(\ref{mad7}), which is obtained after extremizing the action, we can rewrite Eq.
(\ref{em1b17}) as Eq. (\ref{stress6}). Similarly,
if we use the expression (\ref{lh12b}) of the Lagrangian which also relies on
the quantum Hamilton-Jacobi (or Bernoulli)
equation, we can rewrite Eq. (\ref{em1b19}) as Eq. (\ref{stress15}).
In this manner, we recover the
equations of Appendix \ref{sec_stressh} (up to terms that vanish by
integration).

{\it Remark:} The energy density is
\begin{eqnarray}
\label{em1b7w}
T_{00}=\dot S \frac{\partial{\cal L}}{\partial\dot S}-{\cal
L}=\pi\dot S -{\cal L}=-\frac{\rho}{m}\dot S-{\cal L},
\end{eqnarray}
where $\pi=\partial {\cal L}/{\partial\dot S}=-\rho/m$ is
the  conjugate momentum to $S$. This leads to Eqs. (\ref{lh11}) and
(\ref{lh12}). On the other hand, Eq. (\ref{lh12g}) can be rewritten in the form
of Hamilton equations
\begin{eqnarray}
\label{lh12ga}
\frac{\partial S}{\partial t}=\frac{\delta H}{\delta \pi},\qquad \frac{\partial
\pi}{\partial t}=-\frac{\delta H}{\delta S}.
\end{eqnarray}
They are equivalent to the continuity equation (\ref{mad6}) and to the quantum
Hamilton-Jacobi (or Bernoulli) equation (\ref{mad7}). 

\section{Conservation laws for a relativistic complex SF}
\label{sec_stressrelat}

In this Appendix, we establish the local conservation laws of boson number
(charge), impulse and energy for a
relativistic complex SF (see Sec. \ref{sec_ra}). For simplicity, we consider 
flat metric and a static background.

\subsection{Relativistic Lagrangian in terms of the wave function}

The density Lagrangian of a complex SF is [see Eq. (\ref{csf2})]
\begin{eqnarray}
\label{re1}
{\cal L}=\frac{1}{2c^2}\left | \frac{\partial\varphi}{\partial t}\right
|^2-\frac{1}{2}|\nabla\varphi|^2-V_{\rm tot}(|\varphi|^2).
\end{eqnarray}

The components of the current (\ref{j1}) or  (\ref{charge1})
are
\begin{eqnarray}
\label{re7c}
J^0=-\frac{m}{2i\hbar c}\left (\varphi^*\frac{\partial\varphi}{\partial
t}-\varphi\frac{\partial\varphi^*}{\partial t}\right ),
\end{eqnarray}
\begin{eqnarray}
\label{re7d}
J^i=\frac{m}{2i\hbar}\left
(\varphi^*\partial_i\varphi-\varphi\partial_i\varphi^*\right ).
\end{eqnarray}
The local
conservation of the  boson number
(charge) can be written as
\begin{eqnarray}
\label{j4b}
\frac{1}{c}\frac{\partial J^0}{\partial t}+\partial_iJ^i=0.
\end{eqnarray}

The components of the energy-momentum tensor (\ref{em1bb}) are
\begin{eqnarray}
\label{re2}
T^{00}=\frac{1}{2c^2}\left |
\frac{\partial\varphi}{\partial t}\right |^2+\frac{1}{2}|\nabla\varphi|^2+V_{\rm
tot}(|\varphi|^2),
\end{eqnarray}
\begin{eqnarray}
\label{re3}
T^{0i}=-\frac{1}{2c}\frac{\partial\varphi^*}{\partial
t}\partial_i\varphi-\frac{1}{2c}\frac{\partial\varphi}{\partial
t}\partial_i\varphi^*,
\end{eqnarray}
\begin{eqnarray}
\label{re4}
T^{ij}=\frac{1}{2}\partial_i\varphi^*
\partial_j\varphi+\frac{1}{2}\partial_i\varphi\partial_j\varphi^*+\delta_{ij}{
\cal L}.
\end{eqnarray}
We note that
$T^{ii}=|\nabla\varphi|^2+3{\cal L}$. The conservation of the energy-momentum
tensor [see Eq. (\ref{lh2r})] can be
written as 
\begin{eqnarray}
\label{re5}
\frac{1}{c}\frac{\partial
T^{00}}{\partial t}+\partial_i T^{0i}=0,
\end{eqnarray}
\begin{eqnarray}
\label{re6}
\frac{1}{c}\frac{\partial
T^{i0}}{\partial t}+\partial_j T^{ij}=0.
\end{eqnarray}

The Euler-Lagrange equation (\ref{lh2rel}) yields the KG equation
\begin{eqnarray}
\label{re7}
\frac{1}{c^2}\frac{\partial^2\varphi}{\partial
t^2}-\Delta\varphi+2\frac{dV_{\rm tot}}{d|\varphi|^2}\varphi=0
\end{eqnarray}
which displays the d'Alembertian operator
$\square=\frac{1}{c^2}\frac{\partial^2}{\partial
t^2}-\Delta$.

In the homogeneous case, the foregoing equations reduce
\begin{eqnarray}
\label{re8}
{\cal L}=\frac{1}{2c^2}\left | \frac{\partial\varphi}{\partial t}\right
|^2-V_{\rm tot}(|\varphi|^2),
\end{eqnarray}
\begin{eqnarray}
\label{re9}
T^{00}=\frac{1}{2c^2}\left | \frac{\partial\varphi}{\partial t}\right
|^2+V_{\rm tot}(|\varphi|^2),
\end{eqnarray}
\begin{eqnarray}
\label{re10}
T^{0i}=0,
\end{eqnarray}
\begin{eqnarray}
\label{re11}
T^{ij}=\delta_{ij}{\cal L},
\end{eqnarray}
\begin{eqnarray}
\label{re12}
\frac{1}{c^2}\frac{\partial^2\varphi}{\partial
t^2}+2\frac{dV_{\rm tot}}{d|\varphi|^2}\varphi=0.
\end{eqnarray}
The energy-momentum tensor is diagonal $T^{\mu\nu}={\rm
diag}(\epsilon,P,P,P)$. The energy density
and the pressure
are given by
\begin{eqnarray}
\label{re13}
\epsilon=\frac{1}{2c^2}\left | \frac{\partial\varphi}{\partial t}\right
|^2+V_{\rm tot}(|\varphi|^2),
\end{eqnarray}
\begin{eqnarray}
\label{re14}
P=\frac{1}{2c^2}\left | \frac{\partial\varphi}{\partial t}\right
|^2-V_{\rm tot}(|\varphi|^2).
\end{eqnarray}

To obtain the nonrelativistic limit, we first make the Klein transformation
(\ref{klein}) and use Eq. (\ref{csf3}). We get
\begin{eqnarray}
\label{re15}
{\cal L}=\frac{\hbar^2}{2m^2c^2}\left |
\frac{\partial\psi}{\partial t}\right
|^2+\frac {i\hbar}{2m}\left
(\psi^*\frac{\partial\psi}{\partial t}-\psi \frac{\partial\psi^*}{\partial
t}\right )\nonumber\\
-\frac{\hbar^2}{2m^2}
|\nabla\psi|^2-V(|\psi|^2),
\end{eqnarray}
\begin{eqnarray}
\label{re15a}
J^0=-\frac{\hbar}{2icm}\left (\psi^*\frac{\partial\psi}{\partial
t}-\psi\frac{\partial\psi^*}{\partial t}-\frac{2imc^2}{\hbar}|\psi|^2\right ),
\end{eqnarray}
\begin{eqnarray}
\label{re15b}
J^i=\frac{\hbar}{2im}\left
(\psi^*\partial_i\psi-\psi\partial_i\psi^*\right ),
\end{eqnarray}
\begin{eqnarray}
\label{re16}
T^{00}=\frac{\hbar^2}{2m^2c^2}\left |
\frac{\partial\psi}{\partial t}\right
|^2+\frac{\hbar^2}{2m^2}
|\nabla\psi|^2+c^2|\psi|^2+V(|\psi|^2)\nonumber\\
+\frac { i\hbar } { 2m}\left
(\frac{\partial\psi}{\partial t}\psi^*-\psi \frac{\partial\psi^*}{\partial
t}\right ),\qquad 
\end{eqnarray}
\begin{equation}
\label{re17}
T^{0i}=-\frac{\hbar^2}{2m^2c}\left (\frac{\partial\psi^*}{\partial
t}\partial_i\psi+\frac{\partial\psi}{\partial
t}\partial_i\psi^*\right )-\frac{i\hbar c}{2m}\left
(\psi^*\partial_i\psi-\psi\partial_i\psi^*\right ),
\end{equation}
\begin{equation}
\label{re18}
T^{ij}=\frac{\hbar^2}{m^2}{\rm
Re}\left (\partial_i\psi\partial_j\psi^*\right
)+\delta_{ij} {\cal L}.
\end{equation}
If we take the limit $c\rightarrow
+\infty$ in Eq. (\ref{re15}) we recover Eq. (\ref{lh1}). If we divide
Eq. (\ref{re15a}) by $c$ and take the limit $c\rightarrow
+\infty$, we obtain $J^0/c=|\psi|^2$ leading to Eq. (\ref{j2}).  Equation
(\ref{re15b}) is equivalent to Eq. (\ref{j3}). To leading order, Eq.
(\ref{re16}) gives $T^{00}\sim \rho c^2$. If we subtract the
rest mass term $cJ^0$ (see Ref. \cite{landaulifshitz})
and take the limit $c\rightarrow
+\infty$ in Eq. (\ref{re16}), we recover  Eq. (\ref{em1b8}). If we multiply
or divide Eq. (\ref{re17}) by $c$ and consider the
terms
that are independent of $c$  (see Ref. \cite{landaulifshitz}) we get
\begin{equation}
\label{re20}
\frac{T^{0i}}{c}=-\frac{i\hbar}{2m}\left
(\psi^*\partial_i\psi-\psi\partial_i\psi^*\right ),
\end{equation}
\begin{equation}
\label{re21}
T^{0i}c=-\frac{\hbar^2}{2m^2}\left (\frac{\partial\psi^*}{\partial
t}\partial_i\psi+\frac{\partial\psi}{\partial
t}\partial_i\psi^*\right ).
\end{equation}
This returns Eqs. (\ref{em1b9}) and (\ref{em1b10}).  Equation (\ref{re18})
returns Eq. (\ref{em1b11}). Finally, the KG
equation (\ref{re7}) becomes
\begin{equation}
\label{re21b}
i\hbar\frac{\partial\psi}{\partial
t}-\frac{\hbar^2}{2mc^2}\frac{\partial^2\psi}{\partial
t^2}+\frac{\hbar^2}{2m}\Delta\psi-m\frac{dV}{d|\psi|^2}\psi=0.
\end{equation}
In the nonrelativistic limit $c\rightarrow +\infty$, we recover the GP equation
(\ref{mfgp9new}).

{\it Remark:} We note that the energy-momentum tensor is symmetric in
relativity theory ($T^{\mu\nu}=T^{\nu\mu}$) while it is not symmetric in
Newtonian theory ($T_{0i}\neq T_{i0}$). This is because space and time are not
treated on equal footing in Newtonian theory. This is also why we have to
consider the two terms $T^{0i}/{c}$ and $T^{0i}{c}$ in the nonrelativistic
limit [see Eqs. (\ref{re20}) and (\ref{re21})].

\subsection{Relativistic Lagrangian in terms of hydrodynamic variables}

The density Lagrangian of a complex SF in its hydrodynamic representation is
[see Eq. (\ref{db5})]
\begin{eqnarray}
\label{jre1}
{\cal L}=\frac{\hbar^2}{8\rho m^2c^2}\left (\frac{\partial\rho}{\partial
t}\right )^2+\frac{1}{2c^2}\rho \left (\frac{\partial\theta}{\partial
t}\right )^2\nonumber\\
 -  \frac{\hbar^2}{8\rho m^2}(\nabla\rho)^2
-\frac{1}{2}\rho (\nabla\theta)^2-V_{\rm tot}(\rho).
\end{eqnarray}

The components of the current (\ref{charge4}) are
\begin{eqnarray}
\label{jre8}
J^0=-\frac{\rho}{c}\frac{\partial\theta}{\partial t},\qquad
J^i=\rho\partial_i\theta.
\end{eqnarray}

The components of the energy-momentum tensor (\ref{tmunuh}) are
\begin{eqnarray}
\label{jre2}
T^{00}=\frac{\hbar^2}{8\rho m^2c^2}\left
(\frac{\partial\rho}{\partial
t}\right )^2+\frac{1}{2c^2}\rho \left (\frac{\partial\theta}{\partial
t}\right )^2\nonumber\\
 +  \frac{\hbar^2}{8\rho m^2}(\nabla\rho)^2
+\frac{1}{2}\rho (\nabla\theta)^2+V_{\rm tot}(\rho),
\end{eqnarray}
\begin{eqnarray}
\label{jre3}
T^{0i}=-\frac{1}{c}\rho\frac{\partial\theta}{\partial
t}\partial_i\theta-\frac{\hbar^2}{4\rho m^2 c}\frac{\partial\rho}{\partial
t}\partial_i\rho,
\end{eqnarray}
\begin{eqnarray}
\label{jre4}
T^{ij}=\frac{\hbar^2}{4\rho m^2}\partial_i\rho \partial_j\rho
+\rho\partial_i\theta\partial_j\theta+\delta_{ij}{\cal L}.
\end{eqnarray}

The Euler-Lagrange equations (\ref{db6}) and (\ref{db7})  yield the
continuity equation
\begin{eqnarray}
\label{jre5}
\frac{1}{c^2}\frac{\partial}{\partial
t}\left (\rho\frac{\partial\theta}{\partial
t}\right )-\nabla\cdot (\rho \nabla\theta)=0 
\end{eqnarray}
and the quantum Hamilton-Jacobi (or Bernoulli) equation
\begin{eqnarray}
\label{jre6}
\frac{1}{2c^2}\left (\frac{\partial\theta}{\partial
t}\right )^2-\frac{1}{2}(\nabla\theta)^2-\frac{Q}{m}-V'_{\rm tot}(\rho)=0,
\end{eqnarray}
with the quantum potential 
\begin{eqnarray}
\label{jre7}
Q\equiv
\frac{\hbar^2}{2m}\frac{\square\sqrt{\rho}}{\sqrt{\rho}}&=&\frac{\hbar^2}{4\rho
mc^2}\frac{\partial^2\rho}{\partial t^2}-\frac{\hbar^2}{8\rho^2
mc^2}\left (\frac{\partial\rho}{\partial t}\right
)^2\nonumber\\
&-&\frac{\hbar^2}{4\rho
m}\Delta\rho+\frac{\hbar^2}{8\rho^2
m}(\nabla\rho)^2.
\end{eqnarray}
According to Eq. (\ref{db12}), the pseudo energy and the
pseudo
velocity are
\begin{eqnarray}
\label{jre18}
v^0=v_0=-\frac{1}{c}\frac{\partial\theta}{\partial t},\qquad
v^i=-v_i=\partial_i\theta.
\end{eqnarray}
Recalling that $\theta=S_{\rm tot}/m$, we get 
\begin{eqnarray}
\label{jre19}
E_{\rm tot}=-\frac{\partial S_{\rm tot}}{\partial t},\qquad {\bf
v}=\frac{\nabla S_{\rm tot}}{m}.
\end{eqnarray}
We also have 
\begin{eqnarray}
J^0=\frac{\rho E_{\rm tot}}{mc},\qquad {\bf
J}=\rho {\bf v}.
\end{eqnarray}
The continuity equation
(\ref{jre5}) and the quantum Hamilton-Jacobi (or Bernoulli) equation
(\ref{jre6})
can be rewritten as  
\begin{eqnarray}
\label{jre20}
\frac{\partial}{\partial t}\left (\rho\frac{E_{\rm tot}}{mc^2}\right
)+\nabla\cdot (\rho {\bf v})=0 
\end{eqnarray}
and
\begin{eqnarray}
\label{jre21}
\frac{E_{\rm tot}^2}{2m^2c^2}-\frac{{\bf v}^2}{2}-\frac{Q}{m}-V'_{\rm
tot}(\rho)=0.
\end{eqnarray}
Taking the gradient of Eq.
(\ref{jre21}), we obtain the Euler equation 
\begin{eqnarray}
\label{eumar}
\frac{E_{\rm tot}}{mc^2}\frac{\partial {\bf
v}}{\partial
t}+({\bf v}\cdot\nabla){\bf v}=-\frac{1}{m}\nabla Q-\frac{1}{\rho}\nabla P.
\end{eqnarray}

In the homogeneous case, the foregoing equations reduce to
\begin{eqnarray}
\label{jre9}
{\cal L}=\frac{\hbar^2}{8\rho m^2c^2}\left (\frac{d\rho}{d
t}\right )^2+\frac{1}{2c^2}\rho \left (\frac{d\theta}{d
t}\right )^2-V_{\rm tot}(\rho),
\end{eqnarray}
\begin{equation}
\label{jre10}
T^{00}=\frac{\hbar^2}{8\rho m^2c^2}\left
(\frac{d\rho}{d
t}\right )^2+\frac{1}{2c^2}\rho \left (\frac{d\theta}{d
t}\right )^2+V_{\rm tot}(\rho),
\end{equation}
\begin{eqnarray}
\label{jre11}
T^{0i}=0,
\end{eqnarray}
\begin{eqnarray}
\label{jre12}
T^{ij}=\delta_{ij}{\cal L},
\end{eqnarray}
\begin{eqnarray}
\label{jre13}
\frac{d}{d
t}\left (\rho\frac{d\theta}{d t}\right )=0,
\end{eqnarray}
\begin{eqnarray}
\label{jre14}
\frac{1}{2c^2}\left (\frac{d\theta}{d
t}\right )^2-\frac{Q}{m}-V'_{\rm tot}(\rho)=0,
\end{eqnarray}
with the quantum potential 
\begin{equation}
\label{jre15}
Q\equiv
\frac{\hbar^2}{2mc^2\sqrt{\rho}}\frac{d^2\sqrt{\rho}}{dt^2}=\frac{\hbar^2}{
4\rho
mc^2}\frac{d^2\rho}{d t^2}-\frac{\hbar^2}{8\rho^2
mc^2}\left (\frac{d\rho}{d t}\right
)^2.
\end{equation}
The energy-momentum tensor is diagonal $T^{\mu\nu}={\rm
diag}(\epsilon,P,P,P)$. 
 The energy
and the pressure are given by
\begin{equation}
\label{jre16}
\epsilon=\frac{\hbar^2}{8\rho m^2c^2}\left
(\frac{d\rho}{d
t}\right )^2+\frac{1}{2c^2}\rho \left (\frac{d\theta}{d
t}\right )^2+V_{\rm tot}(\rho),
\end{equation}
\begin{eqnarray}
\label{jre17}
P=\frac{\hbar^2}{8\rho m^2c^2}\left (\frac{d\rho}{d
t}\right )^2+\frac{1}{2c^2}\rho \left (\frac{d\theta}{d
t}\right )^2-V_{\rm tot}(\rho).
\end{eqnarray}

To obtain the nonrelativistic limit, we first make the Klein transformation (see
Sec. \ref{sec_nr})
\begin{eqnarray}
\label{jre22}
m\theta=S-mc^2 t,
\end{eqnarray}
and use Eq. (\ref{db10}). We get
\begin{eqnarray}
\label{jre23}
{\cal L}=\frac{\hbar^2}{8\rho m^2c^2}\left (\frac{\partial\rho}{\partial
t}\right )^2+\frac{\rho}{2 m^2c^2}\left (\frac{\partial S}{\partial
t}\right )^2-\frac{\rho}{m}\frac{\partial S}{\partial
t}\nonumber\\
-\frac{\hbar^2}{8\rho m^2}(\nabla\rho)^2
-\frac{\rho}{2m^2}(\nabla
S)^2-V(\rho),\qquad 
\end{eqnarray}
\begin{eqnarray}
\label{vac1}
J^0=-\frac{\rho}{mc}\frac{\partial S}{\partial t}+\rho c,
\end{eqnarray}
\begin{eqnarray}
\label{vac2}
J^i=\rho \frac{\partial_i S}{m}=\rho u^i,
\end{eqnarray}
\begin{eqnarray}
\label{jre24}
T^{00}=\frac{\hbar^2}{8\rho m^2c^2}\left (\frac{\partial\rho}{\partial
t}\right )^2+\frac{\rho}{2 m^2c^2}\left (\frac{\partial S}{\partial
t}\right )^2+\rho c^2-\frac{\rho}{m}\frac{\partial S}{\partial
t}\nonumber\\
+\frac{\hbar^2}{8\rho m^2}(\nabla\rho)^2+\frac{\rho}{2m^2}(\nabla
S)^2+V(\rho),\qquad 
\end{eqnarray}
\begin{equation}
\label{jre25}
T^{0i}=\rho\frac{\partial_iS}{m}c-\frac{\rho}{m^2c}\frac{\partial
S}{\partial t}\partial_iS-\frac{\hbar^2}{4\rho
m^2 c}\frac{\partial\rho}{\partial t}\partial_i\rho,
\end{equation}
\begin{equation}
\label{jre26}
T^{ij}=\frac{\hbar^2}{4\rho
m^2}\partial_i\rho\partial_j\rho+\frac{\rho}{m^2}
\partial_i S\partial_j S+\delta_{ij} {\cal L}. 
\end{equation}
If take the limit $c\rightarrow
+\infty$ in Eq. (\ref{jre23}) we recover Eq. (\ref{lh8}). 
If we divide
Eq. (\ref{vac1}) by $c$ and take the limit $c\rightarrow
+\infty$, we obtain $J^0/c=\rho$ leading to Eq. (\ref{j2bn}). Equation
(\ref{vac2}) is equivalent to Eq. (\ref{j3bn}). To leading order, Eq.
(\ref{jre24}) gives $T^{00}\sim \rho c^2$. If we subtract the
rest mass term $cJ^0$ (see Ref. \cite{landaulifshitz})
and take the limit $c\rightarrow
+\infty$ in Eq. (\ref{jre24}), we recover  Eq. (\ref{em1b14}). If we
multiply or
divide Eq. (\ref{jre25}) by $c$ and consider the terms
that are independent of $c$ (see Ref. \cite{landaulifshitz}) we get 
\begin{equation}
\label{jre28}
\frac{T^{0i}}{c}=\rho\frac{\partial_iS}{m},
\end{equation}
\begin{equation}
\label{jre29}
T^{0i}c=-\frac{\rho}{m^2}\frac{\partial S}{\partial
t}\partial_iS-\frac{\hbar^2}{4\rho
m^2}\frac{\partial\rho}{\partial t}\partial_i\rho.
\end{equation}
This returns Eqs. (\ref{em1b16}) and (\ref{em1b17}).  Equation (\ref{jre26})
returns Eq. (\ref{em1b19}). Finally, the 
continuity equation (\ref{jre5}) and the quantum Hamilton-Jacobi (or Bernoulli)
equation (\ref{jre6}) become
\begin{eqnarray}
\label{jre30}
-\frac{1}{mc^2}\frac{\partial}{\partial
t}\left (\rho\frac{\partial S}{\partial
t}\right )+\frac{\partial \rho}{\partial t}+\nabla\cdot \left (\rho \frac{\nabla
S}{m}\right )=0
\end{eqnarray}
and
\begin{equation}
\label{jre31}
-\frac{1}{2mc^2}\left (\frac{\partial S}{\partial
t}\right
)^2+\frac{\partial
S}{\partial t}+\frac{1}{2m}(\nabla S)^2+Q+mV'(\rho)=0.
\end{equation}
Using the identities
\begin{eqnarray}
S_{\rm tot}=S-mc^2 t,\qquad E_{\rm tot}=E+mc^2,
\end{eqnarray}
\begin{eqnarray}
E=-\frac{\partial S}{\partial t},\qquad {\bf
u}=\frac{\nabla
S}{m},
\end{eqnarray}
\begin{eqnarray}
J^0=\frac{\rho E}{mc}+\rho c, \qquad {\bf J}=\rho {\bf u},
\end{eqnarray}
they can be rewritten
as
\begin{eqnarray}
\label{jre30m}
\frac{1}{mc^2}\frac{\partial}{\partial
t}\left (\rho E\right )+\frac{\partial \rho}{\partial
t}+\nabla\cdot \left (\rho {\bf u}\right )=0
\end{eqnarray}
and
\begin{equation}
\label{jre31m}
-\frac{E^2}{2mc^2}-E+\frac{1}{2}m {\bf u}^2+Q+mV'(\rho)=0.
\end{equation}
Taking the gradient of Eq. (\ref{jre31m}) we obtain the Euler equation
\begin{equation}
\label{jre31b}
\frac{\partial {\bf u}}{\partial t}+({\bf u}\cdot \nabla){\bf
u}=-\frac{1}{\rho}\nabla P-\frac{1}{m}\nabla
Q+\frac{1}{2m^2c^2}\nabla(E^2).
\end{equation}
In the limit $c\rightarrow +\infty$, we recover Eqs. (\ref{mad6}),
(\ref{mad7}) and (\ref{mad10}).

\end{document}